\documentclass[nofootinbib, showkeys, unsortedaddress, superscriptaddress, notitlepage]{revtex4-1}
\usepackage{graphicx}
\usepackage{subcaption}
\usepackage{amsmath}
\usepackage{amsthm}
\usepackage{dsfont}
\usepackage{mathptmx}
\usepackage{geometry}
\usepackage{physics}
\usepackage{csquotes}
\usepackage{xcolor}
\usepackage{MnSymbol}
\usepackage{enumerate}

\usepackage[framemethod=TikZ]{mdframed}
\theoremstyle{break}
\newmdtheoremenv[%
linecolor=black,leftmargin=5,%
rightmargin=5,
backgroundcolor=gray!10,%
innertopmargin=5pt,%
innerbottommargin=5pt,%
ntheorem,%
outerlinewidth=.8]{mythm}{Proposition}[subsection]

\newmdtheoremenv[%
linecolor=black,leftmargin=5,%
rightmargin=5,
backgroundcolor=gray!10,%
innertopmargin=5pt,%
innerbottommargin=5pt,%
ntheorem,%
outerlinewidth=.8]{mylemma}[mythm]{Lemma}

\newtheorem{myrmk}[mythm]{Remark}
\newtheorem{myexpl}[mythm]{Example}
\newenvironment{myproof}[2] {\paragraph*{Derivation of {#1} :}}{
\hfill$\square$}

\newcommand{\LGram}[2]{\left\llangle #1 \middle\vert #2 \right\rrangle}
\newcommand{\EE}{\mathds{E}}
\newcommand{\RR}{\mathds{R}}

\newcommand{\ZZ}{\mathds{Z}}
\newcommand{\NN}{\mathds{N}}
\newcommand{\PP}{\mathds{P}}
\newcommand{\eps}{\varepsilon}
\DeclareMathOperator*{\argmin}{arg\,min}
\DeclareMathOperator{\Det}{Det}

\DeclareMathOperator{\vol}{vol}

\makeatletter
\renewcommand\@makecaption[2]{%
  \par
  \vskip\abovecaptionskip
  \begingroup
   \small\rmfamily
    \begingroup
     \samepage
     \flushing
     \let\footnote\@footnotemark@gobble
     \@make@capt@title{#1}{#2}\par
    \endgroup
  \endgroup
  \vskip\belowcaptionskip
}
\makeatother

\makeatletter
\renewcommand{\p@subsection}{}
\renewcommand{\p@subsubsection}{}
\makeatother

\usepackage{chngcntr}

\usepackage{hyperref}

\begin{document}

\title{Symmetries and zero modes in sample path large deviations}

\author{T. Schorlepp}
\email{Timo.Schorlepp@rub.de}
\affiliation{Institute for Theoretical Physics I, Ruhr-University Bochum,
             Universit{\"a}tsstrasse 150,
             44801 Bochum, Germany}
\author{T. Grafke}
\email{T.Grafke@warwick.ac.uk}
\affiliation{Mathematics Institute, University of Warwick, Coventry CV4 7AL, United Kingdom}
\author{R. Grauer}
\email{grauer@tp1.rub.de}
\affiliation{Institute for Theoretical Physics I, Ruhr-University Bochum,
             Universit{\"a}tsstrasse 150,
             44801 Bochum, Germany}

\date{\today}


\begin{abstract}

Sharp large deviation estimates for stochastic differential equations with small noise,
based on minimizing the Freidlin-Wentzell action functional under appropriate boundary
conditions, can be obtained by integrating certain matrix Riccati differential equations
along the large deviation minimizers or instantons, either forward or backward in time.
Previous works in this direction often
rely on the existence of isolated minimizers with positive definite second variation.
By adopting techniques from field theory and explicitly evaluating the large deviation
prefactors as functional determinant ratios using Forman's theorem, we extend the approach
to general systems where degenerate submanifolds of minimizers exist. The key technique
for this is a boundary-type regularization of the second variation operator.
This extension is particularly relevant if the system possesses continuous
symmetries that are broken by the instantons. We find that removing the
vanishing eigenvalues associated with the zero modes is possible within
the Riccati formulation and amounts to modifying the initial or final conditions and
evaluation of the Riccati matrices. We apply our results in multiple examples
including a dynamical phase transition for the average surface height in short-time
large deviations of the one-dimensional Kardar-Parisi-Zhang equation with flat
initial profile.

\end{abstract}


\maketitle


\tableofcontents


\section{Introduction}

In its classical formulation, large deviation theory (LDT)
is often used to gain access to the limiting behavior of probabilities or
expectations at an approximate, i.e.\ exponential scale, which is the content of
notions such as large deviation principles in general or
Varadhan's lemma (see e.g.~\cite{dembo-zeitouni:2010}). However,
in any practical application where quantitative estimates are required, it
is desirable to refine such an analysis to get absolute and asymptotically
correctly normalized results instead of mere scaling for the probabilities
of rare events, effectively supplementing the \textit{exponential} LDT
estimate by a sub-exponential \textit{prefactor}. Such precise Laplace
asymptotics, which are the subject of this paper for the specific scenario
of stochastic differential equations
(SDEs) subject to small Gaussian noise, have a long
history~\cite{piterbarg-fatalov:1995}.\\

In the past decades, sample path LDT or Freidlin-Wentzell
theory~\cite{freidlin-wentzell:2012} and the related notion of
instanton calculus in theoretical
physics~\cite{coleman:1979,vainshtein-zakharov-novikov-etal:1982} have
been widely applied as a tool to study rare event probabilities in
stochastic dynamical systems, either numerically,
e.g.\ in~\cite{chernykh-stepanov:2001,e-ren-vanden-eijnden:2004,bouchet-laurie-zaboronski:2011,grafke-etal:2015,dematteis-grafke-onorato-etal:2019},
or through analytical analysis of the corresponding minimization
problems,
e.g.\ in~\cite{gurarie-migdal:1996,balkovsky-etal:1997,deuschel-etal-1:2014,*deuschel-etal-2:2014,
  krajenbrink-doussal:2021}.  Reviews of the theory, highlighting
connections of large deviation theory to field-theoretic methods and
optimal fluctuations or instantons in theoretical physics are given
by~\cite{touchette:2009,grafke-grauer-schaefer:2015,grafke-vanden-eijnden:2019}.
For the metastable setup for reversible systems prefactor corrections
are classical~\cite{eyring:1935, kramers:1940}, and recent
generalizations and rigorous progress has been
made~\cite{bovier-eckhoff-gayrard-etal:2004, berglund:2013, berglund-di_gesu-weber:2017,
  bouchet-reygner:2016, landim-seo:2018}. With some notable exceptions
such as~\cite{lehmann-reimann-haenggi:2003,nickelsen-engel:2011,nickelsen-touchette:2022},
however, most of the work for general irreversible systems and extreme
events has focused only on exponential asymptotics using the large
deviation minimizers themselves, solution to a deterministic
optimization problem.  As an additional, concrete motivation to go
beyond such rough estimates in practical applications, it has been
pointed out very recently that for assessing the relative importance
of different instantonic transition paths, knowledge of the LDT
prefactor at leading order may be vital even at comparably small noise
strengths~\cite{kikuchi-adhikari-kappler:2022}.\\

In the last year, there has been a lot of activity to provide generic numerical
tools that also allow for the computation of the leading order term of the large
deviation prefactor for the statistics of final
time \textit{observables} of small noise ordinary SDEs using symmetric
Riccati matrix differential equations, either forward or backward in
time~\cite{schorlepp-grafke-grauer:2021,grafke-schaefer-vanden-eijnden:2021,ferre-grafke:2021,bouchet-reygner:2022}. In an abstract setting, expressions for
prefactors in this context, even at arbitrarily high order, have already been
known rigorously since the
1980's~\cite{ellis-rosen:1981,ellis-rosen:1982,ben-arous:1988,piterbarg-fatalov:1995} and are, not surprisingly, related to a certain operator determinant at
the leading order. The Riccati formalism then allows one to compute such
determinants in a closed form through the solution of an initial value
problem instead of eigenvalue computations
(see~\cite{tong-vanden-eijnden-stadler:2021} for a recent work in the
latter direction, as well as~\cite{psaros-kougioumtzoglou:2020}), much in the
spirit of the classical Gel'fand-Yaglom technique
in quantum mechanics~\cite{gelfand-yaglom:1960} or its later generalization
via Forman's theorem~\cite{forman:1987}. This is advantageous if either, from
a numerical point of view, the spatial dimension of the system is not too
large, with the Riccati matrix being of size $n \times n$ for a $n$-dimensional
SDE, or if an analytical analysis of the resulting equations is desired. Our
first contribution in this paper is to make the connection to functional
determinants more precise and to add to the existing derivations of the Riccati
equations using (i) a WKB analysis of the Kolmogorov backward
equation~\cite{grafke-schaefer-vanden-eijnden:2021} (ii) a discretization
approach of the path integral~\cite{schorlepp-grafke-grauer:2021} or (iii)
the use of the Feynman-Kac formula for Gaussian
fluctuations~\cite{schorlepp-grafke-grauer:2021,bouchet-reygner:2022} a fourth
derivation that makes explicit use of Forman's theorem. Furthermore, in contrast to
previous derivations, we also include the case of It{\^o} SDEs with
multiplicative noise here. In general, we
stress the technical advantage of working with the moment-generating
function (MGF) as the principal quantity of interest here, only later transforming
onto probabilities or probability density functions~(PDFs).\\

This groundwork then opens the way to treat a new class of problems using
Riccati equations compared to the previous works. Notably, all of the cited
previous works on this approach have been limited to unique or at least
isolated large deviation minimizers with positive definite second variation
of the associated functional at the minimizers. In contrast to this, we extend
the Riccati approach to cases where compact submanifolds of minimizers exist.
There, the application of the infinite-dimensional Laplace method requires the
removal of the zero eigenvalues of the corresponding second variation operator,
as discussed in a general setting in~\cite{ellis-rosen:1981} already. The
eigenfunctions corresponding to these zero eigenvalues are usually called
\textit{zero modes}. In the
context of mean transition times in the small noise limit, a paper that deals
with related problems is~\cite{berglund-gentz:2010}. Carrying out the procedure
described above through a boundary-value type regularization that builds
on the work of~\cite{falco-fedorenko-gruzberg:2017} among others, we obtain
Riccati equations with suitably regularized initial or final conditions in this
paper that implicitly remove the divergences that would otherwise be encountered
in the solution of the Riccati equations.\\

Situations where degenerate families of instantons exist are in fact far from
pathological. Importantly, many stochastic dynamical systems, in particular
stochastic partial differential equations (SPDEs) motivated from physics, possess
certain \textit{symmetries}, such that the equations of motion are invariant
e.g.\ under translations, rotations, Galilei transformations and so forth. If,
in addition to the SDE itself, the observable whose statistics are computed has
the same symmetries, then it is possible to search for unique minimizers or
instantons of the large deviation minimization problem obeying the same symmetry.
Generically, however, the global minimum will not be attained this way, but
instead the true minimizer will \textit{break the symmetry} and hence be
comprised of a family of equivalent possible solutions related by the symmetry
group of the system. Of particular interest is the case of a dynamical phase
transition, where this symmetry breaking happens spontaneously with the
extremeness of the rare event under consideration as the control parameter.
Relevant examples of this phenomenon in the context of
sample path LDT include the one-dimensional
Kardar-Parisi-Zhang (KPZ)
equation~\cite{janas-kamenev-meerson:2016,krajenbrink-doussal:2017,smith-kamenev-meerson:2018,hartmann-meerson-sasorov:2021}
for the surface height at one point in space and with two-sided Brownian motion
initial condition (leading to discrete mirror symmetry breaking),
the two-dimensional~\cite{falkovich-lebedev:2011} and
three-dimensional~\cite{schorlepp-etal:2022}
incompressible Navier-Stokes equations and a Lagrangian turbulence
model~\cite{alqahtani-grafke-grigorio:2021} (all with rotational symmetry breaking).
In all of these cases, due to
the underlying symmetries, it turns out that it suffices to integrate a
single Riccati equation, corresponding to a single reduced functional
determinant evaluation, which thereby allows for a generalization of
earlier results~\cite{schorlepp-grafke-grauer:2021,grafke-schaefer-vanden-eijnden:2021,ferre-grafke:2021,bouchet-reygner:2022} without increasing the
computational costs. In addition to the examples listed above, further systems
where the methods and results of this paper could be applied are those within
the scope of the macroscopic fluctuation theory~\cite{bertini-etal:2015}, e.g.\
the Kipnis-Marchioro-Presutti model on a ring where a dynamical phase
transition for the current due to translational symmetry breaking is known to
occur~\cite{hurtado-garrido:2011,zarfaty-meerson:2016}.\\

Regarding limitations of this paper, we consider only systems where the drift
term of the SDE has a unique, stable fixed point. Further, we do not explicitly
discuss the extension to infinite time intervals which could be done through an
appropriate geometric parameterization~\cite{heymann-vanden-eijnden:2008} that
could be incorporated similar to~\cite{grafke-schaefer-vanden-eijnden:2021}. We
formulate our general results only for ordinary stochastic differential equations
in $\RR^n$, and leave the (at least on a purely formal level) simple extension
towards stochastic partial differential equations to the reader, treating this
extension only by means of an example in this paper. The presentation throughout, which
is based on stochastic path integrals, is not rigorous in favor of
intuition and brevity, while still using a structure in terms of propositions,
lemmas and derivations for clarity.\\

This paper is organized as follows: In section~\ref{sec:prefac-nondegen},
we start with the rederivation of known Riccati matrix results for unique
large deviation minimizers with positive definite second variation. We introduce
the general setup in subsection~\ref{subsec:setup} and give the main results for
prefactors of MGFs in
subsection~\ref{subsec:prefac-mgf-nondegen}. The transformation onto PDF
prefactors is carried out in subsection~\ref{subsec:prefac-nondegen-pdf-trafo}.
Afterwards, section~\ref{sec:prefac-degen} follows the same structure for the
zero mode case. In subsection~\ref{subsec:prefac-degen-motivation}, we briefly
motivate degenerate Laplace asymptotics in finitely many dimensions and then
derive analogous results to subsections~\ref{subsec:prefac-mgf-nondegen}
and~\ref{subsec:prefac-nondegen-pdf-trafo} in
subsection~\ref{subsec:prefac-mgf-degen} and~\ref{subsec:prefac-degen-pdf-trafo}.
Afterwards, we consider four specific examples with degenerate instantons in
section~\ref{sec:examples} and compare the result of our leading order
degenerate Laplace expansion to known theoretical results or direct sampling of
the SDEs at hand. In addition to three finite-dimensional systems, we also deal
with a dynamical phase transition in an irreversible one-dimensional stochastic
partial differential equation (SPDE) in this section, namely the KPZ equation
where we investigate the probability distribution of the average surface height
at short times with flat initial condition.
We conclude the paper with
a discussion of the results and comments on future extensions in
section~\ref{sec:discussion-outlook}.
Appendix~\ref{app:forman-lagrange-to-hamilton} contains the general statement
of Forman's theorem for second order ordinary differential operators as well
as general Lagrangian and Hamiltonian formulations of the theorem for
second variation operators. Appendix~\ref{app:mgf-ldt-estimate} states a general
expression for the MGF prefactor in the non-degenerate case for an arbitrary
continuous time Markov process satisfying a large deviation principle as a reference.
Finally, appendix~\ref{app:kpz-prefac-hom} deals with an analytical computation
for the LDT prefactor in the KPZ equation when expanding around the spatially
homogeneous instantons of subsection~\ref{subsec:pde-example}.


\section{Prefactor in the nondegenerate case}
\label{sec:prefac-nondegen}

\subsection{Freidlin-Wentzell theory setup}
\label{subsec:setup}

For $n \in \NN$ and $\eps > 0$, we consider the It{\^o} SDE
\begin{align}
\dd X^\eps_t = b\left(X^{\eps}_t\right) \dd t + \sqrt{\eps} \sigma\left(X^{\eps}_t\right)
\,\dd B_t\,, \quad X^\eps_0 = x \in \RR^n
\label{eq:sde-general}
\end{align}
on the finite time interval $[0,T]$, $T>0$, with multiplicative
Gaussian noise. We assume that the process starts deterministically at
$x \in \RR^n$.  The drift $b : \RR^n \mapsto \RR^n$ is not necessarily
gradient. We assume it to be sufficiently smooth and to possess only a
single fixed point $x_* \in \RR^n$ which is stable.  The process $B =
\left( B_t \right)_{t \in [0,T]}$ is a standard $n$-dimensional
Brownian motion, and the diffusion matrix $a := \sigma \sigma^\top
\colon \RR^n \to \RR^{n \times n}$, also assumed to be sufficiently
smooth, as well as nonvanishing at $x$, is not necessarily diagonal or
invertible\footnote{We do not attempt to give mathematically strict
conditions on the drift field~$b$, diffusion matrix~$a$ and
observable~$f$ in this paper, which, beyond the existence and
uniqueness of solutions of~\eqref{eq:sde-general}, would also
guarantee the rigorous applicability of the results of the following
sections.  For the case of component projections as observables and
unique instantons, we refer the reader
e.g.\ to~\cite{deuschel-etal-1:2014,*deuschel-etal-2:2014} for works
in this direction.}.\\

We are interested in obtaining precise estimates, as the noise
strength~$\epsilon$ tends to zero, for the PDF~$\rho_f^\eps:\RR
\mapsto [0, \infty)$ of a random
variable $f(X^\eps_T)$ where $f:\RR^{n} \to \RR$ is a possibly
nonlinear \textit{observable} of the process~$X^\eps$ at final time~$T$.
Typically, we are interested in situations where~$n$ is large, as
in the (semi-)discretization of an SPDE,
and~$f$ corresponds to the observation of a real-valued
physical quantity that is characteristic for a process described by an
SPDE, either at a single point in space or averaged over the spatial volume.
In the limit~$\eps \downarrow 0$, it
is intuitive that trajectories $\left(X^\eps_t\right)_{t \in
[0,T]}$ concentrate around the deterministic trajectory~$\phi_0$ solving
\begin{align}
\dot{\phi}_0 = b(\phi_0)\,, \quad \phi_0(0) = x\,.
\end{align}
LDT tells us
that this concentration happens exponentially fast in $\eps$,
and deviations
from this deterministic behavior correspond to rare events.\\

The Freidlin-Wentzell rate (or action) functional that governs the
concentration of the path measure on~$\phi_0$ is given
by~\cite{freidlin-wentzell:2012}
\begin{align}
  S[\phi] = \begin{cases}
    \int_0^T \underbrace{\frac{1}{2} \left<\dot{\phi} - b(\phi), a^{-1}(\phi)
      \left[\dot{\phi} - b(\phi) \right] \right>_n}_{=:L(\phi,
      \dot{\phi})} \dd t \,,
    \quad & \phi \in AC\left([0,T], \RR^n \right),\,
    \dot{\phi} - b(\phi) \in \text{im}(a(\phi))
    \text{ a.e.},\, \phi(0) = x\\
    + \infty, & \text{else}\,,
  \end{cases}
  \label{eq:fw-action}
\end{align}
where $a^{-1}$ is the Moore-Penrose inverse of $a$, $\left< \cdot,
\cdot \right>_{n}$ is the standard Euclidean inner product on~$\RR^n$
and~$AC\left([0,T], \RR^n \right)$ is the space of absolutely continuous paths
$\phi \colon [0,T] \to \RR^n$.
Note that we will treat~$a$ as invertible below, but no final result
will contain any inverse of~$a$, and all results remain valid if the
limit to singular diffusion matrices is considered carefully.  The
asymptotic LDT estimate for the PDF~$\rho_f^\eps$ as~$\eps \downarrow
0$ reads
\begin{align}
  \lim_{\eps \downarrow 0} \eps \log \rho_f^\eps (z) = -
  \inf_{\substack{\phi(0) = x\\f(\phi(T)) = z}} S[\phi] = -S[\phi_z]
  =: - I_f(z)\,.
\label{eq:ldt-rough}
\end{align}
We call~$I_f$ the \textit{rate function} of the observable. The
minimizer~$\phi_z$, also termed the \emph{instanton}, is a
solution to the constrained minimization problem~(\ref{eq:ldt-rough}),
and thus satisfies the first order necessary conditions in Hamiltonian
form (cf.\ the derivation of Proposition~\ref{thm:mgf-prefactor})
\begin{equation}
  \begin{cases}
    \dot \phi_z = b \left(\phi_z \right) + a(\phi_z) \theta_z\,,
    & \phi_z(0) = x\,, \quad f\left(\phi_z(T)\right) = z\\
    \dot \theta_z = - \nabla b\left(\phi_z \right)^\top \theta_z - \tfrac{1}{2} \left \langle \theta_z, \nabla a(\phi_z) \theta_z \right \rangle_n\,,
    & \theta_z(T) = \lambda_z \nabla f\left(\phi_z(T)\right)\,,
  \end{cases}
  \label{eq:instanton-eq-z}
\end{equation}
where $\theta_z = \partial L(\phi_z, \dot \phi_z)
/ \partial \dot{\phi}$ is the conjugate momentum of the instanton
$\phi_z$, and $\lambda_z
\in \RR$ is a Lagrange multiplier, suitably chosen to enforce the
final time constraint $f\left(\phi_z(T)\right) = z$.
Comparing~\eqref{eq:instanton-eq-z} to the SDE~\eqref{eq:sde-general}
indicates that $\eta_z = \sigma^\top(\phi_z) \theta_z$ can be
interpreted as the optimal (in the sense of most likely)
forcing realization that drives the system towards the
outcome $f(X^\eps_T) = z$.\\


The mere exponential scaling estimate from Freidlin-Wentzell theory,
as given in~(\ref{eq:ldt-rough}), can be refined to next order to
obtain a prefactor estimate in the small noise limit. These
refinements rely on the fact that a sample path large deviation
estimate formally
corresponds to an infinite dimensional application of Laplace's
method, and higher order estimates can then be obtained by integrating
the Gaussian integral of the second variation around the minimizer to
obtain a ratio of determinants as prefactor. In this section, we will
rederive the results of~\cite{schorlepp-grafke-grauer:2021,
grafke-schaefer-vanden-eijnden:2021} following this strategy, including
the explicit evaluation of the appearing functional determinants using Forman's
theorem. Importantly, we only consider the case of unique instantons and
positive definite second variations in this section.\\

In section~\ref{sec:prefac-degen}, we will then demonstrate that the
approach can be generalized to SDEs and observables with
\emph{degenerate} instantons which are rendered non-unique due to an
underlying symmetry of the system. While an extension towards multiple
isolated global minimizers of the action functional is trivially
achieved by simply summing over the contributions of each individual
minimizer, we here consider the case of a degenerate family of
instantons that define an $r$-dimensional submanifold ${\cal M}^r_z$
with $r \in \left\{1, \dots, n \right\}$ in the space of all permitted
paths $\phi:[0,T] \mapsto \RR^n$ that fulfill the boundary conditions
$\phi(0) = x$, $f(\phi(T)) = z$, such that the action functional $S$
is globally minimized and constant on ${\cal M}^r_z$. In order to
formally derive an analogue procedure in this case, we will rely on
well-known tools from field theory, where the spontaneous symmetry
breaking of instantons is known to generate zero- or Nambu-Goldstone
modes that need to be explicitly integrated out. The small noise
expansion for sample path large deviations then necessitates removing
zero eigenvalues from the second variation of the action at the
instanton.

\subsection{Moment-generating function prefactor estimates for Freidlin-Wentzell theory with unique instantons}
\label{subsec:prefac-mgf-nondegen}

We define the \textit{moment-generating function} (MGF) of the
real-valued random variable~$f(X^\eps_T)$ as
\begin{align}
A_f^\eps \colon \RR \to [0, \infty]\,, \quad A_f^\eps(\lambda)
:= \EE \left[
\exp\left\{\frac{\lambda}{\eps}
f(X^\eps_T)\right\} \right]\,,
\end{align}
and assume in the remainder of this paper that the scaled
cumulant-generating function
\begin{align}
G_f\colon \RR \to \RR\,, \quad G_f(\lambda) := \lim_{\eps \downarrow 0}
\left[ \eps \log A_f^\eps(\lambda) \right]
\end{align} 
exists in $\RR$ for all $\lambda \in \RR$.
For systems and observables where
this assumption is not fulfilled, a convexification of the rate
function~$I_f$ through a reparameterization of the observable as
in~\citep{alqahtani-grafke:2021} makes our results applicable.\\

We will proceed to derive precise large deviation results for
$A_f^\eps$, which is simpler on a technical level than directly
computing the PDF, and only \textit{afterwards} perform an inverse
Laplace transform onto the PDF, which can again be evaluated by a
saddlepoint approximation as~$\eps \downarrow 0$.\\


\begin{mythm}[Sharp estimates for MGFs via functional determinants]
  \label{thm:mgf-prefactor}
  Denote by $\phi_\lambda$ and $\phi_0$ the instanton and the ``free''
  instanton with conjugate momenta $\theta_\lambda$ and $\theta_0 \equiv 0$,
  unique solutions to the minimization problems
  \begin{equation}
   \argmin_{\phi(0) = x} \left( S[\phi] - \lambda f(\phi(T)) \right) =
    \phi_\lambda \qquad\text{and}\qquad \argmin_{\phi(0) = x}
    S[\phi] = \phi_0
  \end{equation}
  for the Freidlin-Wentzell action~\eqref{eq:fw-action}.
  Further, for variations $\gamma:[0,T] \to \RR^n$, let
  \begin{align}
    \delta^2 S[\phi][\gamma] &= \frac{1}{2} \int_0^T \left \langle \gamma,
    \Omega[\phi] \gamma \right \rangle_n \dd t
  \end{align}
  be the second variation of $S$
  around $\phi$, where the Jacobi operator $\Omega$ is given by
  \begin{align}
  \Omega[\phi] &= \left[- \dv{}{t} - \nabla b(\phi)^\top - \left(\nabla a(\phi)
  \theta\right)^\top \right]
  a^{-1}(\phi) \left[ \dv{}{t} - \nabla b(\phi) - \left(\nabla a(\phi)
  \theta\right) \right] \nonumber \\
  & \quad - \left \langle
  \nabla^2 b(\phi), \theta \right \rangle_n - \frac{1}{2} \left \langle
  \theta, \nabla ^2 a(\phi) \theta \right \rangle_n
  \label{eq:fw-scnd-var-op}
  \end{align}
  and we impose mixed Dirichlet-Robin boundary conditions
  \begin{align}
  {\cal A}_\lambda : \begin{cases}
  \gamma(0) = 0\\
  \zeta(T) = \lambda \nabla^2 f(\phi_\lambda(T))
  \gamma(T) 
  \end{cases} \label{eq:bc-lbda-general}
  \end{align}
  for variations along $\phi = \phi_\lambda$. Here 
  \begin{align}
  \zeta := a^{-1}(\phi) \left[\dv{}{t} - \nabla b(\phi) - \left(\nabla a(\phi)
  \theta\right)\right] \gamma
  \end{align}
  is the conjugate momentum variation associated with~$\gamma$.
  Then we have the
  following sharp asymptotic estimate for $A_f^\eps$:
  \begin{equation}
    A_f^\eps(\lambda)\overset{\eps \downarrow 0}{\sim} R_\lambda
    \exp\left\{-\eps^{-1} \left( S[\phi_\lambda] - \lambda
    f\left(\phi_\lambda(T)\right) \right)\right\}
    \label{eq:mgf-general}
  \end{equation}
  with prefactor
  \begin{align}
  R_\lambda := \left(\frac{\Det_{{\cal A}_\lambda}
  \left(a(\phi_\lambda) \Omega[\phi_\lambda] \right)}
  {\Det_{{\cal A}_0} \left(a(\phi_0) \Omega[\phi_0] \right)}
  \right)^{-1/2} \exp\left\{-\tfrac12 \int_0^T \left(\nabla\cdot
  b(\phi_\lambda) + \trace \left[\nabla a(\phi_\lambda) \theta_\lambda \right] - \nabla\cdot b(\phi_0)\right)\,\dd t\right\}\,.
  \label{eq:ratio-funcdet}
  \end{align}
\end{mythm}


\begin{myrmk}
We set $(\nabla b)_{ij} = \partial_j b_i$ and use the
short-hand notations $\left[\left< \nabla^2 b(\phi),
\theta\right>_{n} \right]_{ij} := \sum_{k=1}^n\partial_i
\partial_j b_k(\phi) \theta_{k}$ as well as
$\left[ \nabla a(\phi) \theta \right]_{ij} = \sum_{k=1}^n
\partial_j a_{ik}(\phi) \theta_k$ and
$\left[\left \langle
\theta, \nabla ^2 a(\phi) \theta \right \rangle_n \right]_{ij}
=\sum_{k=1}^n\sum_{l=1}^n \partial_i \partial_j a_{kl}(\phi)
\theta_k \theta_l$.
The precise meaning of the ratio
of functional determinants
in~(\ref{eq:ratio-funcdet}) will be explained below, where we
will also rederive efficient computational methods in order to evaluate
it. Throughout this paper, we denote functional determinants by~$\Det$
with the boundary conditions under which the determinant is computed
as a subscript, whereas ordinary matrix determinants are written
as~$\det$ with the dimension of the respective matrix as a subscript.
The operator $a$ in the functional determinants
in~(\ref{eq:ratio-funcdet}) is to be understood as pointwise multiplication
with~$a(\phi(t))$ for all $t \in [0,T]$.
\end{myrmk}


\begin{myrmk}
The exponent
\begin{align}
\inf_{\phi(0)=x} \left( S[\phi] - \lambda f(\phi(T))
\right) =  \inf_{z \in \RR} \left(I_f(z) - \lambda z \right) = -G_f(\lambda)
\end{align}
in~\eqref{eq:mgf-general} is (minus) the Legendre-Fenchel transform
of the rate function $I_f$ evaluated at $\lambda$, which yields the
scaled cumulant-generating function and is finite by assumption.
\end{myrmk}


\begin{myproof}{Proposition~\ref{thm:mgf-prefactor}}

We express the MGF $A_f^\eps$ at $\lambda \in \RR$ as a
Wiener path
integral over all realizations of the increments $\eta = \dd B / \dd t$
of the Brownian motion $B$ on $[0,T]$
\begin{align}
A_f^\eps(\lambda) = \frac{\int {\cal D} \eta \; \exp\left\{\frac{\lambda}{\eps}
f(X^\eps_T[\eta])-\frac{1}{2} \int_0^T
\left<\eta, \eta \right>_n \dd t \right\}}{\int {\cal D} \eta \; \exp
\left\{-\frac{1}{2} \int_0^T \left<\eta, \eta \right>_n \dd t \right\}}\,,
\end{align}
where $X^\eps_T[\eta]$ indicates that $X^{\eps}_T$ is a functional of
the realization $\eta$ of the noise, and we divide by the \enquote{free}
path integral $\int {\cal D} \eta \; \exp\left\{-\frac{1}{2} \int_0^T
\left<\eta, \eta \right>_n \dd t \right\}$ to ensure correct normalization
\begin{align}
\EE[1] \overset{!}{=} 1
\end{align}
of the path measure. We now perform a change of variables $\eta \to
X^\eps$ in the path integrals, which necessitates including the
correction terms 
\begin{align}
C[\phi] :=
\exp\left\{-\frac{1}{2} \int_{0}^T \nabla \cdot
b(\phi(t)) + \trace\left[\nabla a(\phi) \theta \right] 
- \frac{\eps}{4} \left[ \nabla^2 \cdot a(\phi) - \left \langle
\nabla \cdot a(\phi), a^{-1}(\phi) \nabla \cdot a(\phi) \right
\rangle_n \right] \dd t \right\}
\end{align}
for a midpoint discretization of the path integral
(see~\cite{langouche-roekaerts-tirapegui:1982,cugliandolo-lecomte:2017}
and in particular~\cite{itami-sasa:2017} for a detailed discussion), so that the rules of
standard calculus apply in the subsequent expansion around the instanton. We obtain
\begin{align}
A_f^\eps(\lambda) = \frac{\int_{\phi(0) = x}
{\cal D} \phi\; C[\phi] \exp\left\{- \frac{1}{\eps} \left( S[\phi] - \lambda f(
\phi(T))\right) \right\}}{\int_{\phi(0) = x}
{\cal D}\phi\; C[\phi] \exp\left\{- \frac{1}{\eps} S[\phi] \right\}}\,,
\label{eq:ratio-pi}
\end{align}
where $S$ is the Freidlin-Wentzell action functional~\eqref{eq:fw-action}.
Both path integrals have a free right boundary and hence consider
all paths that
start at $x$, regardless of their final position at $t = T$. The only
difference is the final time boundary term in the numerator, which imposes
different boundary conditions for the first and second variation of the 
action functional. We apply
an infinite-dimensional version of Laplace's method to both path integrals
in the small noise limit $\eps \downarrow 0$, which leads to the computation
of a ratio of functional determinants for the pre-exponential factor. Note
that the additional terms in the exponent originating from~$C$ are
irrelevant for the determination and expansion around the minimum as
$\eps \downarrow 0$, and will just be evaluated at the expansion point.\\

For the denominator of~\eqref{eq:ratio-pi}, the first variation of
the action around a fixed path $\phi$ becomes
\begin{align}
S[\phi + \sqrt{\eps} \gamma] - S[\phi] = \sqrt{\eps} \left(\int_0^T
\left\langle \gamma,
\left[- \dv{}{t} - \nabla b(\phi)^\top \right] \theta - \tfrac{1}{2} \left \langle \theta, \nabla a(\phi) \theta \right \rangle_n \right\rangle_n
\dd t +
\left. \left\langle \gamma, \theta \right \rangle_n \right|_{0}^{T}\right)
 + {\cal O}\left(\eps \right)\,,
\end{align}
where $\theta$ is the conjugate momentum of $\phi$. Since $\phi(0)
= x$ due to the only boundary condition of the path integral, we
have $\gamma(0) = 0$ for all variations. Demanding that the first
variation around $\phi$ should vanish hence imposes the natural
boundary condition $\theta(T) = 0$
for a stationary path. We conclude that the deterministic trajectory
$\phi_0$ with vanishing momentum $\theta_0(t) \equiv 0$ is
the unique stationary point of the action functional in the
denominator of~\eqref{eq:ratio-pi} with $S[\phi_0] = 0$.
Expanding $S$ around $\phi_0$ to second order as in
appendix~\ref{app:forman-lagrange-to-hamilton}, we see that in
addition to $\gamma(0) = 0$, the variations need to satisfy
$\zeta(T)=0$ for the boundary term $\tfrac{1}{2} \left
\langle \gamma, \zeta \right \rangle_n \rvert_0^T$ to vanish in the
path integral~\cite{kleinert:2009}, i.e.\ we obtain the boundary
conditions~\eqref{eq:bc-lbda-general} for $\lambda = 0$. Hence
\begin{align}
&\int_{\phi(0) = x}
{\cal D} \phi \; C[\phi] \exp\left\{- \frac{1}{\eps} S[\phi] \right\}
\overset{\eps \downarrow 0}{\sim}
\left[\Det_{{\cal A}_0}
\left(a(\phi_0) \Omega\left[\phi_0\right]\right) \right]^{-1/2}
\exp\left\{-\frac{1}{2} \int_0^T \nabla \cdot b(\phi_0) \dd t\right\}\,,
\label{eq:free-laplace}
\end{align}
where we used the expansion
\begin{align}
\phi \to \phi_0 + \sqrt{2 \pi \eps} \sigma(\phi_0) \gamma\,.
\end{align}
Note that, for any
discretization $0 = t_0 < t_1 < \dots < t_K = T$ of the time
interval $[0,T]$ with spacing $\Delta t = T / K$, the Jacobian
of this transformation cancels
the divergent normalization constants of the discrete path measure
\begin{align}
\left(2 \pi \eps \Delta t \right)^{-nK / 2} \prod_{i = 1}^{K}
\frac{\dd^n \phi_i}{\left[ \det a \left( \frac{\phi_i +
\phi_{i-1}}{2} \right) \right]^{1/2}} \,,
\end{align}
and also leads to a second order coefficient of the second variation
operator of $-1$ in the determinant
\begin{align}
\Det_{{\cal A}_0}
\left(\sigma^\top(\phi_0) \Omega\left[\phi_0\right] \sigma(\phi_0) \right) =
\Det_{{\cal A}_0}
\left(a(\phi_0) \Omega\left[\phi_0\right]\right)\,.
\end{align}

For the expansion of the numerator of~\eqref{eq:ratio-pi}, we first
need to determine the
instanton $\phi_\lambda$ (with conjugate momentum $\theta_\lambda$)
which minimizes $S$ under the given boundary conditions. Additionally
expanding the term $-\lambda f(\phi(T)) $ around $\phi_\lambda$
results in the first order necessary
conditions~\eqref{eq:instanton-eq-z} for a stationary
path~$\phi_\lambda$.
The boundary conditions of the fluctuations $\gamma$ are
given by $\gamma(0) = 0$, and, taking into account the additional
boundary term $-\tfrac{\lambda}{2} \left \langle \gamma(T),
\nabla^2 f(\phi_\lambda(T)) \gamma(T) \right
\rangle_n$ as well as the
boundary term~$\tfrac{1}{2} \left \langle \gamma, \zeta \right
\rangle_n \rvert_0^T$ from the general expansion in
appendix~\ref{app:forman-lagrange-to-hamilton},
\begin{align}
\zeta(T) = \lambda \nabla^2 f(\phi_z\lambda(T)) \gamma(T)\,,
\end{align}
i.e.\ the boundary conditions~\eqref{eq:bc-lbda-general}
(cf.~\cite{vilenkin-yamada:2018,di-tucci-lehners:2019} for examples of
path integrals with similar boundary conditions).
Proceeding with the application of Laplace's method to the numerator
in~\eqref{eq:ratio-pi} with these boundary conditions for the fluctuations,
we conclude that
\begin{align}
&\int_{\phi(0) = x}
{\cal D} \phi\; C[\phi] \exp\left\{- \frac{1}{\eps} \left( S[\phi] - \lambda
f(\phi(T))\right) \right\} \overset{\eps \downarrow 0}{\sim}
\left[\Det_{{\cal A}_\lambda}
\left(a(\phi_\lambda) \Omega\left[\phi_\lambda\right] \right) \right]^{-1/2} \times\nonumber\\
& \hspace{2cm}\times
\exp\left\{-\tfrac{1}{2} \int_0^T \nabla \cdot
b(\phi_\lambda) + \trace\left[\nabla a(\phi_\lambda) \theta_\lambda \right]  \dd t
\right\} \exp\left\{- \tfrac{1}{\eps} \left( S[\phi_\lambda] -
\lambda f(\phi_\lambda(T))\right) \right\}\,.
\end{align}
\end{myproof}\\


The functional determinants in Proposition~\ref{thm:mgf-prefactor}
can either be defined as the (divergent) product of all eigenvalues of
the differential operator under the boundary conditions in question when suitable
\textit{ratios} of operator determinants are considered, or individually via
zeta function regularization~\cite{ray-singer:1971};
see e.g.~\cite{dunne:2008} for a short introduction.
Since the top order coefficient
of both operators in Proposition~\ref{thm:mgf-prefactor} is
identical (and equal to -1), the spectra of the two
operators should agree for asymptotically large eigenvalues and we can
expect their determinant ratio to be finite. This idea is made precise
for example by using Forman's theorem~\cite{forman:1987}, which is a
generalization of
the initial work of Montroll~\cite{montroll:1952}, Gel'fand and
Yaglom~\cite{gelfand-yaglom:1960} and others on ratios of functional
determinants of Schr{\"o}dinger operators in quantum mechanics. While
the results of~\cite{forman:1987} are valid for the general case of
elliptic differential operators on Riemannian manifolds, we only need
the special case of second order ordinary differential operators on
finite time intervals as stated in
appendix~\ref{app:forman-lagrange-to-hamilton}. In a Hamiltonian
formulation in terms of fluctuations and momentum
fluctuations, applying the general proposition~\ref{thm:forman-hamiltonian}
to the Freidlin-Wentzell action~\eqref{eq:fw-action} directly
yields the following
proposition in order to evaluate the ratio of functional determinants
in~\eqref{eq:ratio-funcdet}:\\


\begin{mythm}[Hamiltonian formulation of Forman's theorem for the
second variation of the Freidlin-Wentzell Lagrangian]
\label{thm:forman-fw}
Let $\Upsilon_\lambda, \Upsilon_0:[0,T] \to \RR^{2n \times 2n}$
be two fundamental systems of solutions with
arbitrary (invertible) initial conditions
$\Upsilon_\lambda(0), \Upsilon_0(0) \in \RR^{2n \times 2n}$ of the
first order differential equation
\begin{align}
\dv{}{t} \left( \begin{array}{c}
\gamma \\ \zeta
\end{array} \right) &= \Gamma[\phi] \left( \begin{array}{c}
\gamma \\ \zeta
\end{array} \right) \nonumber\\
&= \left(\begin{array}{c|c}
      \nabla b(\phi) + \left(\nabla a(\phi) \theta \right) & a(\phi)\\
      \hline
      - \left \langle \nabla^2 b(\phi), \theta \right
      \rangle_n - \frac{1}{2} \left \langle \theta, \nabla^2 a(\phi) \theta \right
\rangle_{n} & - \nabla b(\phi)^\top - \left(\nabla a(\phi) \theta\right)^\top
    \end{array} \right) \left( \begin{array}{c}
\gamma \\ \zeta
\end{array} \right)
\label{eq:jacobi-fw}
\end{align}
for $\phi = \phi_\lambda$ and $\phi = \phi_0$, respectively.
Fix any matrices $M_\lambda, N_\lambda, M_0, N_0 \in
\RR^{2n \times 2n}$ that realize the boundary conditions ${\cal A}_\lambda$
and ${\cal A}_0$ from~\eqref{eq:bc-lbda-general} via
\begin{align}
M \left( \begin{array}{c}
\gamma(0) \\ \zeta(0)
\end{array} \right) + N \left( \begin{array}{c}
\gamma(T) \\ \zeta(T)
\end{array} \right) = 0\,.
\end{align}
Then the ratio of functional determinants
in~\eqref{eq:ratio-funcdet} can be expressed as
\begin{align}
\frac{\Det_{{\cal A}_\lambda}
  \left(a(\phi_\lambda) \Omega[\phi_\lambda]
  \right)}
  {\Det_{{\cal A}_0} \left(a(\phi_0) \Omega[\phi_0]
  \right)} = \frac{{\det}_{2n} \left(M_\lambda
 \Upsilon_\lambda (0) + N_\lambda \Upsilon_\lambda(T) \right)}
              {{\det}_{2n} \left(M_0
              \Upsilon_0 (0) + N_0
              \Upsilon_0(T) \right)}
              \frac{{\det}_{2n} \Upsilon_0(0)
              }{
              {\det}_{2n} \Upsilon_\lambda(0)} \,.
              \label{eq:fw-variation-forman-general}
\end{align}
\end{mythm}

\begin{myrmk}
\label{rmk-jacobi}

We call~\eqref{eq:jacobi-fw} the (first order) Jacobi equation for the
Freidlin-Wentzell action functional~\eqref{eq:fw-action}.
Expressing it in terms
of $\gamma$ and $\dot{\gamma}$, i.e.\ from a Lagrangian instead of a
Hamiltonian perspective, the Jacobi
equation can equivalently be stated as a second order ordinary differential
equation
\begin{align}
\Omega \left[\phi \right] \gamma = 0\,,
\label{eq:jacobi-fw-second-order}
\end{align}
with the Freidlin-Wentzell Jacobi operator $\Omega$
defined in~\eqref{eq:fw-scnd-var-op}.
This transformation is carried out explicitly for a general action
functional in appendix~\ref{app:forman-lagrange-to-hamilton}.
\end{myrmk}


\begin{myrmk}

A particularly convenient aspect of proposition~\ref{thm:forman-fw} is
the fact that it makes the dependence of the functional determinants
on the boundary conditions very transparent and easy to calculate. We
just need \textit{any} fundamental system of
solutions~$\Upsilon$ for each of the
operators~$\Omega$, which is entirely independent of the imposed
boundary conditions, and then, for given boundary condition matrices
$M$, $N$, we can immediately evaluate the right-hand side
of~\eqref{eq:fw-variation-forman-general} from our knowledge
of the~$\Upsilon$'s. The
separation of the fundamental system of solutions and
boundary condition dependence
is the crucial feature that allows for the treatment of zero
eigenvalues via boundary perturbations later.
\end{myrmk}

\begin{myrmk}

Since $\Gamma[\phi]$ is traceless, $\det \Upsilon_\lambda(t)$ and
$\det \Upsilon_0(t)$ are constant for all $t \in [0,T]$.
\end{myrmk}

\begin{myrmk}

Some examples, treated in~\cite{falco-fedorenko-gruzberg:2017}, for typical
boundary conditions encountered in physics and
their representations in terms of matrices $M,N \in
\RR^{2n \times 2n}$ (which
are unique up to $\text{GL}(2n)$ transformations) are
\begin{enumerate}[(i)]
\item Dirichlet boundary conditions $\gamma(0) = \gamma(T) = 0$: 
\begin{align}
M_{\text{Dirichlet}} = \left(\begin{array}{c|c}
1_{n \times n} & 0_{n \times n}\\
\hline
0_{n \times n} & 0_{n \times n}
\end{array} \right)\,, \quad N_{\text{Dirichlet}} = \left(\begin{array}{c|c}
0_{n \times n} & 0_{n \times n}\\
\hline
1_{n \times n} & 0_{n \times n}
\end{array} \right)\,.
\end{align}
In quantum mechanics, functional determinants of operators with Dirichlet
boundary conditions typically appear in the computation of semi-classical
propagators.
\item Periodic (Antiperiodic) boundary conditions
$\gamma(0) = p \cdot \gamma(T)$, $\zeta(0) = p \cdot \zeta(T)$ with
$p = 1$ ($p = -1$):
\begin{align}
M_p = \left(\begin{array}{c|c}
1_{n \times n} & 0_{n \times n}\\
\hline
0_{n \times n} & 1_{n \times n}
\end{array} \right)\,, \quad N_p = \left(\begin{array}{c|c}
-p \cdot 1_{n \times n} & 0_{n \times n}\\
\hline
0_{n \times n} & -p \cdot 1_{n \times n}
\end{array} \right)\,.
\end{align}
Functional determinants with periodic (antiperiodic) boundary conditions
need to be evaluated for the calculation of partition functions and other
thermal averages of bosons (fermions) in quantum statistical physics and
field theory.
\end{enumerate}
\end{myrmk}


For the boundary conditions~\eqref{eq:bc-lbda-general},
possible choices for $M, N$ are
\begin{align}
M_\lambda = M_0 = \left(\begin{array}{c|c}
1_{n \times n} & 0_{n \times n}\\
\hline
0_{n \times n} & 0_{n \times n}
\end{array} \right)\,, \quad N_\lambda = 
\left( \begin{array}{c|c}
0_{n \times n} & 0_{n \times n}\\
\hline
- \lambda \nabla^2 f(\phi_\lambda(T)) & 1_{n \times n}
\end{array} \right)\,, \quad N_0 = \left(\begin{array}{c|c}
0_{n \times n} & 0_{n \times n}\\
\hline
0_{n \times n} & 1_{n \times n}
\end{array} \right)\,.
\end{align}
Using proposition~\ref{thm:forman-fw} and choosing $\Upsilon_\lambda(0) =
\Upsilon_0(0) = 1_{2n \times 2n}$ the prefactor $R_\lambda$
in~\eqref{eq:ratio-funcdet} simplifies to
\begin{align}
R_\lambda =  \left[ {\det}_n \left( -\lambda \nabla^2
f(\phi_\lambda(T))
\gamma(T) + \zeta(T) \right) \exp\left(\int_0^T \nabla \cdot
b(\phi_\lambda) + \trace \left[\nabla a(\phi_\lambda)
\theta_\lambda \right] \,\dd t\right) \right]^{-1/2}\,,
\label{eq:ratio-simplified}
\end{align}
with $(\gamma, \zeta): [0,T] \to \RR^{2n \times n}$ solving
the Jacobi equation with boundary conditions
\begin{align}
\dv{}{t} \left( \begin{array}{c}
\gamma \\ \zeta
\end{array} \right) = \Gamma[\phi_\lambda] \left( \begin{array}{c}
\gamma \\ \zeta
\end{array} \right)\,, \quad \left( \begin{array}{c}
\gamma(0) \\ \zeta(0)
\end{array} \right) = \left(\begin{array}{c}
0_{n \times n}\\1_{n \times n}
\end{array} \right)\,. \label{eq:jacobi-fw-bc}
\end{align}
As remarked in~\cite{schorlepp-grafke-grauer:2021,
grafke-schaefer-vanden-eijnden:2021}, considering the example of
an Ornstein-Uhlenbeck process with $b(x) = - \beta x$ for $\beta > 0$
and $\sigma(x) \equiv \sqrt{2}$
shows that the equation for $\zeta$ in~\eqref{eq:jacobi-fw-bc} should
naturally be integrated backwards in time due to the appearance of
$-\nabla b(\phi_z)^\top$ on the right-hand side, in contrast to the
formulation above in terms of an initial value problem. For large~$T$,
we consequently expect that the determinant in~\eqref{eq:ratio-simplified}
will diverge to $+ \infty$, whereas the exponential term will tend to~$0$.
The following transformation onto a symmetric matrix Riccati differential
equation mitigates this problem and is hence in particular well suited
for numerical calculations of the prefactor~$R_\lambda$:

\begin{mythm}[MGF prefactor estimate via \textbf{forward} Riccati equation]
\label{thm:gy-old}
We have the following exact expression for the prefactor $R_\lambda$
as defined in~\eqref{eq:ratio-funcdet}:
\begin{align}
R_\lambda = \frac{\exp \left\{ \frac{1}{2} \int_0^T
\trace \left[ \left( \left \langle \nabla^2 b(\phi_\lambda), \theta_\lambda \right
\rangle_n + \tfrac{1}{2} \left \langle \theta_\lambda, \nabla^2 a(\phi_\lambda)
\theta_\lambda \right
\rangle_{n} \right) Q_\lambda \right] \dd t \right\}}{\left[ {\det}_n \left(1_{n \times n}
- \lambda \nabla^2 f(\phi_\lambda(T)) Q_\lambda(T)\right) \right]^{1/2}}\,,
\end{align}
where $Q_\lambda:[0,T] \to \RR^{n \times n}$ solves the
forward symmetric matrix Riccati
differential equation
\begin{align}
\begin{cases}
\dot{Q}_\lambda &= a(\phi_\lambda) + Q_\lambda \left[ \nabla b\left(\phi_\lambda
\right)^\top + \left( \nabla a(\phi_\lambda) \theta_\lambda \right)^\top \right] \\
& \quad+
\left[ \nabla b\left(\phi_\lambda \right) + \left(\nabla a(\phi_\lambda)
\theta_\lambda \right)\right] Q_\lambda + Q_\lambda \left[ \left< \nabla^2
b(\phi_\lambda), \theta_\lambda\right>_n + \tfrac{1}{2} \left \langle
\theta_\lambda, \nabla^2 a(\phi_\lambda) \theta_\lambda \right
\rangle_{n} \right] Q_\lambda \,,\\
Q_\lambda(0) &= 0_{n \times n} \in \RR^{n \times n}\,.
\end{cases}
\label{eq:riccati-fw}
\end{align}
\end{mythm}

This result quantifies the impact of the
Gaussian fluctuations around the instanton in a numerically convenient way.
These fluctuations satisfy the linear SDE
\begin{align}
\dd Y_t = \left[\nabla b(\phi_\lambda(t)) +
\left(\nabla a(\phi_\lambda(t)) \theta_\lambda(t) \right) \right]
Y_t \; \dd t + \sigma(\phi_\lambda(t)) \dd B_t\,, \quad 
Y_0 = 0 \in \RR^n\,,
\end{align}
and from a probabilistic point of view, proposition~\ref{thm:gy-old}
effectively computes the expectation
\begin{align}
R_\lambda = 
&\EE\left[e^{\tfrac{\lambda}{2} \left< Y_T,
\nabla^2 f\left(\phi_\lambda(T) \right), Y_T
\right>_n} e^{\tfrac{1}{2} \int_0^T \left< Y_t,\left[ \left<
\nabla^2 b(\phi_\lambda(t)),
\theta_\lambda(t)\right>_n + \tfrac{1}{2} \left \langle \theta_\lambda(t),
\nabla^2 a(\phi_\lambda(t)) \theta_\lambda(t) \right \rangle_n\right] Y_t \right>_n
\dd t}\right]\,.
\end{align}
Computationally, the inefficient approach
to estimate $A_f^\eps(\lambda)$ for small $\eps$ using Monte Carlo
simulations is thus replaced by the ($\eps$-independent) problem to
minimize the action functional $S$, subject to final time boundary
conditions $\theta_\lambda(T) = \lambda \nabla f(\phi_\lambda(T))$,
plus the numerical integration of an initial value
problem for~$Q_\lambda$. For moderate dimensions $n$ (e.g.\
if the SDE at hand stems from the semi-discretization of a
one-dimensional SPDE), the direct numerical integration of $Q$ poses
no problems.\\


\begin{myproof}{Proposition~\ref{thm:gy-old}}

The transformation of the Jacobi equation~\eqref{eq:jacobi-fw-bc}
to the solution $Q = \gamma \zeta^{-1}$ of the forward Riccati
equation~\eqref{eq:riccati-fw} is explained for a general action
functional in appendix~\ref{app:forman-lagrange-to-hamilton}.
Hence, the proposition is obtained by factoring out $\zeta(T)$
in~\eqref{eq:ratio-simplified} and using $\det = \exp \trace \log$ for 
\begin{align}
&{\det}_n \left(\zeta(T)\right) = \frac{{\det}_n \left(
\zeta(T)\right)}{{\det}_n \left(\zeta(0)
\right)} = \exp \left\{\int_0^T \dv{}{t}\trace \left[\log
\zeta \right] \dd t \right\} = \exp \left\{\int_0^T \trace \left[\dot{\zeta}
\zeta^{-1} \right] \dd t \right\} \nonumber\\
&\overset{\eqref{eq:jacobi-fw}}{=} \exp \left\{-\int_0^T 
\trace \left[ \left( \left \langle \nabla^2 b(\phi_\lambda), \theta_\lambda \right
\rangle_n + \tfrac{1}{2} \left \langle \theta_\lambda, \nabla^2 a(\phi_\lambda)
\theta_\lambda \right
\rangle_{n} \right) Q_\lambda \right] \dd t - \int_0^T
\nabla \cdot
b(\phi_\lambda) + \trace \left[\nabla a(\phi_\lambda)
\theta_\lambda \right]\; \dd t \right\}\,. \nonumber
\end{align}
\end{myproof}


It is also straightforward to derive a representation of the
prefactor~$R_\lambda$ in terms of a backward Riccati differential equation from
Proposition~\ref{thm:forman-fw}:

\begin{mythm}[MGF prefactor estimate via \textbf{backward} Riccati equation]
\label{thm:gy-old-bkwd}
We have the following alternative, exact expression for the prefactor
$R_\lambda$ as defined in~\eqref{eq:ratio-funcdet}:
\begin{align}
R_\lambda = \exp\left\{\frac{1}{2} \int_0^T
\trace \left[a(\phi_\lambda(t)) W_\lambda(t) \right] \dd t\right\}\,,
\end{align}
where $W_\lambda:[0,T] \to \RR^{n \times n}$ solves the backward
symmetric matrix Riccati
differential equation
\begin{align}
\begin{cases}
\dot{W}_\lambda &= -W_\lambda a(\phi_\lambda) W_\lambda -
\left[ \nabla b\left(\phi_\lambda
\right)^\top + \left( \nabla a(\phi_\lambda) \theta_\lambda \right)^\top
\right]W_\lambda \\
& \quad- W_\lambda
\left[ \nabla b\left(\phi_\lambda \right) + \left(\nabla a(\phi_\lambda)
\theta_\lambda \right)\right]- \left< \nabla^2
b(\phi_\lambda), \theta_\lambda\right>_n - \tfrac{1}{2} \left \langle
\theta_\lambda, \nabla^2 a(\phi_\lambda) \theta_\lambda \right
\rangle_{n} \,,\\
W_\lambda(T) &= \lambda \nabla^2 f(\phi_\lambda(T))
\in \RR^{n \times n}\,.
\end{cases}
\label{eq:riccati-bw}
\end{align}
\end{mythm}


\begin{myproof}{Proposition~\ref{thm:gy-old-bkwd}}

The general transformation of the Jacobi equation~\eqref{eq:jacobi-fw-bc}
to the solution $W = \zeta \gamma^{-1}$ of the backward Riccati
equation~\eqref{eq:riccati-bw} can also be found in
appendix~\ref{app:forman-lagrange-to-hamilton}. Instead of the initial
condition $\Upsilon_\lambda(0) = 1_{2n \times 2n}$, we now pick
(assuming for simplicity that $ \nabla^2 f(\phi_\lambda(T))$ has full rank)
\begin{align}
\Upsilon_\lambda(T) = \left(\begin{array}{c|c}
1_{n\times n} & 1_{n\times n}\\
\hline
\lambda \nabla^2 f(\phi_\lambda(T)) & 0_{n \times n}
\end{array} \right)
\end{align}
as final condition of the fundamental system of solutions.
Hence ${\det}_{2n}
\Upsilon_\lambda(T) = {\det}_{n} \left(-\lambda \nabla^2
f(\phi_\lambda(T)) \right)$ and
\begin{align}
{\det}_{2n} \left(M_\lambda
 \Upsilon_\lambda (0) + N_\lambda \Upsilon_\lambda(T) \right)
 = {\det}_{n} \left(-\lambda \nabla^2 f(\phi_\lambda(T)) \right)
 {\det}_{n} \gamma(0),
\end{align}
where $\gamma$ is composed of the upper left block of the
fundamental system of solutions. 
Again computing
\begin{align}
{\det}_n \left(\gamma(0)\right) &= \frac{{\det}_n \left(
\gamma(0)\right)}{{\det}_n \left(\gamma(T)
\right)} = \exp \left\{-\int_0^T \dv{}{t}\trace \left[\log
\gamma \right] \dd t \right\} = \exp \left\{-\int_0^T \trace \left[\dot{\gamma}
\gamma^{-1} \right] \dd t \right\} \nonumber\\
&\overset{\eqref{eq:jacobi-fw}}{=} \exp \left\{-\int_0^T 
\trace \left[a(\phi_\lambda) W_\lambda \right] \dd t - \int_0^T
\nabla \cdot
b(\phi_\lambda) + \trace \left[\nabla a(\phi_\lambda)
\theta_\lambda \right]\; \dd t \right\}\
\end{align}
completes the derivation.
\end{myproof}


\subsection{Probability density function prefactor estimates for Freidlin-Wentzell theory with unique instantons}
\label{subsec:prefac-nondegen-pdf-trafo}

Assuming, as usual, strict convexity of the rate function $z \mapsto I_f(z)$:

\begin{mythm}[PDF prefactor estimate from a sharp LDT result for the MGF]
\label{thm:pdf-saddle}
If an asymptotic estimate
  \begin{equation}
    A_f^\eps(\lambda)\overset{\eps \downarrow 0}{\sim} R_\lambda
    \exp\left\{-\eps^{-1} \left( S[\phi_\lambda] - \lambda
    f\left(\phi_\lambda(T)\right) \right)\right\}
  \end{equation}
  of the MGF $A_f^\eps$ holds, then for any $z \in \RR$, we have
  \begin{align}
  \rho_f^\eps(z) \overset{\eps \downarrow 0}{\sim} \left(2 \pi \eps
  \right)^{-1/2} R_{\lambda_z} \left[\left.\dv{}{\lambda}\right
  \rvert_{\lambda_z} f(\phi_\lambda(T)) \right]^{-1/2} \exp \left\{
  -\frac{1}{\eps} S \left[\phi_{\lambda_z} \right] \right\}\,,
  \end{align}
  with $\lambda_z$ uniquely determined by $f(\phi_{\lambda_z}(T)) = z$.
\end{mythm}


\begin{myrmk}
By Legendre duality,
we have $\lambda_z = I_f'(z)$ for the observable rate
function $I_f(z) = S[\phi_{\lambda_z}]$, so the additional
term in the PDF prefactor in Proposition~\ref{thm:pdf-saddle} compared to the
MGF case of the previous section can be written as
\begin{align}
\left[\left.\dv{}{\lambda}\right
  \rvert_{\lambda_z} f(\phi_\lambda(T)) \right]^{-1/2} = \sqrt{I_f''(z)}\,,
\end{align}
where the second derivative of $I_f$ is positive by our assumption of strict
convexity.
\end{myrmk}


\begin{myproof}{Proposition~\ref{thm:pdf-saddle}}

Since the scaled MGF is a two-sided Laplace transform ${\cal L}$ of the PDF
\begin{align}
A_f^\eps(\lambda) = \EE \left[
\exp\left\{\frac{\lambda}{\eps}
f(X^\eps_T)\right\} \right] = {\cal L}\left[\rho_f^\eps\right]\left(
-\frac{\lambda}{\eps}\right)\,,
\end{align}
it can be inverted by contour integration (with a suitable shift $\alpha \in \RR$ for the contour):
\begin{align}
\rho_f^\eps(z) &= \frac{1}{2\pi i \eps} \int_{\alpha- i \infty}^{
\alpha + i \infty} A_f^\eps(\lambda) \exp\left\{-\frac{\lambda z}{\eps}
\right\}\;\dd \lambda \nonumber\\
&\overset{\eps \downarrow 0}{\sim} \frac{1}{2\pi i \eps}
\int_{\alpha- i\infty}^{\alpha + i \infty} R_\lambda \exp\bigl\{-\frac{1}
{\eps} \underbrace{\left(S[\phi_\lambda] - \lambda (f(\phi_\lambda(T)) - z)
 \right)}_{=:\tilde{S}_z(\lambda)}
\bigr\}\;\dd \lambda \nonumber\\
& \overset{\eps \downarrow 0}{\sim} \frac{R_{\lambda_z}}{\sqrt{2\pi\eps}}
\exp \left\{-\frac{1}{\eps} S \left[\phi_{\lambda_z} \right] \right\}
\frac{1}{i} \int_{- i \infty}^{+ i \infty} \exp\left\{-\pi
\tilde{S}''_z(\lambda_z) \left( \lambda' \right)^2
\right\}\;\dd \lambda'\,,
\end{align}
where we applied a saddlepoint approximation in the last line.
At stationary points of the Lagrange function $\tilde{S}_z$, we demand
that the first derivative
\begin{align}
\tilde{S}_z'(\lambda) &= \int_0^T \bigg \langle \underbrace{\left.
\fdv{S}{\phi} \right \rvert_{\phi_\lambda}}_{=0},
\dv{\phi_\lambda}{\lambda} \bigg \rangle_n \dd t + \bigg \langle \underbrace{
\theta_\lambda(T)}_{\lambda \nabla f(\phi_\lambda(T))},
\dv{\phi_\lambda}{\lambda} (T) \bigg \rangle_n - (f(\phi_\lambda(T))
- z) - \lambda \left \langle\nabla f(\phi_\lambda(T)),
\dv{\phi_\lambda}{\lambda} (T)\right \rangle_n\nonumber \\
&=  - (f(\phi_\lambda(T)) - z)
\end{align}
vanishes, and hence $f(\phi_{\lambda_z}(T)) = z$ at the unique
minimum. Furthermore, we see that $\tilde{S}_z''(\lambda) =
-\dv{}{\lambda} f(\phi_\lambda(T))$, thereby concluding the derivation.
\end{myproof}\\


\begin{myrmk}
Via partial integration, as detailed in~\cite{bleistein:1975}, it is also
straightforward to derive an asymptotic expression for
\textit{tail probabilities} $\PP \left[f(X_T^\eps) > z \right]$ from
Proposition~\ref{thm:pdf-saddle}:
For any~$z \in \RR$ such
  that $S \left[\phi_\cdot\right]$ increases monotonically on~$[z,\infty)$
  with~$\dd S[\phi_z] / \dd z > 0$ (where $\phi_z := \phi_{\lambda_z}$), we have
  \begin{align}
  \PP \left[f\left(X_T^\eps \right) > z \right] \overset{\eps 
  \downarrow 0}{\sim} \left(2 \pi
  \right)^{-1/2} \eps^{1/2} R_{\lambda_z} \left[\left.\dv{}{\lambda}\right
  \rvert_{\lambda_z} f(\phi_\lambda(T)) \right]^{-1/2} \lambda_z^{-1} \exp \left\{
  -\frac{1}{\eps} S \left[\phi_{\lambda_z}\right] \right\}\,,
  \end{align}
  with~$\lambda_z$ uniquely determined by~$f(\phi_{\lambda_z}(T)) = z$.
\end{myrmk}


Expressing the derivative of $f(\phi_\lambda(T))$ with respect
to~$\lambda$ at~$\lambda_z$ in terms of the forward Riccati matrix
$Q_z = Q_{\lambda_z}$ (similarly $\phi_z = \phi_{\lambda_z}$, etc)
finally recovers the full result
of~\cite{schorlepp-grafke-grauer:2021} for the PDF of one-dimensional
observables:


\begin{mythm}[Complete PDF prefactor estimate in terms of \textbf{forward} Riccati matrix]
\label{thm:pdf-riccati-unique}
We have the following asymptotically sharp estimate for the PDF
of $f(X^\eps_T)$ at $z \in \RR$:
\begin{align}
\rho_f^\eps(z) \overset{\eps \downarrow 0}{\sim} (2 \pi \eps)^{-1/2}
\frac{\exp\left\{\frac{1}{2} \int_0^T
\trace \left[ \left( \left \langle \nabla^2 b(\phi_z), \theta_z \right
\rangle_n + \tfrac{1}{2} \left \langle \theta_z, \nabla^2 a(\phi_z)
\theta_z \right
\rangle_{n} \right) Q_z \right] \dd t\right\}}{
\left[{\det}_n \left( U_z \right) \left \langle \nabla f(\phi_z(T)),
Q_z(T) U_z^{-1} \nabla f(\phi_z(T)) \right \rangle_n  \right]^{1/2}}
\exp \left\{-\frac{S\left[\phi_z\right]}{\eps} \right\}
\end{align}
with
\begin{align}
U_z := 1_{n \times n} - \lambda_z \nabla^2 f
\left(\phi_z(T) \right) Q_z(T) \in \RR^{n \times n}\,.
\end{align}
\end{mythm}


\begin{myrmk}
Note that, alternatively, we could have directly evaluated a path
integral expression for the PDF at $z$, which necessitates integrating over
all paths that start at $\phi(0)=x$ and end with $f(\phi(T))=z$. This
results in the boundary conditions
 \begin{align}
  {\cal A}_z : \begin{cases}
  \gamma(0) = 0\\
  \gamma(T) \perp \nabla f(\phi_z(T))\\
  \zeta(T) - \lambda_z \nabla^2 f(\phi_z(T))
  \gamma(T) \parallel \nabla f(\phi_z(T))
  \end{cases}
  \end{align}
for the quadratic fluctuations and functional determinant, thereby making
the application of Forman's theorem and the introduction of the Riccati
matrices more involved. Nevertheless, it would also be possible to derive
the PDF prefactor results in this section using this direct approach.  
\end{myrmk}


\begin{myproof}{Proposition~\ref{thm:pdf-riccati-unique}}

The fluctuation mode $(\dd \phi_\lambda / \dd \lambda, \dd
\theta_\lambda / \dd \lambda)$ satisfies the boundary conditions
\begin{align}
\begin{cases}
\dv{\phi_\lambda}{\lambda} (0) = 0\,,\\
\dv{\theta_\lambda}{\lambda} (T) - \lambda \nabla^2 f(\phi_\lambda(T))
\dv{\phi_\lambda}{\lambda} (T) = \nabla f(\phi_\lambda(T))\,,
\end{cases}
\end{align}
as well as the Jacobi equation~\eqref{eq:jacobi-fw}
along $(\phi_\lambda, \theta_\lambda)$.  Hence, choosing $(\dd
\phi_\lambda / \dd \lambda, \dd \theta_\lambda / \dd \lambda)$ as the
first column of $n$ linearly independent solutions $(\gamma,
\zeta):[0,T] \to \RR^{2n \times n}$ with $\gamma(0) =0$ and $Q =
\gamma \zeta^{-1}$ results in
\begin{align}
\zeta (T) - \lambda \nabla^2 f(\phi_\lambda(T)) \gamma (T) = \left(
Q_\lambda^{-1}(T) - \lambda \nabla^2 f(\phi_\lambda(T)) \right)
\gamma (T) = \left(\nabla f(\phi_\lambda(T)), (*)_{n \times (n-1)} \right)\,,
\end{align}
where $(*)_{n\times (n-1)}$ is a placeholder for the further $n-1$
irrelevant columns. Then
\begin{align}
\dv{\phi_\lambda}{\lambda} (T) = \left(Q_\lambda^{-1}(T) - \lambda
\nabla^2 f(\phi_\lambda(T)) \right)^{-1} \nabla f(\phi_\lambda(T))
= Q_\lambda(T) U_\lambda^{-1} \nabla f(\phi_\lambda(T))\,,
\end{align}
and consequently
\begin{align}
\dv{}{\lambda} f(\phi_\lambda(T)) = \left \langle \nabla
f(\phi_\lambda(T)),  Q_\lambda(T) U_\lambda^{-1} \nabla
f(\phi_\lambda(T))\right \rangle_n\,.
\end{align}
\end{myproof}\\


\section{Prefactor in the presence of zero modes}
\label{sec:prefac-degen}

\subsection{Motivation and finite-dimensional examples}
\label{subsec:prefac-degen-motivation}

\begin{figure}
\centering
\includegraphics[width = \textwidth]{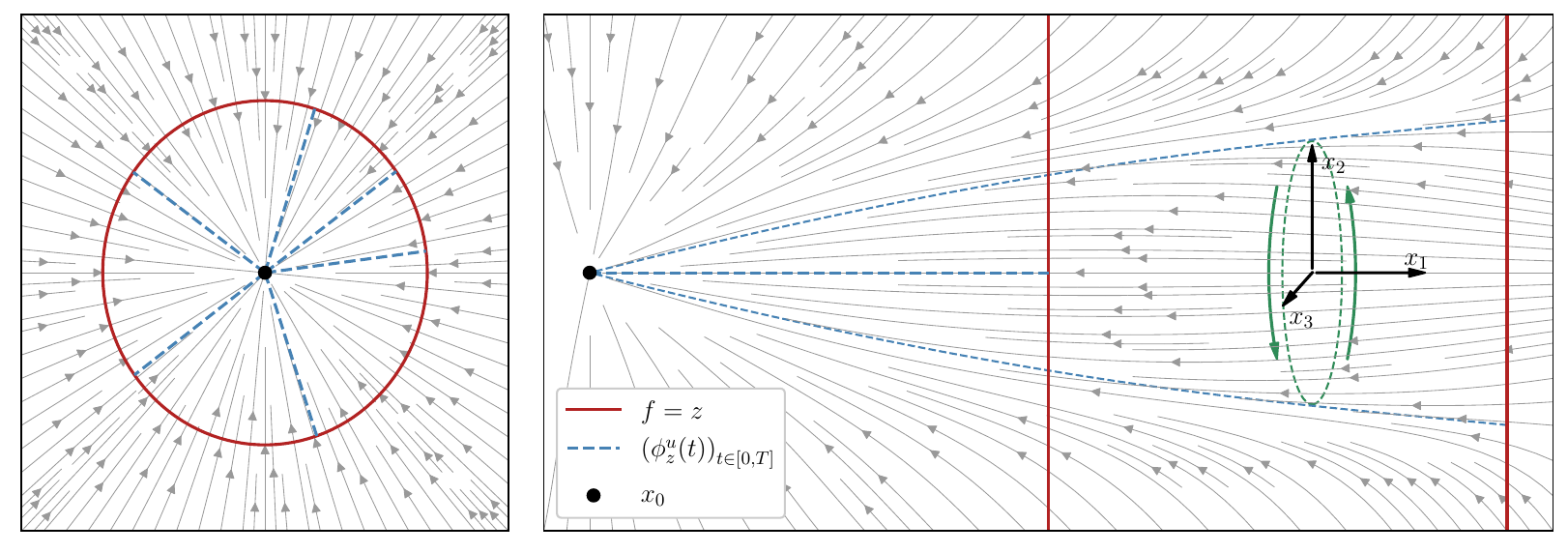}
\caption{Left: Example for the simplest scenario where the instanton has
to break the symmetry of the system including the observable at any~$z$
or~$\lambda$. Here,~$b$ is a radial vector field and $f(x) = \norm{x}_n$.
An example of this kind is discussed in Section~\ref{subsec:n-dimens-ornst}.
Right: Example for a problem with spontaneous symmetry breaking. Suppose
that the whole system is three-dimensional, such that the plot only shows
the $(x_1,x_2)$-plane at $x_3 = 0$, and that the system is
rotationally symmetric about the $x_1$ axis. Then, the instanton
realizing a given value of $z = x_1$ at the final time, as indicated by the
red planes, could, for a suitably constructed drift, break its symmetry
beyond a critical value $z_c$,
thereby transitioning from a solution with $(x_2, x_3) = 0$ along the whole
path to a continuous family of instantons, indicated by the
green arrows symbolizing
out of plane rotation. A toy example for such an instance of
spontaneous symmetry breaking is considered in
Section~\ref{subsec:dynam-phase-trans-ode-potential}.}
\label{fig:inst-sketch}
\end{figure}

\begin{figure}
\centering
\includegraphics[width = \textwidth]{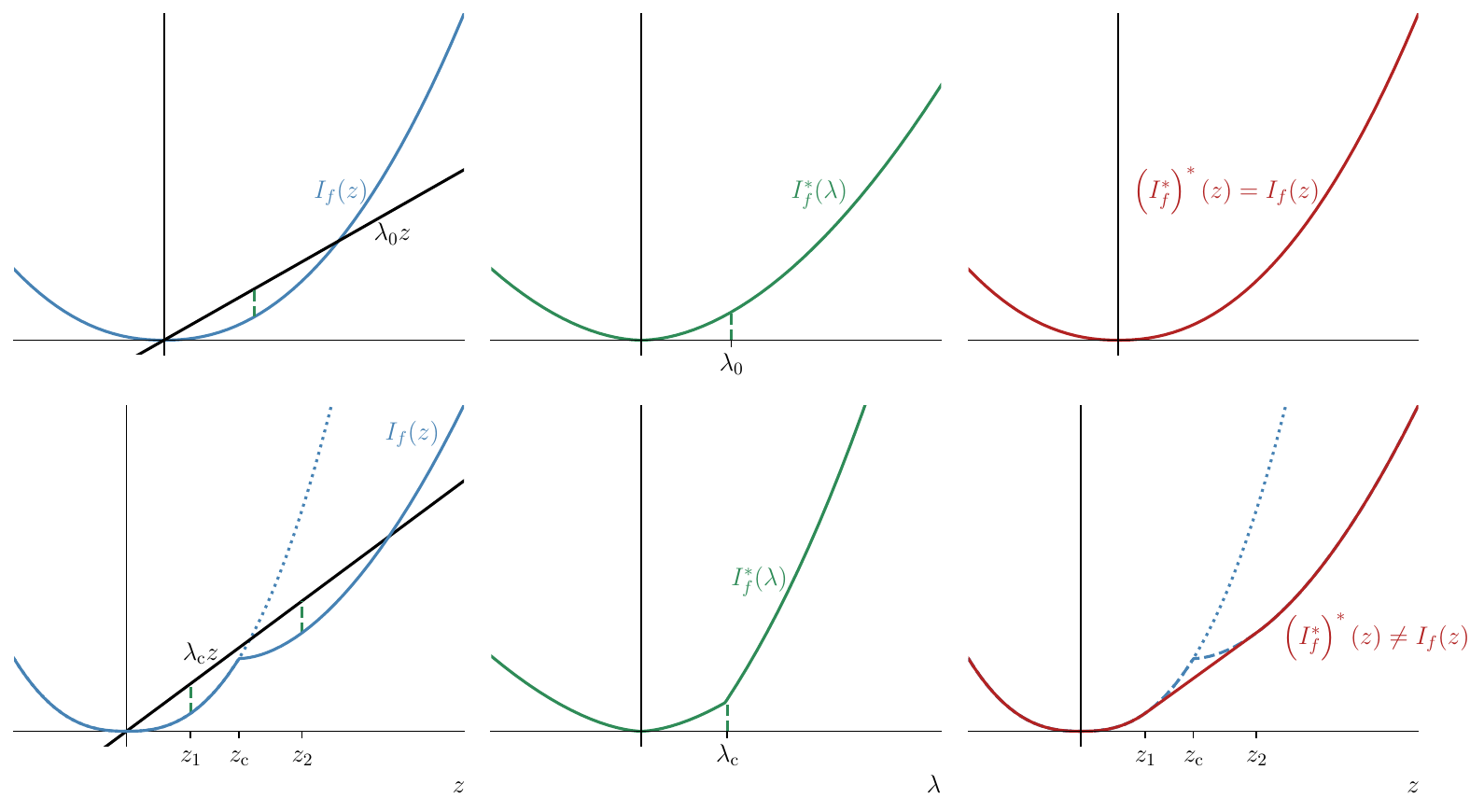}
\caption{Top row: The simplest scenario where the rate function $I_f$
is strictly convex, its Legendre-Fenchel dual $I^*_f$ (that is computed
by minimizing the augmented action functional at fixed $\lambda$) is finite
and differentiable everywhere, and transforming from the MGF to the PDF via
a saddlepoint approximation yields the original rate function. Note that
this situation can still occur if there are zero modes present in the
computation of the prefactor of the MGF $A_f^\eps$.
Bottom row: A different rate function where at a
critical observable value $z_{\text{c}}$, the instanton solution
spontaneously breaks some of its symmetries in order to realize a given
observable value with less action cost than on the symmetric, dotted branch.
In the specific case of a first order phase transition as sketched here,
the rate function is no longer convex, its dual is
non-differentiable at a critical $\lambda_{\text{c}}$, and transforming from
the MGF to the PDF yields the convex hull of the original rate function.}
\label{fig:lf-trafo-rf}
\end{figure}

In this section, we derive in detail analogous statements to the
previous section for situations where an $r$-dimensional continuous
family ${\cal M}^r_z$ of instanton solutions exist for a given
observable value~$z$. We are in particular interested in the case of
dynamical phase transitions due to \textit{spontaneous
  symmetry breaking} of the instanton, where the action functional and
boundary conditions as a whole possess a certain symmetry, the
possible violation of which beyond a \textit{critical observable value
  $z_{\text{c}}$} gives rise to a continuous family of degenerate
instantons and associated flat directions or zero modes in the
function space of all variations.  An alternative to a phase
transition at a critical observable value for zero modes to occur
would be the \enquote{trivial} case where all instantons at any
observable strength must necessarily break the symmetry of the
problem, an example of which is sketched in
Figure~\ref{fig:inst-sketch}. On the level of rate functions, these
two different scenarios roughly look as sketched in
Figure~\ref{fig:lf-trafo-rf}. These examples will be discussed in
sections~\ref{subsec:n-dimens-ornst}
and~\ref{subsec:dynam-phase-trans-ode-potential}.\\

Both of these situations are not only relevant in many examples,
but furthermore convenient from a numerical perspective, since, due to
the underlying symmetry of the entire problem, it
will turn out that it suffices to consider a single, arbitrarily
chosen instanton in ${\cal M}^r_z$ and compute a modified prefactor
for this particular instanton by solving the same Riccati equations as before.
We will again proceed first on the level of MGFs and afterwards transform
onto the PDF. Despite the fact that in the case of spontaneous
symmetry breaking, the rate function can become non-convex
as in Figure~\ref{fig:lf-trafo-rf}, the final
results for the PDF prefactor remain valid in this case as well. The idea
is that even though some instantons might be unobtainable
through minimization
at fixed~$\lambda$~\cite{alqahtani-grafke:2021},
as in Figure~\ref{fig:lf-trafo-rf} with
$z \in (z_1, z_2)$, they can still be computed directly using different
minimization strategies such as penalty
methods~\cite{schorlepp-etal:2022}, and of course correspond to some
value of~$\lambda$ depending on their final time position and momentum, which
can then be used to compute the prefactor. If the rate function branches
are then locally convex individually (or convexified appropriately), then the
corresponding prefactor derivations go through without changes.\\


In order to derive appropriately modified prefactor formulas, we will
use the following, conceptually
simple strategy: First, we split the integration in path space
into components
along the submanifold of degenerate minimizers and the subspace which is
$L^2$-orthogonal to it. For each point on the submanifold, we can then
use Laplace's method on the normal space, where all flat directions
of the second variation of the action are removed by construction.
Then, a boundary-type regularization procedure~\cite{mckane-tarlie:1995,
kleinert-chervykov:1998,falco-fedorenko-gruzberg:2017} is used to
compute functional
determinants with removed zero eigenvalues by integrating a Riccati
equation similar to the non-degenerate case.\\


We start with a brief motivation in finitely many dimensions, as well as
two simple examples: Consider the
Laplace-type integral
\begin{align}
J_\eps = \int_{\RR^n} h(x) e^{-S(x)/\eps} \dd^n x\,,
\end{align}
in the case where there is a family of global minimizers
${\cal M}^r$ of $S:\RR^n \to \RR$, and $h:\RR^n \to \RR$ is any
continuous function. We assume that ${\cal M}^r = \argmin S$ is an
$r$-dimensional submanifold of $\RR^n$ with $0 < r < n$.
Then, we know that
for small $\eps > 0$, the integral $J_\eps$ is dominated by the behavior
of $S$ in an open neighborhood $U_{{\cal M}^r}$ of ${\cal M}^r$, such that
\begin{align}
J_\eps \overset{\eps \downarrow 0}{\sim} \int_{U_{{\cal M}^r}} h(x)
e^{-S(x) / \eps} \dd^n x \overset{\eps \downarrow 0}{\sim}
\int_{{\cal M}^r} \dd^r \mu(y)
\int_{N_y{\cal M}^r} \dd^{n-r} z \  h(y+z) e^{-S(y+z) / \eps}\,,
\end{align}
where the integration was split into the integration along ${\cal
  M}^r$ (with surface measure $\dd^r \mu$) and the (entire, for
$\eps \downarrow 0$) normal space $N_y {\cal M}^r$ perpendicular to
the hypersurface ${\cal M}^r$. This split of integration directions is
usually done formally using the Faddeev-Popov
method~\cite{faddeev-popov:1967} in the physics literature, which
consists of inserting a suitable Dirac $\delta$ function into the
initial integral. For each $y \in {\cal M}^r$, applying Laplace's
method in $z$ yields
\begin{align}
J_\eps \overset{\eps \downarrow 0}{\sim} &(2 \pi \eps)^{(n-r)/2}
\int_{{\cal M}^r} \dd^r \mu(y) \frac{h(y) e^{-S(y) / \eps}}
{\sqrt{ {\det}'_{n-r}(\nabla^2 S(y)) }} \nonumber \\
= &(2 \pi \eps)^{(n-r)/2} e^{-S(y_0) / \eps}
\int_{{\cal M}^r} \dd^r \mu(y) \frac{h(y)}
{\sqrt{ {\det}'_{n-r}(\nabla^2 S(y)) }}\,, \label{eq:finite-dim-zero}
\end{align}
where ${\det}'_{n-r}$ denotes the removal of the $r$ zero eigenvalues
of the matrix $\nabla^2S(y) \in \RR^{n \times n}$ from the determinant
that correspond to
eigenvectors in the tangent space $T_y{\cal M}^r$. In the second line,
we used that $S$ is constant in ${\cal M}^r$ in order to pull the
exponential factor out of the integral, evaluated at any $y_0 \in
{\cal M}^r$. Now, there are two cases: If ${\det}'_{n-r}(\nabla^2 S)$
and $h$ are constant along ${\cal M}^r$, the volume of ${\cal M}^r$
factors out
and we obtain (if this volume is finite; otherwise, the integral is
infinite and needs to be regularized in some way in order to make
sense of it, e.g.\ by normalizing it with respect to the volume)
\begin{align}
J_\eps \overset{\eps \downarrow 0}{\sim} & (2 \pi \eps)^{(n-r)/2}
\vol \left({\cal M}^r \right) \frac{h(y_0)e^{-S(y_0) / \eps}}
{\sqrt{ {\det}'_{n-r}(\nabla^2 S(y_0)) }}\,.
\label{eq:finite-dim-zero-symmetric}
\end{align}
Otherwise, the integral along ${\cal M}^r$ in~\eqref{eq:finite-dim-zero}
needs to be evaluated explicitly. It is easy to find two-dimensional
examples ($n = 2$, $r = 1$) for either case (with $h \equiv 1$):


\begin{enumerate}[(i)]
\item Consider $S:\RR^2 \to \RR$, $S(x,y) = (1+x^4)y^2$. Then the set
  of minimizers of $S$ is given by the $(r=1)$-dimensional manifold
  ${\cal M}^1 = \left\{(x,0) \in \RR^2 \right\}$ with $S|_{{\cal M}^1}
  = 0$, and Hessian $\nabla^2 S(x,0) = \text{diag}(0, 2(1+x^4))$. Since
  the integration along $y$ for each $x$ is already
  Gaussian,~\eqref{eq:finite-dim-zero} yields the exact result
  \begin{align}
    J_\eps = (2 \pi \eps)^{1/2} \int_{-\infty}^\infty \frac{\dd x}
    {\sqrt{{\det}'_1(\nabla S(x,0))}} = \left(\pi \eps \right)^{1/2}
    \int_{-\infty}^\infty \frac{\dd x}{\sqrt{1+x^4}} = \frac{\Gamma
      \left(\frac{1}{4} \right)^2}{2} \sqrt{\eps}\,.
  \end{align}
  Notably, in this case, ${\det}'_{n-r}(\nabla^2 S)$ is not constant
  along the family of minimizers, and the dependency of $\nabla^2 S$
  on $x$ was needed in order to obtain the correct, finite result
  despite the infinite volume of the family of minimizers. Also, in
  this example, while the action on ${\cal M}^1$ is constant (and
  equal to 0) under translations $x \to x + \delta x$, this is not
  true for the action $S$ on all of $\RR^2$.
\item Next, consider $S:\RR^2 \to \RR$, $S(x,y) = \left(x^2 + y^2 -
  a^2\right)^2$ with $a>0$, such that the set of minimizers is the
  $r=1$-dimensional manifold ${\cal M}^1 = \left\{ (x,y) \in \RR^2
  \mid x^2 + y^2 = a^2 \right\}$. Here, the eigenvalues of the
  Hessian at the minimizers are given by $\lambda_0 = 0$ and
  $\lambda_1 = 8 a^2$. In this case, the eigenvalues are independent
  of the position on~${\cal M}^1$, since the entire action is
  rotationally invariant. From~\eqref{eq:finite-dim-zero-symmetric},
  we obtain $J_\eps \overset{\eps \downarrow 0}{\sim} (2 \pi
  \eps)^{1/2} 2 \pi a (8 a^2)^{-1/2} = \pi^{3/2} \eps^{1/2}\,,$ in
  accordance with the $\eps \downarrow 0$ asymptotics of the exact
  result $J_\eps = \pi^{3/2} \eps^{1/2} \left[1 + \erf\left(a^2 /
    \eps\right) \right]/2$.
\end{enumerate}


\subsection{Moment-generating function prefactor estimates for Freidlin-Wentzell theory with zero modes}
\label{subsec:prefac-mgf-degen}

In our setup of sample path large deviation theory, we will only consider
the second scenario where the volume of the manifold factors out and is
finite. Note that in this sense, the volume part in the prefactor can
always be trivially found, such as a sphere or box volume of the
\enquote{equi-observable} hypersurfaces, and the nontrivial part of
our analysis is to find the exact way in which the Riccati approach can
be adjusted when the second variation functional possesses vanishing
eigenvalues.\\

Usually, when solving the instanton
equations~\eqref{eq:instanton-eq-z} for $(\phi_\lambda,\theta_\lambda)$ in
the situation that there is an $r$-dimensional
submanifold, $r \geq 1$, of global minimizers ${\cal M}^r_\lambda$, we will
find a specific \textit{parameterization} of ${\cal M}^r_\lambda$, $u
\mapsto \phi_\lambda^u$ for $u\in D \subseteq \RR^r$. Then, a
basis of the tangent space $T_{\phi_\lambda^u}{\cal M}^r_\lambda$ is given
by the zero modes
\begin{align}
\psi^u_{\lambda,i}
:= \pdv{\phi_\lambda^u}{u_i}:[0,T] \to \RR^n
\end{align}
with $i = 1, \dots, r$. We denote the corresponding momentum fluctuations as
\begin{align}
\xi^u_{\lambda,i} := \pdv{\theta_\lambda^u}{u_i} = a^{-1} \left[\dv{}{t} -
\nabla b(\phi_\lambda^u) - \left(\nabla a(\phi_\lambda^u)
\theta_\lambda^u \right)\right] \psi^u_{\lambda,i}\,.
\end{align}
We make the following two observations:
\begin{itemize}
\item The zero modes $\psi^u_{\lambda,i}$, $i = 1, \dots, r$ satisfy
  the Jacobi equation~\eqref{eq:jacobi-fw} (or,
  equivalently,~\eqref{eq:jacobi-fw-second-order}), since
\begin{align}
\left. \fdv{S}{\phi} \right|_{\phi_\lambda^u} = 0 \; \forall
u \in D \quad \overset{\partial/\partial u_i}{\Rightarrow}
\quad \left. \fdv[2]{S}{\phi} \right|_{\phi_\lambda^u}
\psi^u_{\lambda,i} = \Omega[\phi_\lambda^u] \psi^u_{\lambda,i} = 0 \; \forall u \in D\,,
\end{align}
as well as the boundary conditions~${\cal A}_\lambda^u$ of the second
variation, because
\begin{align}
&\phi^u_\lambda(0) = x \; \forall u \in D
\quad \overset{\partial/\partial u_i}{\Rightarrow} \quad
\psi^u_{\lambda,i}(0) = 0 \nonumber\\[.3cm]
&\theta_\lambda^u(T) = \lambda \nabla f(\phi_\lambda^u(T)) \quad 
\overset{\partial/\partial u_i}{\Rightarrow} \quad \xi^u_{\lambda,i}(T)
= \lambda \nabla^2 f(\phi_\lambda^u(T)) \psi^u_{\lambda,i}(T)\,.
\end{align} 
Hence, each of the zero modes is an admissible eigenfunction of
the Jacobi operator~$\Omega[\phi_\lambda^u]$
under ${\cal A}_\lambda^u$ with eigenvalue
$\lambda^{(0)}_i = 0$ and it follows that $\Det_{{\cal A}_\lambda}
\left(\Omega[\phi_\lambda^u] \right) = 0$.
\item We can immediately conclude that $r \leq n$ since there are at
  most $n$ linearly independent solutions of the first order Jacobi
  equation~\eqref{eq:jacobi-fw}, i.e.
\begin{align}
\dv{}{t} \left(\begin{array}{c}
\gamma\\\zeta
\end{array} \right) = \Gamma\left[\phi_\lambda^u \right]\left(\begin{array}{c}
\gamma\\\zeta
\end{array} \right)
\end{align}
that satisfy the initial condition $\gamma(0) = 0 \in \RR^n$.
\end{itemize}

It is now straightforward to formulate the analogue of
Proposition~\ref{thm:mgf-prefactor} in the presence of zero modes:


\begin{mythm}[Sharp estimates for MGFs via functional
determinants in case of broken symmetries]
  \label{thm:mgf-prefactor-degenerate}
  Denote by $\phi_\lambda^u \in {\cal M}^r_\lambda$, parameterized by $u
  \in D \subset \RR^r$, the elements of the $r$-dimensional submanifold
  of instanton solutions of the minimization problem
  \begin{equation}
   \inf_{\phi(0) = x} \left( S[\phi] - \lambda f(\phi(T)) \right)
  \end{equation}
   and by $\phi_0$ the unique ``free''
  instanton, solution to the minimization problems
  \begin{equation}
   \inf_{\phi(0) = x}
    S[\phi]
  \end{equation}
  for the Freidlin-Wentzell action~\eqref{eq:fw-action}.
  Further, for variations $\gamma:[0,T] \to \RR^n$, let
  \begin{align}
    \delta^2 S[\phi][\gamma] &= \frac{1}{2} \int_0^T \left \langle \gamma,
    \Omega[\phi] \gamma \right \rangle_n \dd t
  \end{align}
  be the second variation of $S$
  around $\phi$, where the linear operator $\Omega$ is given
  by~\eqref{eq:fw-scnd-var-op} and we impose mixed Dirichlet-Robin
  boundary conditions ${\cal A}_\lambda^{u_0}$, defined in~\eqref{eq:bc-lbda-general},
  along $\phi = \phi_\lambda^{u_0}$ for any $u_0 \in D$.
  Then we have the
  following sharp asymptotic estimate for the MGF $A_f^\eps$:
  \begin{equation}
    A_f^\eps(\lambda)\overset{\eps \downarrow 0}{\sim} (2 \pi \eps)^{-r/2} 
    \tilde{R}_\lambda
    \exp\left\{-\eps^{-1} \left( S[\phi_\lambda^{u_0}] - \lambda
    f\left(\phi_\lambda^{u_0}(T)\right) \right)\right\}
    \label{eq:mgf-general-degen}
  \end{equation}
  with
  \begin{align}
  \tilde{R}_\lambda := &\vol\left({\cal M}^r_\lambda \right)
  \left(\frac{\Det'_{{\cal A}_\lambda^{u_0}}
  \left(a(\phi_\lambda^{u_0}) \Omega[\phi_\lambda^{u_0}] \right)}
  {\Det_{{\cal A}_0} \left(a(\phi_0) \Omega[\phi_0] \right)}
  \right)^{-1/2} \times \nonumber \\
  & \hspace{2cm} \times\exp\left\{-\tfrac12 \int_0^T \left(\nabla\cdot
  b(\phi_\lambda^{u_0}) + \trace \left[\nabla a(\phi_\lambda^{u_0})
  \theta_\lambda^{u_0} \right] - \nabla\cdot b(\phi_0)\right)\,\dd t\right\}\,.
  \label{eq:ratio-prime}
  \end{align}
  Here, $\Det'$ denotes the functional
    determinant after removal of all $r$ zero eigenvalues.
\end{mythm}


For the given parameterization $u \mapsto \phi_\lambda^u$,
the volume of ${\cal M}^r_\lambda$ can be
computed as
\begin{align}
\vol\left({\cal M}^r_\lambda \right) = \int_D \sqrt{{\det}_r
\LGram{\psi^u_\lambda}{\psi^u_\lambda}}\; \dd^r u\,,
\end{align}
where $\LGram{\psi^u_\lambda}{\psi^u_\lambda} \in \RR^{r \times r}$ is the
Gram matrix defined via
\begin{align}
\LGram{\psi^u_\lambda}{\psi^u_\lambda}_{ij} := \left
\langle \psi^u_{\lambda,i},
\psi^u_{\lambda,j} \right \rangle_{L^2([0,T],\RR^n)}\,.
\end{align}


In order to be able to compute the ratio
\begin{align}
\frac{\Det'_{{\cal A}_\lambda^{u_0}}
  \left(a(\phi_\lambda^{u_0}) \Omega[\phi_\lambda^{u_0}] \right)}
  {\Det_{{\cal A}_0} \left(a(\phi_0) \Omega[\phi_0] \right)}
\end{align}
in $\tilde{R}_\lambda$ efficiently using Forman's theorem, without having to
compute and multiply all non-zero eigenvalues of both operators,
we use a technique based on boundary perturbations. The concept of
the following treatment is
described in~\cite{falco-fedorenko-gruzberg:2017}, who discuss the case
of an arbitrary number of zero modes with Dirichlet and (anti)-periodic
boundary conditions. A related paper in this regard is
also~\cite{corazza-singh:2022}. Note, however, that these references do
not derive manifestly parameterization-invariant results, and further
discuss neither the boundary conditions specific for
low dimensional observables in sample path large deviations, nor the relation
to efficient numerical prefactor computations using Riccati equations.\\


The idea of the boundary regularization procedure to compute
$\Det_{{\cal A}_\lambda^{u_0}}^\prime \left(a(\phi_\lambda^{u_0})
\Omega[\phi_\lambda^{u_0}] \right)$
is as follows:
We modify the boundary conditions ${\cal A}_\lambda^{u_0}$, realized
through $M_\lambda^{u_0}, N_\lambda^{u_0} \in \RR^{2n \times 2n}$, using a
small perturbation, that is, we replace them by $M_\lambda^{u_0}(\delta),
N_\lambda^{u_0}(\delta) \in \RR^{2n \times 2n}$ with $\delta = (\delta_1,
\dots, \delta_r) \in \RR^r$, such that $M_\lambda^{u_0}(0)=
M_\lambda^{u_0}$ and $N_\lambda^{u_0}(0)=N_\lambda^{u_0}$.
The boundary perturbation has
to be chosen in such a way as to remove all zero eigenvalues of
$\Omega[\phi_\lambda^{u_0}]$. Then we carry out the following three steps:
\begin{enumerate}
\item Explicitly compute the leading order asymptotics of the $r$
nonzero eigenvalues $\lambda_1^{(0)}(\delta), \dots, \lambda_r^{(0)}
(\delta)$ of $\Omega[\phi_\lambda^{u_0}]$ under $M_\lambda^{u_0}(\delta),
N_\lambda^{u_0}(\delta)$ that tend to 0 as $\delta \to 0$.
\item Apply Forman's theorem to evaluate the full, nonzero
determinant $\Det_{{\cal A}_\lambda^{u_0}(\delta)} \left(a(\phi_\lambda^{u_0})
\Omega[\phi_\lambda^{u_0}] \right)$.
\item Evaluate
\begin{align}
\Det_{{\cal A}_\lambda^{u_0}}' \left(a(\phi_\lambda^{u_0})
\Omega[\phi_\lambda^{u_0}] \right)
\overset{\cdot}{=}
\lim_{\delta \to 0} \left[ \frac{\Det_{{\cal A}_\lambda^{u_0}(\delta)} \left(a(\phi_\lambda^{u_0})
\Omega[\phi_\lambda^{u_0}] \right)}{\prod_{i=1}^r \lambda_i^{(0)}(\delta)} \right]\,.
\end{align}
\end{enumerate}
Of course, step 2 and 3 only make sense when considering ratios of
functional determinants; however, since it is irrelevant to the
following discussion, we omit the division by the free determinant
for the time being and denote equalities up to division by the free
determinant via \enquote{$\overset{\cdot}{=}$} as
in~\cite{falco-fedorenko-gruzberg:2017}.\\

In our setup, there are different types of regularization that
can be chosen depending on the assumptions. We start with the case
of a \textit{nonlinear} observable with positive definite matrix
\begin{align}
\bra{\psi^u_\lambda(T)} \nabla^2 f(\phi^u_\lambda(T)) \ket{\psi^u_\lambda(T)}
\in \RR^{r \times r}\,,
\end{align}
where
\begin{align}
\bra{\psi^u_\lambda(T)} \nabla^2 f(\phi^u_\lambda(T)) \ket{\psi^u(T)}_{ij}
= \left \langle \psi^u_{\lambda,i}(T), 
\nabla^2 f(\phi^u_\lambda(T)) \psi^u_{\lambda, j}(T) \right \rangle_n\,.
\end{align}
Importantly, the zero modes $(\psi^u, \xi^u)$ are, due to their
initial conditions $\psi^u(0) = 0$ and $\xi^u(0) \neq 0$, part of the 
$n$ solutions $(\gamma, \zeta)$ that make up the forward Riccati matrix
solution with $Q = \gamma \zeta^{-1}$ and $Q(0) = 0$. Now, since
$\nabla^2 f(\phi^u_\lambda(T))$ is non-degenerate on the space of final time
zero mode states $\psi^u(T)$, we conclude that $\xi^u(T)$ will also be
nondegenerate due to the boundary conditions of the zero modes. Hence, the
forward Riccati differential equation for $Q$ remains well-posed and
$Q(t)$ does not explode as $t \to T$, the only problem being the removal
of zero eigenvalues of ${\det}_n \left(1_{n \times n} - \lambda \nabla^2 f
\left(\phi_\lambda(T) \right)
Q_\lambda(T) \right)$ in Proposition~\ref{thm:gy-old}.\\

In this case, the problem can be regularized using the perturbation
\begin{align}
N_\lambda^{u_0} = \left(\begin{array}{c|c}
0_{n \times n} & 0_{n \times n}\\
\hline
-\lambda \nabla^2f(\phi_\lambda^u(T)) & 1_{n \times n}
\end{array} \right) \rightarrow N_\lambda^{u_0}(\delta) :=
 \left(\begin{array}{c|c}
0_{n \times n} & 0_{n \times n}\\
\hline
-\lambda \nabla^2f(\phi_\lambda^u(T)) & 1_{n \times n}
+ \sum_{i=1}^r \delta_i
\cdot \tilde{\xi}^{u_0}_{\lambda,i}
\otimes \tilde{\xi}^{u_0}_{\lambda,i}
\end{array} \right)
\label{eq:boundary-regular}
\end{align}
where $\left\{ \tilde{\xi}^{u_0}_{\lambda,1},
\dots,\tilde{\xi}^{u_0}_{\lambda,r}
\right\}$ is any (oriented)
orthonormal basis of the vector space $\text{span} \left\{
\xi^{u_0}_{\lambda,1}(T), \dots,
\xi^{u_0}_{\lambda,1}(T) \right\} \subset \RR^n$
spanned by the zero mode momenta at $t=T$.
Let us denote by $\psi^{u_0}_{\lambda,i}(\delta)$ the eigenfunctions of
$\Omega[\phi_\lambda^{u_0}]$ under these boundary conditions
${\cal A}_\lambda^{u_0}(\delta)$ that tend to the zero modes
$\psi^u_{\lambda,i}$
as $\delta \to 0$. Then we have the following leading order asymptotics
of $\prod_{i=1}^r \lambda_i^{(0)}(\delta)$ for step 1 with this
particular regularization:


\begin{mylemma}[Leading order behavior of the quasi-zero 
eigenvalues]\label{thm:eigval-asmyptot}
For the boundary regularization~\eqref{eq:boundary-regular},
the asymptotic behavior of the regularized zero eigenvalues
of $\Omega[\phi_\lambda^{u_0}]$ is
\begin{align}
\prod_{i=1}^r \lambda^{(0)}_i(\delta) \overset{\delta \to 0}{\sim}
\frac{{\det}_r \bra{\psi^{u_0}_\lambda(T)} \lambda \nabla^2 f(
\phi^{u_0}_\lambda(T)) \ket{\psi^{u_0}_\lambda(T)}}{{\det}_r
\LGram{\psi^{u_0}_\lambda}{\psi^{u_0}_\lambda}} \prod_{i=1}^r \delta_i\,.
\end{align}
\end{mylemma}


\begin{myproof}{Lemma~\ref{thm:eigval-asmyptot}}

The modified boundary conditions at $t = T$ read
\begin{align}
\left \langle \zeta(T) - \lambda  \nabla^2f(\phi_\lambda^u(T))
\gamma(T),\tilde{\xi}^{u_0}_{\lambda,i} \right \rangle_n = -
\delta_i \left \langle \zeta(T), \tilde{\xi}^{u_0}_{\lambda,i} \right
\rangle_n\,, \quad i = 1, \dots, r\,.
\end{align}
For any $i,j \in \left\{1, \dots, r \right\}$, we compute
\begin{align}
&\left \langle \psi^{u_0}_{\lambda,i}, \Omega[\phi_\lambda^{u_0}]
\psi^{u_0}_{\lambda,j}(\delta)  \right \rangle_{L^2([0,T],\RR^n)}
\overset{\text{eigenvalue}}{=} \lambda^{(0)}_j(\delta) \left
\langle \psi^{u_0}_{\lambda,i}, \psi^{u_0}_{\lambda,j}(\delta)  \right
\rangle_{L^2([0,T],\RR^n)} \nonumber\\
&\overset{\text{adjoining } \Omega}{=} \left. \left \langle
\xi^{u_0}_{\lambda,i}, \psi^{u_0}_{\lambda,j}(\delta)  \right \rangle_n
\right \vert^T_0 - \left. \left \langle \psi^{u_0}_{\lambda,i},
\xi^{u_0}_{\lambda,j}(\delta) \right \rangle_n \right \vert^T_0 +
\underbrace{\left \langle \Omega[\phi_\lambda^{u_0}] \psi^{u_0}_{\lambda,i},
\psi^{u_0}_{\lambda,j}(\delta)  \right \rangle_{L^2([0,T],\RR^n)
}}_{=0}\nonumber\\
&\overset{\text{boundary conditions}}{=} -\left \langle
\psi^{u_0}_{\lambda,i}(T), \xi^{u_0}_{\lambda,j}(\delta)(T) - \lambda
\nabla^2 f(\phi^{u_0}_\lambda(T)) \psi^{u_0}_{\lambda,j}(\delta)(T) \right
\rangle_n \nonumber\\
&= -\sum_{k = 1}^r \left \langle \psi^{u_0}_{\lambda,i}(T),
\tilde{\xi}^{u_0}_{\lambda,k} \right \rangle_n \underbrace{\left \langle
\tilde{\xi}^{u_0}_{\lambda,k}, \xi^{u_0}_{\lambda,j}(\delta)(T) -
\lambda \nabla^2 f(\phi^{u_0}_\lambda(T)) \psi^{u_0}_{\lambda,j}(
\delta)(T) \right
\rangle_n}_{= - \delta_k \left \langle \tilde{\xi}^{u_0}_{\lambda,k},
\xi^{u_0}_{\lambda,j}(\delta)(T) \right \rangle_n} \nonumber\\
&= \left(\bra{\psi^{u_0}_{\lambda}(T)}\ket{\tilde{\xi}^{u_0}_{\lambda}}
\text{diag}_r(\delta) \bra{\tilde{\xi}^{u_0}_{\lambda}}\ket{\xi^{u_0}_{\lambda}
(\delta)(T)} \right)_{ij} \nonumber\\
&\overset{\text{cf.\ first line}}{=} \left(\LGram{\psi^{u_0}_{\lambda}}{
\psi^{u_0}_{\lambda}(\delta)} \text{diag}_r(\lambda^{(0)}(\delta))
\right)_{ij}\,.
\end{align} 
Computing the determinant of these expressions yields
\begin{align}
\prod_{i=1}^r \lambda^{(0)}_i(\delta) = \frac{{\det}_r
\bra{\psi^{u_0}_\lambda(T)}\ket{\xi^{u_0}_\lambda(\delta)(T)}}{{\det}_r
\LGram{\psi^{u_0}_\lambda}{\psi^{u_0}_\lambda(\delta)}} \prod_{i=1}^r
\delta_i \overset{\delta \to 0}{\sim} \frac{{\det}_r
\bra{\psi^{u_0}_\lambda(T)}\ket{\xi^{u_0}_\lambda(T)}}{{\det}_r
\LGram{\psi^{u_0}_\lambda}{\psi^{u_0}_\lambda}} \prod_{i=1}^r \delta_i\,.
\end{align}
In the last step, note that it will not be true in general
that $\psi^{u_0}_{\lambda,i}(\delta) \to \psi^{u_0}_{\lambda,i}$ as
$\delta \to 0$ for each $i = 1, \dots, r$ individually
(cf.~\cite{falco-fedorenko-gruzberg:2017}), but due to
linearity, the transformation matrices from $\lim_{\delta \to 0}
\psi^{u_0}_{\lambda}(\delta)$ to $\psi^{u_0}_{\lambda}$ and from $\lim_{\delta
\to 0} \xi^{u_0}_{\lambda}(\delta)$ to $\xi^{u_0}_{\lambda}$ will coincide and
their determinants therefore cancel in the last step.
\end{myproof}


\begin{mylemma}[Forman's theorem for the perturbed boundary conditions]
\label{thm:forman-perturbed-1}
For the boundary regularization~\eqref{eq:boundary-regular} and any
$\delta \in \RR^r$, the functional determinant of $\Omega[\phi_z^{u_0}]$
under ${\cal A}_z^{u_0}(\delta)$ can be expressed as
\begin{align}
\Det_{{\cal A}_\lambda^{u_0}(\delta)} \left(a(\phi_\lambda^{u_0})
\Omega[\phi_\lambda^{u_0}] \right) &\overset{\cdot}{=}
{\det}_{n-r}' \left( \zeta(T) - \lambda \nabla^2 f(\phi^{u_0}_\lambda(T)
\gamma(T) \right) \times \nonumber \\ 
&\hspace{3cm}\times \frac{\sqrt{{\det}_r \bra{\xi^{u_0}_
\lambda(T)}\ket{\xi^{u_0}_\lambda(T)}}}{{\det}_r
\left(\bra{\tilde{\xi}^{u_0}_\lambda}\ket{
\xi^{u_0}_\lambda(0)} \right)} \cdot \left( \prod_{i=1}^r
\delta_i \right)\,,
\end{align}
where $(\gamma, \zeta):[0,T] \to \RR^{2n \times n}$ is the solution of
\begin{align}
\dv{}{t} \left(\begin{array}{c}
\gamma\\\zeta
\end{array} \right) = \Gamma \left[\phi_\lambda^{u_0}
\right] \left(\begin{array}{c}
\gamma\\\zeta
\end{array} \right)\,, \quad \left(\begin{array}{c}
\gamma(0)\\\zeta(0)
\end{array} \right) = \left(\begin{array}{c}
0_{n \times n}\\
\left(\xi^{u_0}_{\lambda,1}(0), \dots, \xi^{u_0}_{\lambda,r}(0),
v_1, \dots, v_{n-r} \right)
\end{array}\right)\,.
\end{align}
\end{mylemma}


\begin{myproof}{Lemma~\ref{thm:forman-perturbed-1}}

We pick an orthonormal basis of $\RR^n$ by extending $\left\{
\tilde{\xi}^{u_0}_{\lambda,1}, \dots,\tilde{\xi}^{u_0}_{\lambda,r}\right\}$
by $n - r$ additional unit vectors~$v_1, \dots, v_{n-r}$.
In this basis, the right boundary
matrix $N_\lambda^{u_0}(\delta)$ from~\eqref{eq:boundary-regular} becomes
\begin{align}
N_\lambda^{u_0}(\delta) :=
\left(\begin{array}{c|c}
0_{n \times n} & 0_{n \times n}\\
\hline
-\lambda \nabla^2f(\phi_\lambda^u(T)) & \begin{array}{c|c} 1_{r \times r}
+\text{diag}_r \left(\delta\right)
& 0_{r \times (n-r)}\\\hline
0_{(n-r) \times r} & 1_{(n-r) \times (n-r)}\end{array}
\end{array} \right)\,.
\end{align}
For the fundamental system of solutions $\Upsilon$,
we choose the initial condition
\begin{align}
\Upsilon(0) = \left(\begin{array}{c|c}
1_{n \times n} & 0_{n \times n}\\
\hline
0_{n \times n} & \begin{array}{c|c}
\bra{\tilde{\xi}^{u_0}_{\lambda}}\ket{\xi^{u_0}_\lambda(t=0)}
& 0_{r \times (n-r)}\\
\hline
\bra{v}\ket{\xi^{u_0}_\lambda(t=0)} & 1_{(n-r) \times (n-r)}
\end{array}
\end{array} \right)
\end{align}
such that
\begin{align}
{\det}_{2n} \Upsilon(0) = {\det}_n \zeta(0) = {\det}_n
\left(\xi^{u_0}_{\lambda,1}(0), \dots, \xi^{u_0}_{\lambda,r}(0), v_1,
\dots, v_{n-r} \right) = {\det}_r \left(\bra{\tilde{\xi}^{u_0}_\lambda}\ket{
\xi^{u_0}_\lambda(t=0)} \right)
\end{align}
and
\begin{align}
&{\det}_{2n} \left(M^{u_0}_\lambda \Upsilon(0) + N^{u_0}_\lambda(\delta)
\Upsilon(T) \right) \nonumber \\
&= {\det}_{2n} \left(\begin{array}{c|c}
1_{n \times n} & 0_{n \times n}\\
\hline
(*)_{n \times n}
& \begin{array}{c|c}
\text{diag}_r \left(\delta\right) \bra{\tilde{\xi}^{u_0}_{
\lambda}}\ket{\xi^{u_0}_\lambda(t=T)} & (*)_{r \times (n-r)}\\
\hline
0_{(n-r) \times r} & \left[\zeta(T) - \lambda \nabla^2(
\phi^{u_0}_\lambda(T)) \gamma(T)\right]_{\perp}
\end{array}
\end{array} \right)\nonumber\\
&= \left( \prod_{i=1}^r \delta i \right) \sqrt{{\det}_r
\bra{\xi^{u_0}_z(t=T)}\ket{\xi^{u_0}_z(t=T)}} {\det}_{n-r}' \left(
\zeta(T) - \lambda \nabla^2 f(\phi^{u_0}_\lambda(T) \gamma(T) \right)\,.
\end{align}
\end{myproof}\\


Combining the previous two lemmas with
Proposition~\ref{thm:mgf-prefactor-degenerate} and observing that
for the solutions $(\gamma, \zeta)$ of the Jacobi equation in
Lemma~\ref{thm:forman-perturbed-1}, we have
\begin{align}
&{\det}_{n-r}' \left( \zeta(T) - \lambda \nabla^2 f(\phi^{u_0}_\lambda(T))
\gamma(T) \right) = {\det}_{n-r}' \left( 1_{n \times n} - \lambda
\nabla^2 f(\phi^{u_0}_\lambda(T)) Q(T) \right) \frac{\det_n \zeta(T)}
{\sqrt{{\det}_r \bra{\xi^{u_0}_\lambda(T)}\ket{\xi^{u_0}_\lambda(T)}}}\,,
\end{align}
which yields the following
concrete formula to evaluate the MGF prefactor in the presence of zero modes
for nondegenerate, nonlinear observables:


\begin{mythm}[MGF prefactor with zero modes via \textbf{forward}
Riccati equation for nondegenerate, nonlinear observables]
\label{thm:riccati-forward-nonlin-obs}
The prefactor $\tilde{R}_\lambda$ in~\eqref{eq:ratio-prime}
can be computed as
\begin{align}
\tilde{R}_\lambda &= \frac{\exp \left\{\frac{1}{2} \int_0^T
\trace \left[ \left( \left \langle \nabla^2 b(\phi_\lambda^{u_0}),
\theta_\lambda^{u_0} \right
\rangle_n + \tfrac{1}{2} \left \langle \theta_\lambda^{u_0},
\nabla^2 a(\phi_\lambda^{u_0})
\theta_\lambda^{u_0} \right
\rangle_{n} \right) Q_\lambda^{u_0} \right] \dd t\right\}}{
\left[{\det}_{n-r}' \left(1_{n \times n} - \lambda \nabla^2
f(\phi^{u_0}_\lambda(T)) Q^{u_0}_\lambda(T)
\right)\right]^{1/2}} \times \nonumber \\
&\hspace{2cm} \times \int_D \sqrt{{\det}_r \bra{\psi^{u}_\lambda(T)}
\lambda \nabla^2 f(\phi^{u}_\lambda(T)) \ket{\psi^{u}_\lambda(T)}}\;
\dd^r u
\label{eq:prefac-riccati-degen}
\end{align}
for any $u_0 \in \RR^r$, where $Q^{u_0}_\lambda : [0,T]
\to \RR^{n \times n}$ solves the forward Riccati equation
\begin{align}
\begin{cases}
\dot{Q}_\lambda^{u_0} = a(\phi_\lambda^{u_0}) + Q_\lambda^{u_0} \left[ \nabla b\left(\phi_\lambda^{u_0}
\right)^\top + \left( \nabla a(\phi_\lambda^{u_0}) \theta_\lambda^{u_0} \right)^\top \right] \\
\quad +
\left[ \nabla b\left(\phi_\lambda^{u_0} \right) + \left(\nabla a(\phi_\lambda^{u_0})
\theta_\lambda^{u_0} \right)\right] Q_\lambda^{u_0} + Q_\lambda^{u_0} \left[ \left< \nabla^2
b(\phi_\lambda^{u_0}), \theta_\lambda^{u_0}\right>_n + \tfrac{1}{2} \left \langle
\theta_\lambda^{u_0}, \nabla^2 a(\phi_\lambda^{u_0}) \theta_\lambda^{u_0} \right
\rangle_{n} \right] Q_\lambda^{u_0} \,,\\
Q_\lambda^{u_0}(0) = 0_{n \times n} \in \RR^{n \times n}\,.
\end{cases}
\end{align}
\end{mythm}


The second case that we consider is when the matrix 
\begin{align}
\bra{\psi^u_\lambda(T)} \nabla^2 f(\phi^u_\lambda(T)) \ket{\psi^u_\lambda(T)}
\in \RR^{r \times r}
\end{align}
is not positive definite, which is in particular relevant for the
important case of linear observables. Here, the
regularization procedure of the previous proposition will not
work and the solution of the Riccati matrices with unmodified initial or
final conditions can diverge since the zero modes  can provide solutions of
the Jacobi equation~\eqref{eq:jacobi-fw}
with $\gamma(0) = 0$ and $\zeta(T) = 0$.
We will instead suppose in the following that the
matrix
\begin{align}
\bra{\psi^u_\lambda(T)} 1_{n \times n} + \lambda
\nabla^2 f(\phi^u_\lambda(T)) \ket{\psi^u_\lambda(T)}
\in \RR^{r \times r}
\end{align}
is positive definite and regularize
the final time boundary condition as
\begin{align}
N_\lambda^{u_0} = \left(\begin{array}{c|c}
0_{n \times n} & 0_{n \times n}\\
\hline
-\lambda \nabla^2f(\phi_\lambda^{u_0}(T)) & 1_{n \times n}
\end{array} \right) \rightarrow N_\lambda^{u_0}(\delta) :=
 \left(\begin{array}{c|c}
0_{n \times n} & 0_{n \times n}\\
\hline
-\lambda \nabla^2f(\phi_\lambda^{u_0}(T)) + \sum_{i=1}^r \delta_i
\cdot \tilde{\psi}^{u_0}_{\lambda,i}
\otimes \tilde{\psi}^{u_0}_{\lambda,i} & 1_{n \times n}
\end{array} \right)\,,
\label{eq:boundary-regular-alternative}
\end{align}
where $\left\{ \tilde{\psi}^{u_0}_{\lambda,1},
\dots,\tilde{\psi}^{u_0}_{\lambda,r}
\right\}$ is any
orthonormal basis of the vector space $\text{span} \left\{
\psi^{u_0}_{\lambda,1}(T), \dots,
\psi^{u_0}_{\lambda,r}(T) \right\} \subset \RR^n$
spanned by the zero modes at $t=T$. Going through a similar
calculation as above results in the following
proposition~\ref{thm:prefac-riccati-degen-bkwd}, now with
\begin{align}
\prod_{i=1}^r \lambda^{(0)}_i(\delta) \overset{\delta \to 0}{\sim}
\frac{{\det}_r \bra{\psi^{u_0}_\lambda(T)}\ket{\psi^{u_0}_\lambda(T)}}{{\det}_r
\LGram{\psi^{u_0}_\lambda}{\psi^{u_0}_\lambda}} \prod_{i=1}^r \delta_i
\end{align}
for the quasi-zero eigenvalue behavior as $\delta \to 0$, and final condition
\begin{align}
\Upsilon(T) = \left(\begin{array}{c|c}
1_{n \times n} & 1_{n \times n}\\
\hline
\lambda \nabla^2 f\left(\phi_\lambda^{u}(T)\right) & \begin{array}{c|c}
-1_{r \times r}
& \bra{\tilde{\psi}_\lambda^{u_0}}\lambda \nabla^2
f\left(\phi_\lambda^{u}(T)\right) \ket{v}\\
\hline
\bra{v}\lambda \nabla^2 f\left(\phi_\lambda^{u}(T)\right)
\ket{\tilde{\psi}_\lambda^{u_0}} & 1_{(n-r) \times (n-r)}
\end{array}
\end{array} \right)
\end{align}
for the fundamental system of solutions $\Upsilon$
in an orthonormal basis $\left\{
\tilde{\psi}^{u_0}_{\lambda,1}, \dots,\tilde{\psi}^{u_0}_{\lambda,r}, 
v_1, \dots, v_{n-r}\right\}$:


\begin{mythm}[MGF prefactor with zero modes via \textbf{backward}
Riccati equation]
\label{thm:prefac-riccati-degen-bkwd}
The prefactor $\tilde{R}_\lambda$ in~\eqref{eq:ratio-prime}
for a linear observable $f:\RR^n \to \RR$ can be computed as
\begin{align}
\tilde{R}_\lambda = \exp \left\{\frac{1}{2} \int_0^T
\trace \left[a(\phi_\lambda^{u_0}) W_\lambda^{u_0}\right] \dd t\right\}
\int_D \sqrt{{\det}_r \bra{\psi^u_\lambda(T)} 1_{n \times n}
+ \lambda \nabla^2 f(\phi^u_\lambda(T)) \ket{\psi^u_\lambda(T)}}
\; \dd^r u
\label{eq:prefac-riccati-degen-bkwd}
\end{align}
for any $u_0 \in \RR^r$, where $W^{u_0}_\lambda : [0,T]
\to \RR^{n \times n}$ solves the backward Riccati equation
\begin{align}
\begin{cases}
\dot{W}_\lambda^{u_0} &= -W_\lambda^{u_0} a(\phi_\lambda^{u_0}) W_\lambda^{u_0} -
\left[ \nabla b\left(\phi_\lambda^{u_0}
\right)^\top + \left( \nabla a(\phi_\lambda^{u_0}) \theta_\lambda^{u_0} \right)^\top
\right]W_\lambda^{u_0} \\
& \quad- W_\lambda^{u_0}
\left[ \nabla b\left(\phi_\lambda^{u_0} \right) + \left(\nabla a(\phi_\lambda^{u_0})
\theta_\lambda^{u_0} \right)\right]- \left< \nabla^2
b(\phi_\lambda^{u_0}), \theta_\lambda^{u_0}\right>_n - \tfrac{1}{2} \left \langle
\theta_\lambda^{u_0}, \nabla^2 a(\phi_\lambda^{u_0}) \theta_\lambda^{u_0} \right
\rangle_{n} \,,\\
W_\lambda^{u_0}(T) &= \lambda \nabla^2 f(\phi^{u_0}_\lambda(T))
- \ket{\tilde{\psi}^{u_0}_{\lambda}} \bra{\tilde{\psi}^{u_0}_{\lambda}}
1_{n \times n} + \lambda \nabla^2 f(\phi^{u_0}_\lambda(T))
\ket{\tilde{\psi}^{u_0}_{\lambda}} \bra{\tilde{\psi}^{u_0}_{\lambda}}   
\,.
\end{cases}
\end{align}
\end{mythm}


\begin{myrmk}
The final condition of the backward Riccati matrix
in Proposition~\ref{thm:prefac-riccati-degen-bkwd} is to be
understood as
\begin{align}
W_\lambda^{u_0}(T) = \lambda \nabla^2
f (\phi^{u_0}_\lambda(T)) - \sum_{i = 1}^r \sum_{j = 1}^r
\bra{\tilde{\psi}^{u_0}_{\lambda}}
1_{n \times n} + \lambda \nabla^2 f(\phi^{u_0}_\lambda(T))
\ket{\tilde{\psi}^{u_0}_{\lambda}}_{ij}
\tilde{\psi}^{u_0}_{\lambda,i} \otimes \tilde{\psi}^{u_0}_{\lambda,j}
\in \RR^{n \times n}
\end{align}
in index notation, with
\begin{align}
\bra{\tilde{\psi}^{u_0}_{\lambda}}
1_{n \times n} + \lambda \nabla^2 f(\phi^{u_0}_\lambda(T))
\ket{\tilde{\psi}^{u_0}_{\lambda}}_{ij} = 
\left \langle \tilde{\psi}^{u_0}_{\lambda, i},\left[ 1_{n \times n}
+ \lambda \nabla^2 f(\phi^{u_0}_\lambda(T)) \right] \tilde{\psi}^{u_0}_{\lambda, j}  \right \rangle_n \in \RR
\end{align}
as usual. For linear observables~$f$, it reduces to
\begin{align}
W_\lambda^{u_0}(T) &= -\sum_{i=1}^r \tilde{\psi}^{u_0}_{\lambda,i}
\otimes \tilde{\psi}^{u_0}_{\lambda,i}\,.
\end{align}
\end{myrmk}


\subsection{Probability density function prefactor estimates for Freidlin-Wentzell theory with zero modes}
\label{subsec:prefac-degen-pdf-trafo}

Again performing an inverse Laplace transform leads to a proposition
for PDF prefactors in the presence of zero modes. This is the main
result of the paper. It constitutes a complete recipe for the
computation of the PDF when zero modes are present, since every
quantity can be evaluated numerically, after numerically integrating a
Riccati equation along the symmetry broken instanton.

\begin{mythm}[PDF prefactor estimate with zero modes]
\label{thm:pdf-orefac-degen-final}
For any $z \in \RR$ and with $r$ zero modes, we have
  \begin{align}
  \rho_f^\eps(z) \overset{\eps \downarrow 0}{\sim} \left(2 \pi \eps
  \right)^{-\frac{r+1}{2}} \tilde{R}_{z}
  \left[\left.\dv{}{\lambda}\right
  \rvert_{\lambda_z} f\left(\phi_\lambda^{u_0}(T)\right) \right]^{-1/2} \exp \left\{
  -\frac{1}{\eps} S \left[\phi_{z}^{u_0} \right] \right\}\,,
  \label{eq:pdf-prefac-general-degen}
  \end{align}
  with $\lambda_z$ determined by $f(\phi_{\lambda_z}(T)) = z$ and
  \begin{enumerate}[(i)]
\item For nonlinear observables with positive definite matrix
\begin{align}
\bra{\psi^u_z(T)} \nabla^2 f(\phi^u_z(T)) \ket{\psi^u_z(T)}
\in \RR^{r \times r}
\end{align}
 the prefactor can be computed as
 \begin{align}
\tilde{R}_z &= \frac{\exp \left\{\frac{1}{2} \int_0^T
\trace \left[ \left( \left \langle \nabla^2 b(\phi_z^{u_0}),
\theta_z^{u_0} \right
\rangle_n + \tfrac{1}{2} \left \langle \theta_z^{u_0},
\nabla^2 a(\phi_z^{u_0})
\theta_z^{u_0} \right
\rangle_{n} \right) Q_z^{u_0} \right] \dd t\right\}}{
\left[{\det}_{n-r}' \left(1_{n \times n} - \lambda_z \nabla^2
f(\phi^{u_0}_z(T)) Q_z^{u_0}(T)
\right)\right]^{1/2}} \times \nonumber \\
&\hspace{2cm}\times \int_D \sqrt{{\det}_r \bra{\psi^{u}_z(T)}
\lambda_z \nabla^2 f(\phi^{u}_z(T)) \ket{\psi^{u}_z(T)}}
\; \dd^r u
\label{eq:prefac-pdf-degen-forward}
\end{align}
for any $u_0 \in \RR^r$, where $Q^{u_0}_z : [0,T]
\to \RR^{n \times n}$ solves the forward Riccati equation
\begin{align}
\begin{cases}
\dot{Q}_z^{u_0} = a(\phi_z^{u_0}) + Q_z^{u_0} \left[ \nabla b\left(\phi_z^{u_0}
\right)^\top + \left( \nabla a(\phi_z^{u_0}) \theta_z^{u_0}
\right)^\top \right] \\
\quad +
\left[ \nabla b\left(\phi_z^{u_0} \right) + \left(\nabla a(\phi_z^{u_0})
\theta_z^{u_0} \right)\right] Q_z^{u_0} + Q_z^{u_0} \left[ \left< \nabla^2
b(\phi_z^{u_0}), \theta_z^{u_0}\right>_n + \tfrac{1}{2} \left \langle
\theta_z^{u_0}, \nabla^2 a(\phi_z^{u_0}) \theta_z^{u_0} \right
\rangle_{n} \right] Q_z^{u_0} \,,\\
Q_z^{u_0}(0) = 0_{n \times n} \in \RR^{n \times n}\,.
\end{cases}
\end{align}
\item For observables with positive definite matrix
\begin{align}
\bra{\psi^u_z(T)} 1_{n \times n} + \lambda \nabla^2
f(\phi^u_z(T)) \ket{\psi^u_z(T)}
\in \RR^{r \times r}
\end{align}
the prefactor can be computed as
\begin{align}
\tilde{R}_z &= \exp \left\{\frac{1}{2} \int_0^T
\trace \left[a(\phi_z^{u_0}) W_z^{u_0}\right] \dd t\right\} \times \nonumber \\
& \hspace{2cm}\times \int_D
\sqrt{{\det}_r \bra{\psi^u_z(T)} 1_{n \times n} + \lambda
\nabla^2 f(\phi^u_z(T)) \ket{\psi^u_z(T)}}
\; \dd^r u
\label{eq:prefac-pdf-degen-backward}
\end{align}
for any $u_0 \in \RR^r$, where $W^{u_0}_z : [0,T]
\to \RR^{n \times n}$ solves the backward Riccati equation
\begin{align}
\begin{cases}
\dot{W}_z^{u_0} &= -W_z^{u_0} a(\phi_z^{u_0}) W_z^{u_0} -
\left[ \nabla b\left(\phi_z^{u_0}
\right)^\top + \left( \nabla a(\phi_z^{u_0}) \theta_z^{u_0} \right)^\top
\right]W_z^{u_0} \\
& \quad- W_z^{u_0}
\left[ \nabla b\left(\phi_z^{u_0} \right) + \left(\nabla a(\phi_z^{u_0})
\theta_z^{u_0} \right)\right]- \left< \nabla^2
b(\phi_z^{u_0}), \theta_z^{u_0}\right>_n - \tfrac{1}{2} \left \langle
\theta_z^{u_0}, \nabla^2 a(\phi_z^{u_0}) \theta_z^{u_0} \right
\rangle_{n} \,,\\
W_z^{u_0}(T) &= \lambda_z \nabla^2 f(\phi^{u_0}_z(T)) -
\ket{\tilde{\psi}^{u_0}_z} \bra{\tilde{\psi}^{u_0}_z}
1_{n \times n} + \lambda_z \nabla^2 f(\phi^{u_0}_z(T))
\ket{\tilde{\psi}^{u_0}_z} \bra{\tilde{\psi}^{u_0}_z}   
\,.
\end{cases}
\end{align}
  \end{enumerate}
\end{mythm}


Alternatively, the regularization on the left boundary
\begin{align}
M_z^{u_0} = \left(\begin{array}{c|c}
1_{n \times n} & 0_{n \times n}\\
\hline
0_{n \times n} & 0_{n \times n}
\end{array} \right) \rightarrow M_z^{u_0}(\delta) :=
\left(\begin{array}{c|c}
1_{n \times n} & \sum_{i=1}^r \delta_i \cdot \tilde{\xi}^{u_0}_{z,i}
\otimes \tilde{\xi}^{u_0}_{z,i}\\
\hline
0_{n \times n} & 0_{n \times n}
\end{array} \right)
\label{eq:boundary-regular-left}
\end{align}
leads to the following expression for the PDF prefactor using the
same techniques as outlined above:


\begin{mythm}[PDF prefactor with zero modes via \textbf{forward}
Riccati equation with modified initial condition]
\label{thm:riccati-zero-left-boundary}
The prefactor $\tilde{R}_z$ in the asymptotic estimate
\begin{align}
\rho_f(z) \overset{\eps \downarrow 0}{\sim} (2 \pi \eps)^{
-\frac{r+1}{2}} \tilde{R}_z \exp \left\{-\frac{1}{\eps} 
S\left[\phi^{u_0}_z \right] \right\}
\end{align}
for the PDF $\rho_f$ in the presence of $r$ zero modes can be computed as
\begin{align}
\tilde{R}_z = \frac{\exp \left\{\frac{1}{2} \int_0^T
\trace \left[ \left( \left \langle \nabla^2 b(\phi_z^{u_0}),
\theta_z^{u_0} \right
\rangle_n + \tfrac{1}{2} \left \langle \theta_z^{u_0},
\nabla^2 a(\phi_z^{u_0})
\theta_z^{u_0} \right
\rangle_{n} \right) Q_z^{u_0} \right] \dd t\right\}}{
\left[(-1)^r{\det}_n U_z^{u_0} \; \left \langle \nabla f (\phi_z^{u_0}(T))
, Q_z^{u_0}(T) \left(U_z^{u_0}\right)^{-1} \nabla f (\phi_z^{u_0}(T))
\right \rangle_n\right]^{1/2}} \cdot \vol\left(\theta_z(0) \right)
\label{eq:prefac-riccati-degen-mod}
\end{align}
for any $u_0 \in \RR^r$, where $Q^{u_0}_z : [0,T] \to \RR^{n \times n}$ solves
\begin{align}
\begin{cases}
\dot{Q}_z^{u_0} = a(\phi_z^{u_0}) +
\left[ \nabla b\left(\phi_z^{u_0}
\right)^\top + \left( \nabla a(\phi_z^{u_0}) \theta_z^{u_0}
\right)^\top \right] \\
\quad +
\left[ \nabla b\left(\phi_z^{u_0} \right) + \left(\nabla a(\phi_z^{u_0})
\theta_z^{u_0} \right)\right] Q_z^{u_0} + Q_z^{u_0} \left[ \left< \nabla^2
b(\phi_z^{u_0}), \theta_z^{u_0}\right>_n + \tfrac{1}{2} \left \langle
\theta_z^{u_0}, \nabla^2 a(\phi_z^{u_0}) \theta_z^{u_0} \right
\rangle_{n} \right] Q_z^{u_0} \,,\\
Q_z^{u_0}(0) = \sum_{i = 1}^r \tilde{\xi}^{u_0}_{z,i}
\otimes \tilde{\xi}^{u_0}_{z,i}\,,
\end{cases}
\end{align}
as in the non-degenerate case
and $\vol\left(\theta_z(0) \right)$ is the $r$-dimensional volume
of $\{\theta_z^{u}(t=0)\mid u \in D\}$ that can be computed as
\begin{align}
\vol\left(\theta_z(0) \right) = \int_D \sqrt{{\det}_r
\bra{\xi^{u}_z(0)}\ket{\xi^{u}_z(0)}} \dd^r u\,.
\end{align}
\end{mythm}


\begin{myrmk}
Note that, again, the initial conditions were modified in a suitable
way as to remove divergences from the Riccati equation and render the 
determinants in the denominator non-zero. While this result is convenient
in that it can be used regardless of whether the Hessian $\nabla^2
f(\phi_z^{u_0}(T))$ is non-singular, it may be inconvenient for taking
the stationary limit $T \to \infty$. As an example, consider an SDE with
additive noise and initial position $x = x_*$ at the fixed point. Then
$\vol\left(\theta_z(0)
\right)$ will tend to $0$ in this case for $T \to \infty$. Similarly,
the Riccati matrix $Q$
will \enquote{forget} its regularizing initial condition and instead
tend to its stationary solution $Q_*$ determined by the Lyapunov equation
\begin{align}
0 = a + \nabla b(x_*) Q_* + Q_* \nabla b(x_*)^\top\,.
\end{align}
\end{myrmk}

\begin{myrmk}
We observe that the determinant of the $L^2$-scalar products of the
zero modes in~\eqref{eq:ratio-prime} cancels in each of the expressions
which we have derived via boundary regularization, and we are always left
only with integrations over the zero modes at the initial or final time~$T$.
This is a generic feature of the regularization procedure as remarked
already in~\cite{falco-fedorenko-gruzberg:2017}.
\end{myrmk}

 
\section{Examples}
\label{sec:examples}

In this section we illustrate the application of the propositions to
compute PDF prefactors in the presence of zero modes in four
instructive examples. We start with the arguably simplest case in
subsection~\ref{subsec:n-dimens-ornst}: A multidimensional
Ornstein-Uhlenbeck process with a purely radial, linear vector field
as drift and the norm of the process as the observable as sketched in
Figure~\ref{fig:inst-sketch} (left). Here, all results on both finite
and infinite time horizons $T$ can be found analytically. In
subsection~\ref{subsec:2d-swirl}, we consider again a diffusion
process in a rotationally symmetric vector field with the radius as
our observable. Here the vector field is constructed to be non-linear
and to possess an angular component to break the detailed balance
property of the process. In the limit $T \to \infty$, the problem can
again be solved exactly, and, in addition to this limiting case, we
compare the numerical solution of the instanton and Riccati equations
to direct sampling of the SDE for finite times.  Third, in
subsection~\ref{subsec:dynam-phase-trans-ode-potential}, we analyze a
three-dimensional diffusion process in a potential landscape of the
type sketched in Figure~\ref{fig:inst-sketch} (right). This is the
first concrete example with a dynamical phase transition that is
considered in this paper, and, restricting ourselves to the infinite
time limit $T \to \infty$ for clarity, we show that the Riccati
formalism correctly predicts the PDF prefactor in the quadratic
approximation and compare it to the full prefactor at different finite
noise strengths~$\eps > 0$. Finally, in
subsection~\ref{subsec:pde-example}, we show by means of the
one-dimensional KPZ equation with a dynamical phase transition for the
average surface height that the formalism developed in this paper
remains formally applicable and numerically feasible for
out-of-equilibrium systems with infinitely many spatial degrees of
freedom. Numerical applications to spatially extended systems in fluid
dynamics and turbulence theory are left as a subject of future,
separate publications.


\subsection{$n$-dimensional Ornstein-Uhlenbeck process with radius as observable}
\label{subsec:n-dimens-ornst}

We consider the case of an $n$-dimensional Ornstein-Uhlenbeck process with
$n \geq 2$, as sketched in Figure~\ref{fig:inst-sketch} (left) for $n = 2$,
\begin{equation}
  \dd X_t^\eps = -\beta X_t^\eps\:\dd t + \sqrt{2\eps}\:\dd B_t\,,\qquad X_0^\eps = 0\,.
\end{equation}
We take $b(x) = - \beta x$ for the drift with $\beta > 0$, $a = 2 \cdot
1_{n \times n}$ for the diffusion matrix and $f(x) = \norm{x}_n$ for the
observable. In this case, the radial symmetry will always necessarily be
broken by the instanton at any $z > 0$ and
generate $n - 1$ zero modes. As a reference, the PDF $\rho^\eps$ of
$X^\eps_T$ is always Gaussian for any $T > 0$ with
\begin{align}
\rho^\eps(x) = \left(2 \pi \eps \right)^{-n/2} \left[\frac{\beta}{1 -
\exp \left\{ -2 \beta T\right\}} \right]^{n/2} \exp \left\{-\frac{1}{\eps
} \frac{\beta \norm{x}_n^2}{2 \left(1 - \exp \left\{ -2 \beta T\right\}
\right)} \right\}\,.
\end{align}
Note that the prefactor of the full PDF,
given by $\left(2 \pi \eps \right)^{-n/2} \left[\beta / (1 -
\exp \left\{ -2 \beta T\right\})\right]^{n/2}$, is just a constant in $x$,
such that the reference radial PDF
\begin{align}
\rho^\eps_f(z) = \left(2 \pi \eps \right)^{-n/2} \vol_{n-1}
\left(S^{n-1}\right) \left[\frac{\beta}{1 - \exp \left\{ -2
\beta T\right\}} \right]^{n/2} z^{n-1} \exp \left\{-\frac{1}{\eps}
\frac{\beta z^2}{2 \left(1 - \exp \left\{ -2 \beta T\right\}\right)}
\right\}
\label{eq:pdf-ou}
\end{align}
with $\vol_{n-1}\left(S^{n-1}\right) = 2 \pi^{n/2} /
\Gamma(n/2)$ merely acquires a $z$-dependent prefactor
through the multiplication with a hypersphere volume.
Here, $\Gamma$ denotes the gamma function.
Furthermore we can evaluate the MGF $A_f^\eps$ for
$\lambda \geq 0$ using the probability density and applying Laplace's method:
\begin{align}
A_f^\eps(\lambda) &= \int_0^\infty \dd z \; \rho^\eps_f(z) \exp
\left\{\frac{\lambda z}{\eps} \right\} \nonumber\\
&\overset{\eps \downarrow 0}{\sim} (2 \pi \eps)^{-\frac{n-1}{2}}
\vol_{n-1}\left(S^{n-1}\right)   \left[\frac{\beta}{1 - \exp
\left\{ -2 \beta T\right\}} \right]^{-\frac{n-1}{2}} \lambda^{n-1}
\exp \left\{\frac{\lambda^2}{\eps}  \frac{1 - \exp \left\{ -2 \beta
T\right\}}{2 \beta} \right\}\,.
\label{eq:mgf-ou}
\end{align}
Starting with the computation of the MGF using instantons, for
any unit vector $e_u \in \RR^n$ and with
$\nabla f(x) = x / \norm{x}_n$, a valid solution of the instanton
equations is
\begin{align}
\begin{cases}
\phi_\lambda^u(t) = \frac{\lambda}{\beta} \left(\exp\left\{\beta(t -
T) \right\} - \exp\left\{-\beta(t + T) \right\} \right) e_u\,,\\
\theta_\lambda^u(t) = \lambda \exp\left\{\beta(t - T) \right\} e_u\,,
\end{cases}
\end{align}
with corresponding action
\begin{align}
S\left[\phi_\lambda^u \right] = \frac{\lambda^2}{2 \beta} \left(1
- \exp \left\{ -2 \beta T\right\} \right)\,,
\end{align}
so that
\begin{align}
\exp\left\{-\eps^{-1} \left( S[\phi_\lambda^{u_0}] - \lambda
    f\left(\phi_\lambda^{u_0}(T)\right) \right)\right\} = \exp
    \left\{\frac{\lambda^2}{\eps}  \frac{1 - \exp \left\{ -2 \beta
    T\right\}}{2 \beta} \right\}
\end{align}
as expected.\\

For the prefactor, we note that with $n - 1$ zero modes
corresponding to angles on the hypersphere, the $\eps$-scaling of the
prefactor of the MGF in~\eqref{eq:mgf-general-degen} is correct. We first
evaluate the prefactor $\tilde{R}_\lambda$ according
to~\eqref{eq:prefac-pdf-degen-forward}, i.e.\ using the forward Riccati
equation with unmodified initial condition: The solution of the forward
Riccati equation
\begin{align}
\dot{Q}_\lambda^{u} = 2 (1_{n \times n} - \beta Q_\lambda^{u})\,,
\quad  Q_\lambda^{u}(0) = 0_{n \times n}
\end{align}
is
\begin{align}
Q_\lambda^{u}(t) = \frac{1 - \exp \left\{ -2 \beta t\right\}}{\beta}
1_{n \times n}\,,
\end{align}
and with $\nabla^2f(x) = \text{pr}_{x^\perp} / \norm{x}_n$,
where~$\text{pr}_{x^\perp}$ denotes the orthogonal projection onto
the subspace~$x^\perp \subset \RR^n$, we obtain
\begin{align}
1_{n \times n} - \lambda \nabla^2f(\phi_\lambda^u(T)) Q(T)
= e_u \otimes e_u\,.
\end{align}
Hence, $n-1$ eigenvalues are $0$ and
\begin{align}
{\det}_{n-(n-1)}'\left(1_{n \times n} - \lambda \nabla^2f(\phi_\lambda^u(T))
Q(T) \right) = 1\,.
\end{align}
Since $\nabla^2 b = 0$, we are left with evaluating
\begin{align}
\tilde{R}_{\lambda} &= \int_D \sqrt{{\det}_r \bra{\psi^{u}_\lambda(T)}
\lambda \nabla^2 f(\phi^{u}_\lambda(T)) \ket{\psi^{u}_\lambda(T)}}\;
\dd^r u = \int_D \sqrt{{\det}_{n-1} \bra{\psi^{u}_\lambda(T)}\ket{
\xi^{u}_\lambda(T)}}\; \dd^r u \nonumber\\
&= \lambda^{n-1}  \left[\frac{1 - \exp \left\{ -2 \beta T\right\}}{
\beta} \right]^{\frac{n-1}{2}}\vol_{n-1}\left(S^{n-1}\right)\,,
\end{align}
thereby correctly reproducing the MGF~\eqref{eq:mgf-ou} including
the prefactor. In order to get the PDF~\eqref{eq:pdf-ou} using
Proposition~\ref{thm:pdf-orefac-degen-final}, all we have to do
is note that
\begin{align}
\lambda_z = \frac{\beta}{1 - \exp \left\{ -2 \beta T\right\}} \; z\,,
\end{align}
which immediately leads to~\eqref{eq:pdf-ou}
via~\eqref{eq:pdf-prefac-general-degen}.\\

Alternatively, we can use the backward Riccati
approach~\eqref{eq:prefac-pdf-degen-forward}, i.e.\
using the backward Riccati equation with modified final condition.
Then, the volume term becomes
\begin{align}
&\int_D \sqrt{{\det}_r \bra{\psi^{u}_\lambda(T)}
1_{n \times n} + \lambda \nabla^2 f(\phi^{u}_\lambda(T))
\ket{\psi^{u}_\lambda(T)}}\;
\dd^r u \nonumber\\
&=  \left[1 + \frac{1 - \exp \left\{- 2 \beta T \right\}}{\beta}
\right]^{\frac{n-1}{2}} \left[\frac{1 - \exp \left\{- 2 \beta T
\right\}}{\beta} \right]^{\frac{n-1}{2}} \lambda^{n-1}
\vol_{n-1} \left(S^{n-1}\right)
\end{align}
and solving the Riccati equation
\begin{align}
\dot{W}_\lambda^u = -2 \left(W_\lambda^u\right)^2 + 2
\beta W_\lambda^u\,, \quad W_\lambda^u(T) = -
\left(1_{n \times n} - e_u \otimes e_u \right)
\end{align}
to get
\begin{align}
W_\lambda^u(t) = -
\frac{\exp
\left\{2 \beta (t - T) \right\}}{1 + \frac{1 - \exp
\left\{2 \beta (t - T) \right\}}{\beta}}
\left(1_{n \times n} - e_u \otimes e_u \right)
\end{align}
leads to
\begin{align}
\exp \left\{\int_0^T \trace \left[W_\lambda^u \right] \dd t
\right\} = \left[1 + \frac{1 - \exp \left\{- 2 \beta T \right\}}{\beta}
\right]^{-\frac{n-1}{2}}\,,
\end{align}
thereby correctly reproducing the full prefactor.\\

Finally, we compute the prefactor using
Proposition~\ref{thm:riccati-zero-left-boundary} with a forward
Riccati equation with modified initial condition. This is instructive
in that it demonstrates the singular limits of the individual terms
as~$T\to\infty$. We note that $\tilde{R}_{\lambda}$ \textit{and} its
constituents in the previous paragraphs have a well-behaved limit
as~$T \to \infty$, which is in contrast to the PDF prefactor computation
via Proposition~\ref{thm:riccati-zero-left-boundary} presented
here. First
\begin{align}
\vol\left(\theta_z(0) \right) = \int_D \sqrt{{\det}_r
\bra{\xi^{u}_z(0)}\ket{\xi^{u}_z(0)}} \dd^r u = \vol_{n-1}\left(
S^{n-1}\right) \left( \frac{\beta \exp \left\{ - \beta T\right\}}{1 -
\exp \left\{ -2 \beta T\right\}} \right)^{n-1} z^{n-1}
\end{align}
tends to $0$ as $T \to \infty$, whereas, since with
\begin{align}
Q_z^u(0) = \sum_{i = 1}^r \tilde{\xi}^{u}_{z,i}
\otimes \tilde{\xi}^{u}_{z,i} = 1_{n \times n} - e_u \otimes e_u
\end{align}
and
\begin{align}
U_z^u = e_u \otimes e_u - \frac{\beta \exp \left\{ - 2 \beta T
\right\}}{1 - \exp \left\{ -2 \beta T\right\}} \left(1_{n \times n} -
e_u \otimes e_u\right)\,,
\end{align}
we get
\begin{align}
{\det}_n U_z^{u} = (-1)^{n-1} \left(\frac{\beta \exp \left\{ -2 \beta
T \right\}}{1 - \exp \left\{ -2 \beta T\right\}} \right)^{n-1}
\end{align}
and
\begin{align}
\left \langle \nabla f (\phi_z^{u}(T))
, Q_z^{u}(T) \left(U_z^{u}\right)^{-1} \nabla f (\phi_z^{u}(T))
\right \rangle_n = \frac{1 - \exp \left\{ -2 \beta T\right\}}{\beta}\,,
\end{align}
such that the regularized denominator from
Proposition~\eqref{thm:riccati-zero-left-boundary}
\begin{align}
\left[(-1)^r{\det}_n U_z^{u} \; \left \langle \nabla f (\phi_z^{u}(T))
, Q_z^{u}(T) \left(U_z^{u}\right)^{-1} \nabla f (\phi_z^{u}(T))
\right \rangle_n\right]^{1/2} \nonumber\\
= \left(\frac{\beta}{1 - \exp \left\{ -2 \beta T\right\}} \right)^{
\frac{n}{2}-1} \exp \left\{ - (n-1) \beta T \right\}
\end{align}
also tends to zero as $T \to \infty$ and only their quotient
$\tilde{R}_\lambda$ remains finite.

\subsection{Rotationally symmetric two-dimensional vector field with swirl}
\label{subsec:2d-swirl}

\begin{figure}
\centering
\includegraphics[width = \textwidth]{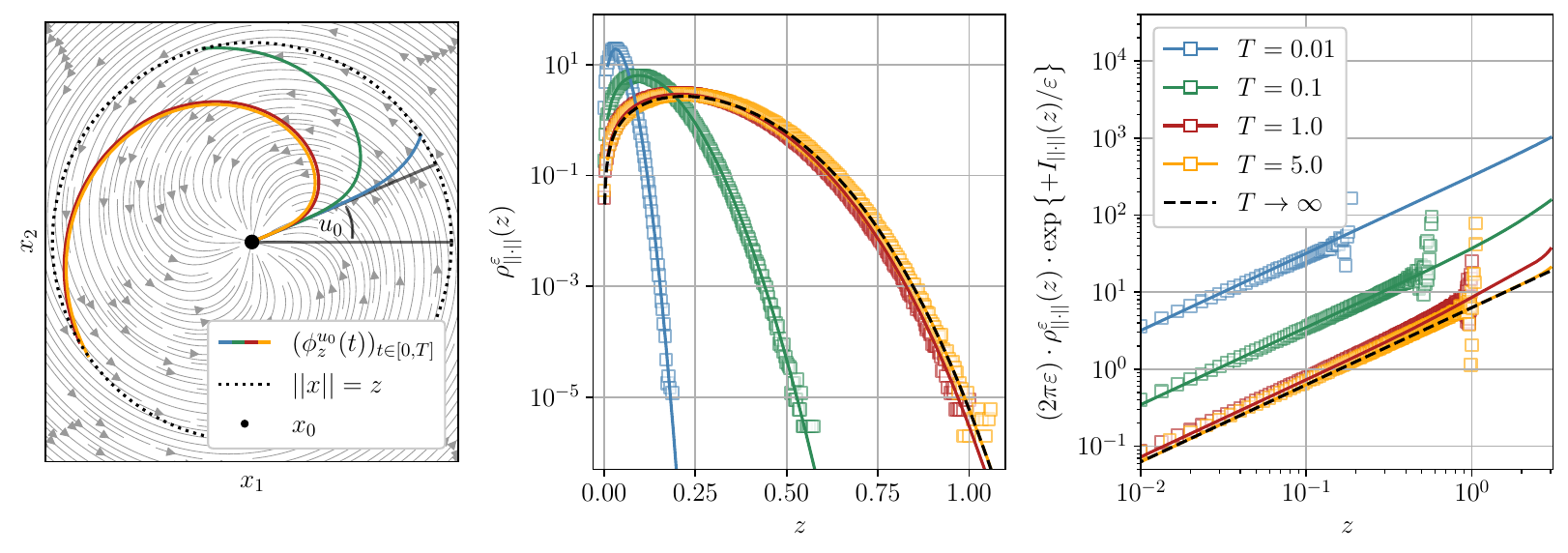}
\caption{Left: Sketch of instanton trajectories $(\phi^{u_0}_z(t))_{t
\in [0,T]}$ for different $T \in \{0.01, 0.1, 1, 5\}$ from the example in
Section~\ref{subsec:2d-swirl} with the specific
potential and swirl~\eqref{eq:swirl-example-vl}
at a fixed value
of the observable~$f(x) = \norm{x} =  z = 3$. The coloring is the same across
all subfigures and explained in the right panel. All numerically obtained
trajectories have been rotated to the same initial angle~$u_0$. For small~$T$,
the instanton is almost radial, and for large~$T$, it follows the angular
component~$\ell$ of the vector field~$b$, with a purely radial momentum
$(\theta^{u_0}_z(t))_{t \in [0,T]}$ acting against the radial force~$-\nabla V$.
Center and right:  Comparison of the instanton and Riccati results
from Section~\ref{subsec:2d-swirl} to Monte Carlo 
simulations of the SDE with~\eqref{eq:swirl-example-vl}.
For fixed~$\eps = 0.05$ and the different final times $T$, we
obtained $10^8$ samples $\norm{X^\eps_T}$ each through
Euler-Maruyama integration of the SDE in order
to estimate the PDF $\rho^\eps_{\norm{\cdot}}$. The resulting PDF estimate,
indicated by the squares, is compared to the theoretical PDF
asymptotics~\eqref{eq:pdf-prefac-general-degen}, which were obtained as
detailed in Section~\ref{subsec:2d-swirl} and are shown (without any free
parameters) by the solid lines. The right subplot compares the full
prefactor of the PDF, defined via $\rho^\eps_{\norm{\cdot}} \cdot \exp
\left\{+ I_{\norm{\cdot}} / \eps \right\}$ and obtained through direct sampling
(with the same data as in the center) to the theoretical result
for the quadratic approximation (solid lines).}
\label{fig:2d-swirl}
\end{figure}

As a second example, we slightly modify the situation of the previous subsection
to a nonlinear radial vector field, to which we then also add a rotationally symmetric
nonlinear swirl. Restricting ourselves to a spatial dimension $n = 2$, we consider
the following drift vector field in polar coordinates~$(r, \varphi)$:
\begin{align}
b(r,\varphi) = -V_r'(r) e_r + l(r) e_\varphi\,,
\label{eq:drift-swirl}
\end{align}
with unit coordinate vectors $e_r = x / \norm{x} = (\cos \varphi, \sin \varphi)$
and $e_\varphi = (-\sin \varphi, \cos \varphi)$. We again consider a diffusion
process~$(X_t^\eps)_{[0,T]}$ in this vector field starting at~$x_0 = 0$ with
final-time observable $f(X_T^\eps) = \norm{X_T^\eps}$, and the radial symmetry of
this problem will generate one zero mode in this case. Even though the drift is
not gradient, the leading order behavior of the PDF $\rho_\eps^f$ in $\eps$ as
$T \to \infty$, i.e. in the stationary case, can be found analytically here. The
reason for this is that the drift given in~\eqref{eq:drift-swirl} is already specified
in terms of its
\textit{transverse decomposition}~\cite{freidlin-wentzell:2012,zhou-etal:2012}
\begin{align}
b = \nabla V + \ell\,, \quad \left \langle \nabla V(x), \ell(x) \right \rangle_2
= 0 \; \forall x \in \RR^2\,,
\end{align}
where $V$ is the quasi-potential. In our example, we have $V(x) = V_r(\norm{x})$
and $\ell(r,\varphi) = l(r) e_\varphi$. The stationary PDF of the process itself
is given by~\cite{grafke-schaefer-vanden-eijnden:2021,bouchet-reygner:2022}
\begin{align}
\rho^\eps_\infty(x) \overset{\eps \downarrow 0}{\sim} (2 \pi \eps)^{-1} \left[
{\det}_2 \nabla^2 V(x_0) \right]^{1/2} \exp \left\{- \int_0^\infty \nabla \cdot
\ell(\phi_x(t)) \dd t \right\} \exp \left\{-\frac{1}{\eps} \left(V(x) -
V(x_0) \right) \right\}\,.
\end{align}
Since the transverse vector field $\ell$ in our example is divergence-free,
we conclude that the PDF $\rho_f^\eps$ of $f(X^\eps_T)$ as $T \to \infty$ and
$\eps \downarrow 0$
will be given by
\begin{align}
\rho_f^\eps(z) \overset{\eps \downarrow 0}{\sim} (2 \pi \eps)^{-1}  \left[
{\det}_2 \nabla^2 V(x_0) \right]^{1/2} \cdot \left(2 \pi z \right) \cdot \exp
\left\{-\frac{1}{\eps} \left(V_r(z) - V_r(0) \right) \right\}\,.
\end{align}
For finite times, no easy analytical solution is available, so we have to solve
the instanton and (forward) Riccati equations numerically in order to obtain the
precise small noise asymptotics of the PDF~$\rho_f^\eps$. For the specific example
\begin{align}
V_r(r) = \frac{1}{4} r^4 + \frac{1}{2}
r^2 \,, \quad l(r) = r^5\,, \label{eq:swirl-example-vl}
\end{align}
we compare the results of this numerical procedure to Monte Carlo sampling at
a fixed, small noise level $\eps$ for different times~$T$ in Figure~\ref{fig:2d-swirl}.
For $T \in \{0.01, 0.1, 1., 5.\}$, instanton solutions $(\phi_z^{u_0},
\theta_z^{u_0}, \lambda_z)$
were computed directly for different, equidistantly spaced $z \in [0,3]$ using the
augmented Lagrangian method for the final time constraint and the L-BFGS algorithm
using adjoints as detailed in~\cite{schorlepp-etal:2022}, with $n_t = 4000$ time
discretization points in all cases and Heun time steps. Here, $u_0 \in [0, 2 \pi)$
is the arbitrary angle characterizing the numerically found instantons. Afterwards,
for each instanton, the forward Riccati equation
from Proposition~\ref{thm:pdf-orefac-degen-final} $(i)$
was solved numerically with the same time
discretization and time stepping. In order to evaluate
the prefactor~\eqref{eq:prefac-pdf-degen-forward}, the $\det'$ expression was
computed by only taking into account the single positive eigenvalue of
$1_{2 \times 2} - \lambda_z \nabla^2
f(\phi^{u}_z(T)) Q_z^{u}(T)$ (the other eigenvalue being close to zero).
The zero mode volume prefactor is
\begin{align}
\int_0 ^{2 \pi} \sqrt{\bra{\psi^{u}_z(T)}
\lambda_z \nabla^2 f(\phi^{u}_z(T)) \ket{\psi^{u}_z(T)}} &= 2 \pi \sqrt{\left
\langle\psi^{u_0}_z(T),\xi^{u_0}_z(T) \right \rangle_2} \nonumber \\
&= 2 \pi \sqrt{\left \langle\phi^{u_0}_z(T),\theta^{u_0}_z(T) \right \rangle_2} =
2 \pi \sqrt{\lambda_z \cdot z}\,,
\end{align}
where, in the last line, we used that due to rotational symmetry, the scalar
product of the tangent vectors is the same as for the original instanton, as well
as $\theta_z^u(T) = \lambda_z \nabla f(\phi_z^u(T)) = \lambda_z \phi_z^u(T)
/ \norm{\phi_z^u(T)}$. The last ingredient for the
prefactor~\eqref{eq:pdf-prefac-general-degen}, the derivative
$\dd \lambda_z \ / \dd z$, was simply computed by numerical differentiation of
the obtained map $z \mapsto \lambda_z$ from the instanton computations. As
Figure~\ref{fig:2d-swirl} shows, both the limiting case $T \to \infty$, as well as
the Monte Carlo data at smaller $T$ and $\eps = 0.05$ are well reproduced.


\subsection{Dynamical phase transition in a three-dimensional gradient system}
\label{subsec:dynam-phase-trans-ode-potential}

\begin{figure}
\centering
\includegraphics[width = \textwidth]{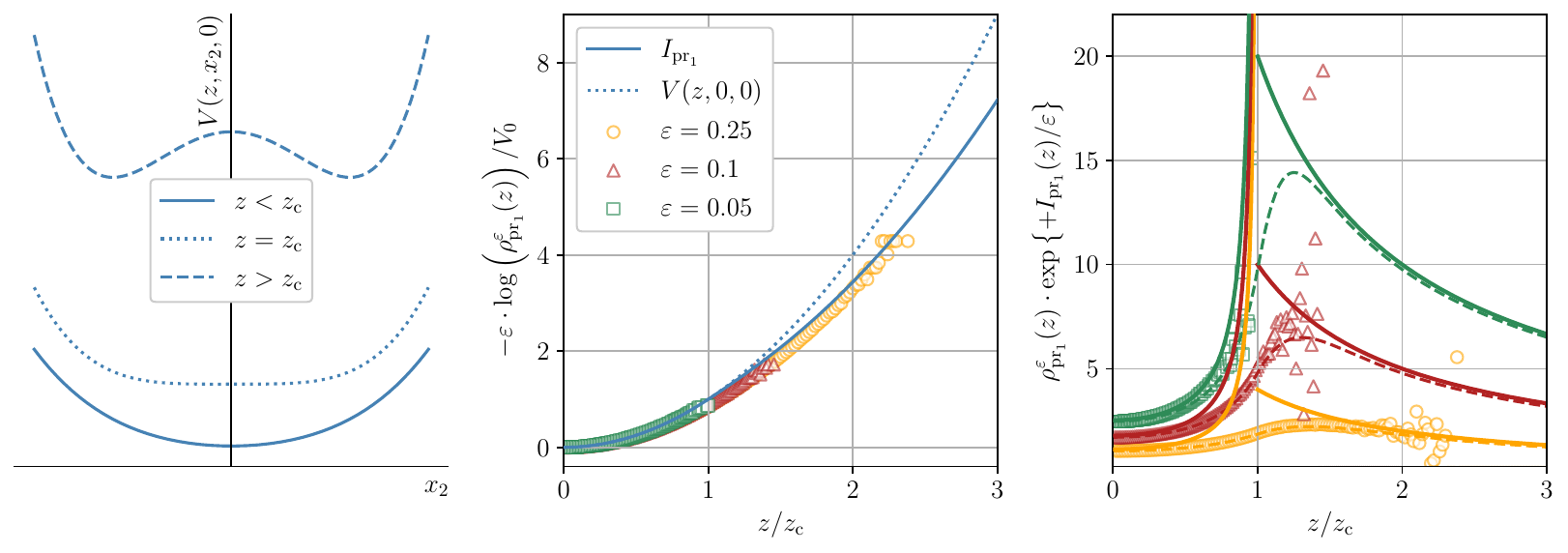}
\caption{Left: Sketch of the potential~$V$ from the example in
Section~\ref{subsec:dynam-phase-trans-ode-potential} at different values
of the observable~$x_1 = z$. The figure shows a cut through the~$(x_2,x_3)$
plane at $x_3 = 0$. Note the double well/Mexican hat structure
beyond~$z_{\text{c}}$, which is responsible for the dynamical phase
transition. Center and right:  Comparison of the theoretical results
from Section~\ref{subsec:dynam-phase-trans-ode-potential} to Monte Carlo 
simulations of the SDE~\eqref{eq:grad-sde} with the specific
potential~\eqref{eq:potential-example} (with~$V_0 = z_{\text{c}} = a = 1$)
as an example. For each~$\eps \in \left\{0.25, 0.1, 0.05 \right\}$,
about~$1.3 \cdot 10^9$ samples from the invariant measure of the SDE
were obtained (with Euler-Maruyama integration steps of
length~$\Delta t = 10^{-3}$ and sampling after each time unit) in order
to estimate the PDF $\rho^\eps_{\text{pr}_1}$. Center: Comparison of the
Monte Carlo results on a logarithmic scale, where~$-\eps
\cdot \log \rho^\eps_{\text{pr}_1}$ collapses onto the rate function for
small~$\epsilon$, as expected. Note in particular that the lower branch
corresponding to the symmetry-broken minima of the potential correctly
matches the Monte Carlo results. Right: Comparison of the full
prefactor of the PDF, defined via $\rho^\eps_{\text{pr}_1} \cdot \exp
\left\{+ I_{\text{pr}_1} / \eps \right\}$, obtained either through
$(i)$ direct sampling (circles, triangles and squares) with the same data
as in the center, or $(ii)$ numerical evaluation of the
integral~\eqref{eq:grad-pdf-marginal-def} for
the marginal PDF (dashed lines), to the theoretical result
for the quadratic approximation (solid lines). Note that the prefactor is
correctly approximated both below and above $z_\text{c}$.}
\label{fig:potential-sys}
\end{figure}

For a system dimension of $n = 3$, we consider a first instructive
example exhibiting spontaneous symmetry breaking beyond a critical
observable value~$z_{\text{c}} > 0$ as sketched in the right subplot
of Figure~\ref{fig:inst-sketch}. Choosing a gradient system
\begin{align}
\dd X_t^{\eps} = - \nabla V \left( X_t^{\eps} \right) \dd t
+ \sqrt{2 \eps} \:\dd B_t\,, \quad X_0^{\eps} = x_0
\label{eq:grad-sde}
\end{align}
on the time interval $[0,T]$ and focusing on the stationary limit
$T \to \infty$ allows us to treat this case in an exact manner.
We assume that the potential has a unique global minimum at $x_0 = (0,0,0)$
with $\nabla^2 V(x_0)$ positive definite. Furthermore, $V$ should be
symmetric in the first component $x_1$, i.e.~$V(-x_1,x_2,x_3) =
V(x_1,x_2,x_3)$, and rotationally symmetric in $(x_2,x_3)$ for any
$x_1$, i.e.~for all $x_1 \in \RR$ and $b \geq 0$,  $V(x_1, b \cos u,
b \sin u)$ is constant in $u \in [0, 2 \pi)$. We assume that there
exists $z_{\text{c}} > 0$, such that for all $x_1 = z \in \RR$ with
$\abs{z} < z_{\text{c}}$, the function $V(z,\cdot,\cdot):\RR^2 \to \RR$
has a unique, nondegenerate global minimum at $(x_2,x_3) = (0,0)$, and
for all $z$ with $\abs{z}> z_{\text{c}}$, $V(z,\cdot,\cdot)$ has a
continuous family of global minima at $(\bar{x}(z) \cos u, \bar{x}(z)
\cos u)$ with $\bar{x}(z) > 0$ and $u \in [0,2 \pi)$, as
sketched in Figure~\ref{fig:potential-sys} (left). A specific example
of such a potential is
\begin{align}
V(x_1, x_2, x_3) = V_0 \left[ \left(\frac{x_1}{z_{\text{c}}}
\right)^2 \frac{\left(x_2^2 + x_3^2 \right)^2}{a^4} +
\left(1 - \left(\frac{x_1}{z_{\text{c}}} \right)^2 \right)
\frac{x_2^2 + x_3^2}{a^2} + \left(\frac{x_1}{z_{\text{c}}}
\right)^2 \right]\,,
\label{eq:potential-example}
\end{align}
with constants $V_0, a, z_{\text{c}} > 0$,
which indeed exhibits a Mexican hat-like structure in the $x_2$-$x_3$ plane
for $x_1 = z > z_{\text{c}}$ with minima at radius
\begin{align}
\bar{x}(z) := \frac{a}{\sqrt{2}}
\sqrt{1 - \left(\frac{z_{\text{c}}}{z} \right)^2}\,.
\end{align}

As our (linear) observable, we take
\begin{align}
f = \text{pr}_1:\RR^3 \to \RR\,, \quad (x_1, x_2, x_3) \mapsto x_1\,,
\end{align}
which allows us to test the backward Riccati equation for the
prefactor from Proposition~\ref{thm:prefac-riccati-degen-bkwd}
in the limit $T \to \infty$. Since the system is gradient,
we known that the stationary
PDF~$\rho^\eps_\infty:\RR^3 \to [0, \infty)$ of~$X^\eps$ is given by
\begin{align}
\rho^\eps_\infty(x) = Z_\eps^{-1} \exp \left\{-\frac{1}{\eps} V(x) \right\}
\end{align}
with normalization constant
\begin{align}
Z_\eps = \int_{\RR^3} \exp \left\{-\frac{1}{\eps} V(x)
\right\} \dd^3 x\,.
\end{align}
Applying Laplace's method on the PDF of the
marginal distribution
\begin{align}
\rho^\eps_{ \text{pr}_1}(z) = Z_\eps^{-1} \int_{\RR^2} \exp
\left\{-\frac{1}{\eps} V(z,x_2,x_3) \right\} \dd^2(x_2,x_3)
\label{eq:grad-pdf-marginal-def}
\end{align}
of the first component $X^\eps_1$
(approximating both~$Z_\eps$ and the~$(x_2,x_3)$-integral) yields
\begin{align}
\rho^\eps_{ \text{pr}_1}(z) \overset{\eps \downarrow 0}{\sim} \begin{cases}
(2 \pi \eps)^{-1/2} \left[\frac{{\det}_3 \nabla^2 V(x_0)}{{\det}_2
\nabla^2 V(z,0,0)} \right]^{1/2} \exp \left\{-\frac{1}{\eps}
\left(V(z,0,0) - V(x_0) \right) \right\}\,, \quad &\abs{z} < z_{\text{c}}\\
(2 \pi \eps)^{-1} \left[\frac{{\det}_3 \nabla^2 V(x_0)}{{\det}_1'
\nabla^2 V(z,\bar{x}(z) \cos u_0, \bar{x}(z) \sin u_0)}
\right]^{1/2} 2 \pi \bar{x}(z) \; \times \\
\quad \times \exp
\left\{-\frac{1}{\eps} \left(V(z,\bar{x}(z) \cos u_0,
\bar{x}(z) \sin u_0) - V(x_0) \right)
\right\}\,, \quad &\abs{z} > z_{\text{c}}
\end{cases}
\label{eq:grad-pdf-ref}
\end{align}
for any $u_0 \in D = [0, 2 \pi)$. Here, ${\det}_2$ denotes the
restriction onto the $(x_2,x_3)$-plane, and~${\det}_1'$ reduces to
the single nonzero eigenvalue of the matrix in the~$(x_2,x_3)$-plane
corresponding to the radial eigenvector.
For the specific example~\eqref{eq:potential-example}, the result is
\begin{align}
\rho^\eps_{ \text{pr}_1}(z) \overset{\eps \downarrow 0}{\sim}\begin{cases}
(2 \pi \eps)^{-1/2} \frac{\sqrt{2 V_0}}{z_{\text{c}}}
\frac{1}{1-\left(z/z_{\text{c}} \right)^2} \exp \left\{-
\frac{V_0}{\eps} \left(\frac{z}{z_{\text{c}}}\right)^2 \right\}\,,
\quad &\abs{z} < z_{\text{c}}\\
(2 \pi \eps)^{-1} \frac{2 \pi V_0}{z}
\exp \left\{-\frac{V_0}{\eps}
\frac{3 \left(z/z_{\text{c}} \right)^2 - \left(z_{\text{c}}/z \right)^2
+ 2}{4} \right\}\,, \quad &\abs{z} > z_{\text{c}}
\end{cases}
\end{align}
as a reference result, with discontinuous second derivative
of the rate function at $z = z_{\text{c}}$ and divergent prefactors
as $z \uparrow z_{\text{c}}$.\\


In order to reproduce this result using sample path large deviations,
we first note that the unique (for the chosen potential) solution to
the instanton equations for any endpoint~$x=(x_1,x_2,x_3) \in \RR^3$
\begin{equation}
  \begin{cases}
    \dot \phi_x = - \nabla V \left(\phi_x \right) + 2 \theta_x\,,
    & \phi_x(0) = x_0\,, \quad \phi_x(T) = x\\
    \dot \theta_x = \nabla^2 V\left(\phi_x \right)^\top \theta_x\,,
    & \theta_x(T) =: \lambda_x \in \RR^3\,,
  \end{cases}
  \label{eq:instanton-eq-gradient}
\end{equation}
is given by 
\begin{align}
\dot{\phi}_x = \nabla V(\phi_x) = \theta_x
\label{eq:grad-inst-solu}
\end{align}
as $T \to \infty$, i.e.\ by time-reversed deterministic dynamics, such that
\begin{align}
S\left[\phi_x\right] = \int_0^\infty \norm{\theta_x}_3^2 \dd t =
\int_0^\infty \left \langle \nabla V(\phi_x), \dot{\phi}_x \right
\rangle_3 \dd t = V(x) - V(x_0)\,.
\end{align}
By the contraction principle, i.e.\ by minimizing this result
over all $(x_2,x_3) \in \RR^2$ for a given $x_1 = z \in \RR$,
we obtain the correct rate function
\begin{align}
I_{\text{pr}_1}(z) = \left. \begin{cases}
S\left[\phi_{(z,0,0)}\right]\,, \; & \abs{z} \leq z_{\text{c}}\\
S\left[\phi_{(z,\bar{x}(z) \cos u_0,\bar{x}(z)
\sin u_0)}\right]\,, \; & \abs{z} > z_{\text{c}}\,.
\end{cases} \right\} = \begin{cases}
V(z,0,0) - V(x_0)\,, v & \abs{z} \leq z_{\text{c}}\\
V(z,\bar{x}(z) \cos u_0,\bar{x}(z) \sin u_0) - V(x_0)\,,
\; & \abs{z} > z_{\text{c}}\,.
\end{cases}
\end{align}
with any $u_0 \in D = [0,2\pi)$. For the prefactor in the nondegenerate case
$\abs{z} < z_{\text{c}}$, we first evaluate
$\exp \left\{\int_0^\infty \trace\left[W_z\right] \right\}$
following~\cite{grafke-schaefer-vanden-eijnden:2021}:
The backward Riccati matrix $W_z$ solves
\begin{align}
\dot{W}_z = -2 W_z^2 + \nabla^2 V(\phi_z) W_z + W_z \nabla^2
V(\phi_z) + \dv{}{t} \left(\nabla^2 V(\phi_z) \right)\,,
\quad W_z(\infty) = 0\,.
\label{eq:riccati-bkwd-gradient}
\end{align}
Defining $W_z = C_z^{-1} \dot{C}_z$ with $C_z(\infty) =
1_{3 \times 3}$, $\dot{C}_z(\infty) = 0_{3 \times 3}$, we have,
on the one hand,
\begin{align}
  \label{eq:det3Cz0}
{\det}_3 C_z(0) = {\det}_3 C_z(\infty)
\exp \left\{-\int_0^\infty \trace \left[ W_z\right] \dd t \right\}\,,
\end{align}
and on the other hand, from~\eqref{eq:riccati-bkwd-gradient},
\begin{align}
\dv{}{t} \left(\dot{C}_z - C_z \nabla^2 V(\phi_z)\right) = -
\left(\dot{C}_z - C_z \nabla^2 V(\phi_z)\right) W_z,,
\end{align}
so
\begin{align}
\frac{{\det}_3 \left(\dot{C}_z(\infty) - C_z(\infty) \nabla^2
V(\phi_z(\infty))\right)}{{\det}_3 \left(\dot{C}_z(0) - C_z(0)
\nabla^2 V(\phi_z(0))\right)}
= \exp \left\{-\int_0^\infty \trace \left[ W_z\right] \dd t \right\}\,.
\end{align}
Using the boundary conditions and equation~\eqref{eq:det3Cz0} as well
as noting that necessarily $\dot{C}_z(0) = 0_{3 \times 3}$ in the
stationary limit, we obtain
\begin{align}
\exp \left\{\int_0^\infty \trace \left[ W_z\right] \dd t
\right\} = \left[\frac{{\det}_3 \left(\nabla^2 V(x_0)\right)}{
{\det}_3 \left(C_z(\infty) \nabla^2 V(z,0,0) -\dot{C}_z(\infty)
\right)} \right]^{1/2} = \left[\frac{{\det}_3 \left(\nabla^2
V(x_0)\right)}{{\det}_3 \left( \nabla^2 V(z,0,0)\right)} \right]^{1/2}\,.
\end{align}
The second ingredient for the PDF prefactor is
\begin{align}
\left[ \left.\dv{}{\lambda}\right|_{\lambda_z} \text{pr}_1(
\phi_z(\infty)) \right]^{-1/2} &= \left[\left( \nabla_{\theta_x(
\infty)} \phi_x(\infty) \right)_{11}\right]^{-1/2} = \left[\left(
\nabla_{\phi_x(\infty)} \theta_x(\infty) \right)^{-1}_{11}
\right]^{-1/2} \nonumber\\
&\overset{\eqref{eq:grad-inst-solu}}{=} \left[\left( \nabla^2
V(z,0,0) \right)^{-1}_{11} \right]^{-1/2} =  \left[\frac{{\det}_3
\left( \nabla^2 V(z,0,0) \right)}{{\det}_{2} \left( \nabla^2
V(z,0,0) \right)} \right]^{1/2}\,,
\end{align}
thereby correctly reproducing the reference result below the
critical observable value~$z_{\text{c}}$ from~\eqref{eq:grad-pdf-ref}
via Propositions~\ref{thm:gy-old-bkwd} and~\ref{thm:pdf-saddle}.\\


Above the critical observable value~$z_{\text{c}}$, the final condition
for the backward Riccati equation becomes
\begin{align}
W_z(\infty) = - \tilde{\psi}^{u_0}_z \otimes \tilde{\psi}^{u_0}_z\,,
\end{align}
where~$\tilde{\psi}^{u_0}_z$ is, in particular, a unit
eigenvector corresponding to the single vanishing eigenvalue
of the Hessian $\nabla^2 V(z,\bar{x}(z) \cos u_0, \bar{x}(z)
\sin u_0)$. Setting $\dot{C}_z(\infty) = - \tilde{\psi}^{u_0}_z
\otimes \tilde{\psi}^{u_0}_z$ in the computation above yields
\begin{align}
\exp \left\{\int_0^\infty \trace \left[ W_z^{u_0}\right] \dd t
\right\} = \left[\frac{{\det}_3 \left(\nabla^2
V(x_0)\right)}{{\det}_3 \left( \nabla^2 V(z,\bar{x}(z) \cos u_0,
\bar{x}(z) \sin u_0) + \tilde{\psi}^{u_0}_z \otimes
\tilde{\psi}^{u_0}_z \right)} \right]^{1/2}\,.
\end{align}
Hence, as desired, the modified initial condition renders the fraction
well defined by replacing the single zero eigenvalue of the matrix
in the denominator by~$1$. Furthermore, we have
\begin{align}
\left[ \left.\dv{}{\lambda}\right|_{\lambda_z} \text{pr}_1(
\phi_z(\infty)) \right]^{-1/2} &= \left[\left( \nabla_{\theta_x(
\infty)} \phi_x(\infty) \right)_{11}\right]^{-1/2} =
\left[\left( \left. \left( \nabla_{\theta_x(
\infty)} \phi_x(\infty) \right) \right|_{\left(\tilde{\psi}^{u_0}_z
\right)^\perp}\right)_{11}\right]^{-1/2} \nonumber\\
 &= \left[\left( \left. \left( \nabla^2 V(z,\bar{x}(z) \cos u_0,
\bar{x}(z) \sin u_0) \right) \right|_{\left(\tilde{\psi}^{u_0}_z
\right)^\perp}\right)_{11}^{-1}
\right]^{-1/2} \nonumber\\
&= \left[\frac{{\det}_3 \left( \nabla^2 V(z,\bar{x}(z) \cos u_0,
\bar{x}(z) \sin u_0) + \tilde{\psi}^{u_0}_z \otimes
\tilde{\psi}^{u_0}_z \right)}{{\det}_{1}' \left( \nabla^2
V(z,\bar{x}(z) \cos u_0,
\bar{x}(z) \sin u_0) \right)} \right]^{1/2}\
\end{align}
by restricting to the invariant subspace $\left(
\tilde{\psi}^{u_0}_z\right)^\perp$ of the Hessian $\nabla^2
V(z,\bar{x}(z) \cos u_0, \bar{x}(z)
\sin u_0)$ on which it is invertible for the computations, and
afterwards reintroducing the full matrix including the modified
eigenvalue~$1$. All in all, we have thus correctly reproduced the
PDF prefactor above the critical value in~\eqref{eq:grad-pdf-ref}.
For the specific example potential~\eqref{eq:potential-example}, the
situation considered here is sketched and compared to the results
of Monte Carlo simulations of the SDE~\eqref{eq:grad-sde} in
Figure~\ref{fig:potential-sys}.\\


\subsection{Average surface height for the one-dimensional KPZ equation with flat initial condition}
\label{subsec:pde-example}

The KPZ equation~\cite{kardar-parisi-zhang:1986}, an SPDE describing nonlinear surface
growth, and in particular its large deviation statistics have been the subject of
various studies. Here, particularly noteworthy works are~\cite{janas-kamenev-meerson:2016,krajenbrink-doussal:2017,smith-kamenev-meerson:2018,hartmann-meerson-sasorov:2021}
for an investigation of a short time dynamical phase transition for the distribution
of the surface height at
one point in space, starting from a stationary surface. Furthermore, recently,
in~\cite{krajenbrink-doussal:2021}, an exact computation of the rate function
for the same observable with
general deterministic initial condition has been carried out; and for the flat
initial condition, the exact distribution of
the height at one point in space for all times has already been found
in~\cite{calabrese-le-doussal:2011}. A systematic short-time expansion for
the height distribution at one point and droplet and Brownian initial
conditions, which goes beyond the rate function and includes subleading
prefactor terms, can be found in~\cite{krajenbrink-doussal-prolhac:2018}.
All of the works listed above deal
with the KPZ equation on an unbounded spatial domain.
Here, we proceed in the spirit of~\cite{janas-kamenev-meerson:2016,krajenbrink-doussal:2017,smith-kamenev-meerson:2018,hartmann-meerson-sasorov:2021},
but modify the setup to study continuous
symmetry breaking instead of only a discrete mirror symmetry.
Accordingly choosing the spatially averaged surface height as an
observable necessitates considering a bounded spatial domain. For
such a domain, the large deviation statistics of the surface height
at one point have been computed in detail in~\cite{smith-meerson-sasorov:2018}, with the analysis of the spatially averaged surface height
left as a future
task there and predicted to display a second order dynamical phase transition.
Here, we will confirm this prediction and compute the leading order
PDF prefactors for both phases numerically. Furthermore, we analytically
compute the PDF prefactor when the spatially homogeneous
instanton dominates, which, in particular,
allows us to determine the critical observable value~$z_{\text{c}}$.
We will focus on a single choice
of the only parameter of the system, the non-dimensionalized
domain size~$l$, and use~$l = \pi$ throughout this paper.
We remark that it would be
an interesting future work to systematically study the large deviation
properties of the system for different domain sizes~$l$ using the methods
developed here, and to derive a complete phase diagram in
the~$(l, z)$ plane for the system, similar to~\cite{smith-meerson-sasorov:2018}.\\

To be more precise,
we consider the KPZ equation in one spatial dimension on a bounded interval in
space $[0,L]$ with periodic boundary conditions for the surface height $H \colon [0, L]
\times [0,T] \to \RR$,
\begin{align}
\partial_t H(x,t) = \nu \partial_{xx} H(x,t) + \frac{\lambda}{2}
\left(\partial_x H(x,t) \right)^2 + \sqrt{D} \eta(x,t)\,,
\label{eq:kpz-with-units}
\end{align}
starting from a flat initial profile $H(\cdot, 0) = H_0 \equiv 0$, and are
interested in precise asymptotic estimates for the probability distribution
(and in particular its tails) of
the spatially averaged surface height at time $T$,
\begin{align}
f(H(\cdot, T)) := \frac{1}{L} \int_0^L H(x,T) \; \dd x\,,
\end{align}
for small $T$.
In~\eqref{eq:kpz-with-units}, we denote by $\nu > 0$ the diffusivity,
by~$\lambda > 0$ (the choice of sign is without loss of generality)
the strength of the nonlinearity, and by~$D > 0$ the noise strength.
The noise term~$\eta$ is assumed to be space-time white Gaussian noise
with
\begin{align}
\EE \left[\eta(x,t) \right] = 0\,, \quad \EE \left[\eta(x,t)
\eta(x', t') \right] = \delta(x-x') \delta(t-t') \,. 
\end{align}
The non-dimensionalization
$t \to t T$, $x \to \sqrt{\nu T} x$ ,
$H \to  2 \nu H / \lambda$
and $\eta \to \left( \nu T^3 \right)^{-1/4} \eta$ leads to
the following model that we will
consider
for all computations in the following: For a dimensionless noise
strength~$\eps = D \lambda^2 T^{1/2} / (4 \nu^{5/2}) > 0$,
we consider $H^{\eps} \colon [0,l]
\times [0,1] \to \RR$ with $l = L / \sqrt{\nu T}$ the solution of
\begin{align}
\begin{cases}\partial_t H^\eps = \partial_{xx} H^\eps +
\left(\partial_x H^\eps\right)^2 + \sqrt{\eps} \eta\,,\\
H^\eps(\cdot,0) = H_0 \equiv 0\,,\\
H^\eps(0,t) = H^\eps(l,t)\,, \quad \partial_x H^\eps(0,t) =
\partial_x H^\eps(l,t) \quad \forall t \in [0,1]\,.
\end{cases}
\label{eq:kpz-wo-units}
\end{align}
and are interested in estimating the PDF of the mean surface height
\begin{align}
f(H^\eps(\cdot, 1)) := \frac{1}{l} \int_0^l H^\eps(x,1) \, \dd x
\label{eq:mean-dens-obs}
\end{align}
at the final time as $\eps \downarrow 0$. The small noise limit in these
dimensionless variables can be seen to directly correspond to either of the
limits $D \downarrow 0$ or $\lambda \downarrow 0$ in the physical variables.
Additionally, as mentioned above, we choose a fixed
and finite non-dimensionalized domain size $l = \pi$
in all of our numerical computations , so the usual
short-time limit
$T \downarrow 0$ considered in KPZ large deviations actually corresponds to
simultaneously taking~$T \downarrow 0$ and~$\nu \propto T^{-1} \uparrow \infty$
in this setup if the physical domain size remains constant.\\

For spatially white noise, the KPZ equation~\eqref{eq:kpz-wo-units}
is only well-posed after \textit{renormalization}, the noise being
too rough for the nonlinearity $- \tfrac{1}{2} \left(\partial_x H^\eps \right)^2$
to make sense otherwise~\cite{quastel:2011,hairer:2013}.
While this is not an issue on the level of instanton
computations, the solutions of which are expected to be classically
differentiable, renormalization
is necessary when dealing with the random
fluctuations around the instanton. We interpret~\eqref{eq:kpz-wo-units}
as the result of applying a Cole-Hopf transformation to the field
${\cal Q}^\eps \colon [0, l] \times [0,1] \to (0,\infty)$, solving the
well-posed stochastic heat equation (SHE) with multiplicative noise in the It{\^o} sense
\begin{align}
\begin{cases}
\partial_t {\cal Q}^\eps = \partial_{xx} {\cal Q}^\eps
+ \sqrt{\eps} {\cal Q}^\eps \eta\,,\\
{\cal Q}^\eps(\cdot, 0) = 1\,.
\end{cases}
\label{eq:stoch-heat-sde}
\end{align}
Then, the height field of the KPZ equation~\eqref{eq:kpz-wo-units} is given by
\begin{align}
H^{\eps} = \log {\cal Q}^\eps\,,
\label{eq:cole-hopf-trafo}
\end{align}
and a formal application of It{\^o}'s lemma shows that the Cole-Hopf transformation
generates a counter-term $-\delta(0)$, where $\delta$ is Dirac's delta function,
on the right-hand side of~\eqref{eq:kpz-wo-units}
that intuitively cancels the divergences in the original KPZ equation. We will compute
the contribution of the Gaussian fluctuations to the distribution of the
observable~\eqref{eq:mean-dens-obs} within this interpretation of the KPZ equation, i.e.\
actually consider the observable
\begin{align}
F({\cal Q}^\eps(\cdot, 1)) := \frac{1}{l} \int_0^{l}
\log {\cal Q}^\eps(x,1) \, \dd x
\end{align}
for the SHE.\\


The instanton equations~\eqref{eq:instanton-eq-z} for the
example~\eqref{eq:kpz-wo-units} and~\eqref{eq:mean-dens-obs} that determine
the instanton $(h_z, \tilde{h}_z, \lambda_z)$ written in terms of the original
field and its conjugate momentum read (see~\cite{fogedby:1999} for an early reference that derives these equations)
\begin{equation}
  \begin{cases}
    \partial_t h_z = \partial_{xx} h_z + \left(\partial_x h_z\right)^2 + \tilde{h}_z\,,
    & h_z(\cdot, 0) \equiv 0\,, \quad f \left( h_z(\cdot, 1) \right)
    = \frac{1}{l} \int_0^{l} h_z(x,1) \, \dd x = z\\
    \partial_t \tilde{h}_z = - \partial_{xx} \tilde{h}_z + 2
    \partial_x \left(\tilde{h}_z \partial_x h_z \right)\,,
    & \tilde{h}_z(\cdot, T) = \lambda_z \nabla f\left(h_z(\cdot, T)\right)
    \equiv \frac{\lambda_z}{l}\,.
  \end{cases}
  \label{eq:instanton-eq-kpz}
\end{equation}


In terms of the SHE, the instanton equations for the fields
$(q_z, p_z, \lambda_z)$ with
\begin{align}
q_z = \exp \left\{h_z\right\}\,, \quad p_z = \tilde{h}_z \exp \left\{-h_z \right\}
\label{eq:ch-trafo-inst}
\end{align}
become
\begin{equation}
  \begin{cases}
    \partial_t q_z = \partial_{xx} q_z + q_z^2 p_z\,,
    & q_z(\cdot, 0) \equiv 1\,, \quad F \left( q_z(\cdot, 1) \right)
    = \frac{1}{l} \int_0^{l} \log q_z(x,1) \, \dd x = z\\
    \partial_t p_z = - \partial_{xx} p_z - q_z p_z^2\,,
    & p_z(\cdot, T) = \lambda_z \nabla F\left(q_z(\cdot, T)\right)
    \equiv \frac{\lambda_z}{l q_z(\cdot, T)}\,.
  \end{cases}
  \label{eq:instanton-eq-kpz-ch}
\end{equation}

\begin{figure}
\centering
\includegraphics[width = \textwidth]{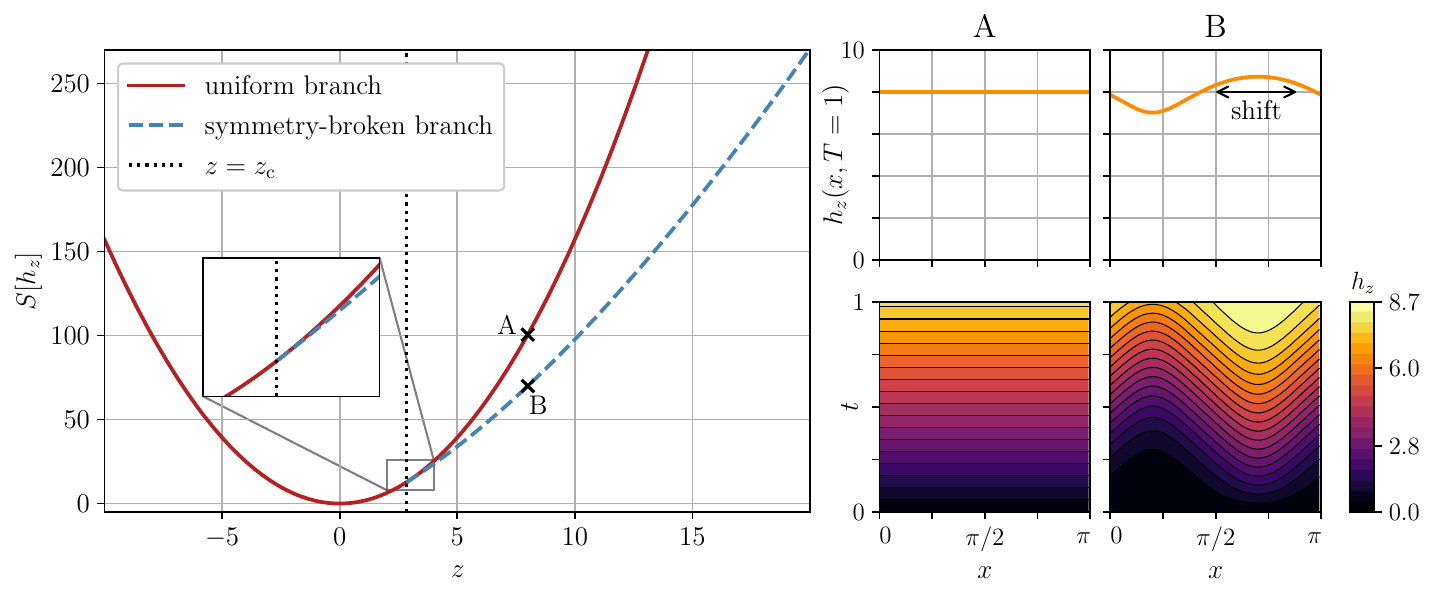}
\caption{Results of the numerical instanton computations for the non-dimensionalized
one-dimensional KPZ equation~\eqref{eq:kpz-wo-units} and average surface height
observable~\eqref{eq:mean-dens-obs} for different heights $z$.
Left: The action $S[h_z]$ of the two different branches that were found;
on the one hand,
a Gaussian branch~\eqref{eq:gaussian-rf} stemming from the critical
point~\eqref{eq:gaussian-inst} which is uniform in space and shown as a solid line,
and on the other hand, a branch given by localized and spatially non-uniform instantons
that splits off at $z_{\text{c}} \approx 2.8259$, indicated by the dotted
vertical line, as derived analytically in
Appendix~\ref{app:kpz-prefac-hom}. The inset shows a zoom onto the critical
point, suggesting a second-order phase transition (see the left
subplot of Figure~\ref{fig:kpz-prefac} for the first
derivative of the rate function). Two specific instanton configurations,
both with $z = 8$, are labeled by A and B and visualized in the
right half
of the figure. Center and right: Each column shows one instanton height profile $h_z$ as
marked in the left subplot; the top row shows the height profile at the final
instance in time and the bottom row shows the full space-time history of the instanton
height profiles. Both columns use the same axis scaling and color map normalization.}
\label{fig:kpz-minimizers}
\end{figure}

The idea is now that a trivial spatially homogeneous critical point $(h_z^{\text{hom}},
\tilde{h}_z^{\text{hom}}, \lambda_z^{\text{hom}})$ of the action
functional for the average
height observable, i.e. a solution of~\eqref{eq:instanton-eq-kpz},
is always given by
\begin{align}
h_z^{\text{hom}}(x, t) = z t\,,
\quad \tilde{h}_z^{\text{hom}}(x, t) = z\,,
\quad \lambda_z^{\text{hom}} = l z\,,
\label{eq:gaussian-inst}
\end{align}
with corresponding SHE instantons
\begin{align}
q_z^{\text{hom}}(x, t) = \exp \left\{z t\right\}\,,
\quad p_z^{\text{hom}}(x, t) = z \exp \left\{-z t\right\}\,,
\quad \lambda_z^{\text{hom}} = l z\,,
\label{eq:gaussian-inst-ch}
\end{align}
leading to the Gaussian rate function
\begin{align}
I_f^{\text{hom}}(z) = S\left[ h_z^{\text{hom}}\right] = \frac{1}{2}l z^2 
\label{eq:gaussian-rf}
\end{align}
for all such $z \in \RR$ for which this critical point realizes the global minimum
of the action under the boundary condition $f \left( h_z(\cdot, 1) \right) = z$.
However, one might expect (with reference to the typical growth patterns of the
KPZ equation due to the nonlinearity and diffusion, as
sketched in~\cite{kardar-parisi-zhang:1986}, as well as
the results and scaling estimates of~\cite{smith-meerson-sasorov:2018})
that for sufficiently large $z > z_{\text{c}}$
in the right tail of
the distribution of $f \left( H^\eps(\cdot, 1) \right)$, the KPZ nonlinearity
will favor a nonuniform surface growth in order to achieve a large average height,
such that the rate function displays a non-equilibrium phase
transition to a continuous family of spatially localized global minimizers
$\left\{ \left(h_z^{\text{loc}, u_0}, \tilde{h}_z^{\text{loc}, u_0},
\lambda_z^{\text{loc}}\right) \big \mid u_0 \in [0, l) \right\}$
of the instanton equations. This intuitive picture
is indeed confirmed by our numerical computations of instantons for this example,
performed directly for~\eqref{eq:instanton-eq-kpz}.
The corresponding results for the rate function as well as the space-time
evolution of typical instantons are shown in Figure~\ref{fig:kpz-minimizers}. For these
instanton computations, we used a pseudo-spectral
discretization in terms of $n_x = 128$ Fourier modes in space $[0, l]$ with $l = \pi$ and a second-order
explicit Runge-Kutta integrator in time $[0, 1]$ with an integrating factor for the diffusion
terms with $n_t = 2 \cdot 10^4$ equidistant time steps of
size $\Delta t = 5 \cdot 10^{-5}$.
The comparably high resolution in time turned out to be necessary for the
subsequent Riccati equation integrations, for which the instantons serve as an input,
as detailed below.
In order to directly compute instantons for different and given observable
values $z$, equidistantly spaced in $[-10, 20]$, we
use a penalty-type method, and minimized the
action using L-BFGS steps with exact discrete adjoint gradient evaluations in order
to reduce the $L^2$-norm of the action gradient by a factor of $10^6$
in each subproblem. For details on the optimization procedure, we 
refer the reader to~\cite{schorlepp-etal:2022}.\\


\begin{figure}
\centering
\includegraphics[width = \textwidth]{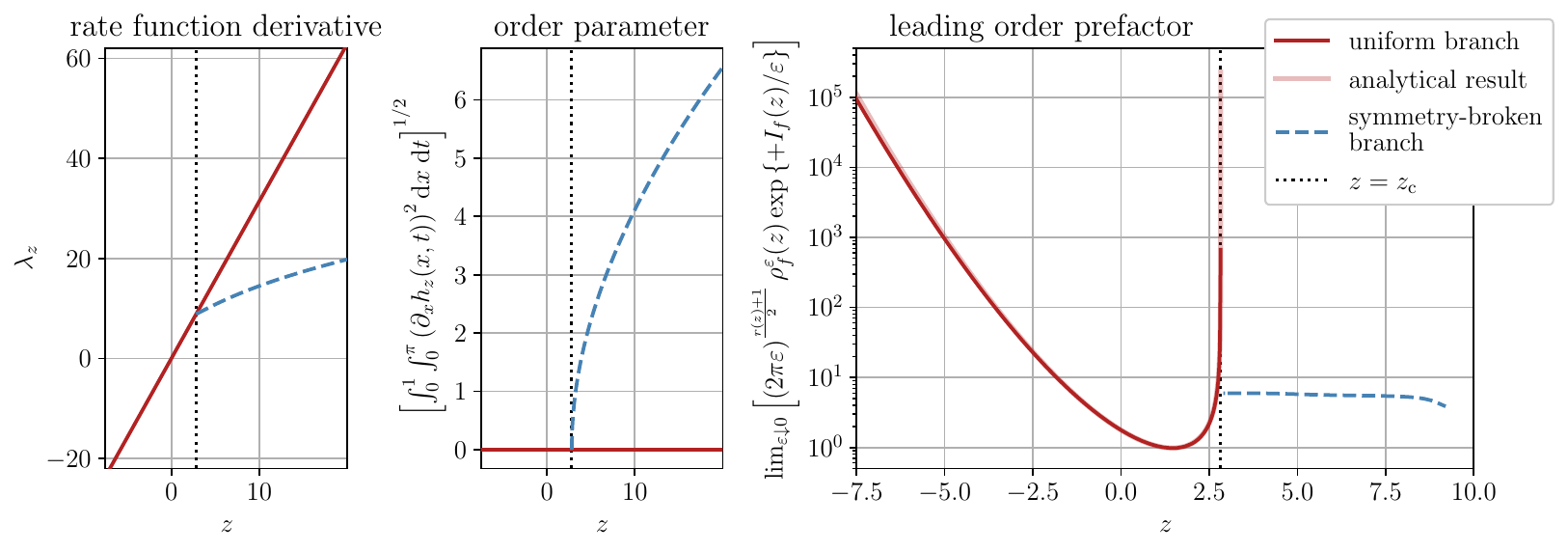}
\caption{Left: Lagrange multiplier~$\lambda_z$, which, by Legendre
duality, is equal to the derivative of the rate function $I_F$, for
the KPZ instantons. At the critical observable value~$z = z_{\text{c}}$,
the first derivative~$I_F'$ is continuous but not differentiable,
hence the phase transition is second order.
Center: The~$L^2$ norm of the derivative~$\partial_x h_z$ for
the KPZ instantons as an order parameter.
Right: Results of the numerical prefactor computations for the non-dimensionalized
one-dimensional KPZ equation~\eqref{eq:kpz-wo-units} and average surface height
observable~\eqref{eq:mean-dens-obs} for different heights $z$. The solid dark red line shows
$R_z \left[ \dd \lambda_z^{\text{hom}} / \dd z\right]^{1/2}$ as obtained from
numerical solution of the backward Riccati equation~\eqref{eq:riccati-she} with
numerical parameters as detailed in the main text, whereas the light red line
indicates the corresponding analytical result from Appendix~\ref{app:kpz-prefac-hom},
eqs.~\eqref{eq:she-res-rz} and~\eqref{eq:prefac-cases-pos-z}
(with the appearing products evaluated until $k = 100$ for this figure).
Beyond $z = z_{\text{c}}$, visualized by the dotted black line,
the dashed blue line shows the prefactor $\tilde{R}_{z} \big[\dd
\lambda_z^{\text{loc}} / \dd z\big]^{1/2}$,
computed by solving the backward Riccati equation
with modified final condition~\eqref{eq:w-final-kpz-zero-mode}. Higher temporal resolution of the Riccati equation close to $t=1$ would allow
to extend the results to $z > 9$.}
\label{fig:kpz-prefac}
\end{figure}

From the results of the instanton computations, we see that this
constitutes an example of a dynamical
phase transition in an irreversible SPDE where the associated symmetry that is
broken is continuous, thereby allowing us to apply the methods developed in the
previous section in order to compute not only the large deviation rate function, given
by the pointwise minimum of the two branches in Figure~\ref{fig:kpz-minimizers}, but
also a more refined, asymptotically sharp prefactor estimate. The phase transition
is second order, as can be seen from the derivative of the rate function in the left subplot of
Figure~\ref{fig:kpz-prefac}, and we also show the $L^2$ norm of $\partial_x h_z$
for the instantons as an order parameter for the different phases in
the center subplot of Figure~\ref{fig:kpz-prefac}. Since the KPZ equation is a
non-equilibrium system, in contrast to the previous
example~\ref{subsec:dynam-phase-trans-ode-potential}, the complete Riccati formalism
and the corresponding numerical integration of a Riccati partial differential equation
with regularized boundary data is now required to get the leading order prefactor.\\


When the spatially homogeneous instanton dominates, the rate function of the
average surface height in the small noise limit is Gaussian with
\begin{align}
\rho_F^\eps(z) \overset{\eps \downarrow 0}{\sim} \left(2 \pi \eps
\right)^{-1/2}  \underbrace{l^{1/2}}_{= \left[ \dv{}{z}
\lambda_z^{\text{hom}}\right]^{1/2}}\, R_z \, \exp \bigg\{-\frac{1}{\eps}
\underbrace{\tfrac{1}{2}l z^2}_{= I_f^{\text{hom}}(z)} \bigg\}\,,
\label{eq:kpz-gauss-pdf}
\end{align}
but the prefactor component $R_z$ can still depend nontrivially on $z$.
The only restriction on the function $R_\cdot$ is that at $z = 0$,
we have $R_0 = 1$ for correct normalization of the PDF
as $\eps \downarrow 0$.
In the case of the spatially homogeneous instanton, the prefactor
component $R_z$ can be found analytically using probabilistic methods without
explicit reference to the functional integration methods developed here,
which is carried out in detail in Appendix~\ref{app:kpz-prefac-hom}. The analysis
of $R_z$ for the homogeneous instantons in particular yields the prediction that
the critical observable value $z_{\text{c}}$ for the
second order phase transition, where the $(k = 1)$-contribution to the
prefactor is found to
diverge, is the smallest nontrivial real solution of the equation
\begin{align}
\tan \left(\frac{2 \pi}{l} \sqrt{2 z_{\text{c}}(l) -
\left(\frac{2 \pi}{l} \right)^2} \right) + \left(\frac{2 \pi}{l}
\right)^{-1} \sqrt{2 z_{\text{c}}(l) - \left(\frac{2 \pi}{l}
\right)^2} = 0\,,
\end{align}
for $l = \pi$ and hence $z_{\text{c}}(l = \pi) \approx 2.8259$
as sketched in Figure~\ref{fig:kpz-minimizers},
which matches the numerical results
of the instanton computations quite well.\\


Now, we turn to the numerical prefactor computation in the SHE formulation
using Riccati fields. We use the backward Riccati formalism\footnote{The system
at hand is an example where, regardless of the spontaneous symmetry breaking
and indeed already for the spatially homogeneous instanton, the forward Riccati
equation can be ill-posed for certain observable values, whereas the backward equation
remains well-posed for the same observable values. Conceptually, we conjecture
that this is due to the fact that divergences of the backward Riccati
matrix~$W = \zeta \gamma^{-1}$
are related to conjugate points and violations of the positive definiteness of
the second variation at the instanton, whereas divergences of the forward Riccati
matrix~$Q = \gamma \zeta^{-1}$ can appear when the momentum passes through zero
without ``physical'' consequences. In the example of this subsection, one can
find parameters for which the solution of the forward Riccati equation
in~\eqref{eq:ric-alpha-beta} passes through a singularity in~$(0, T)$,
prohibiting forward numerical integration, while the analytical
result~\eqref{eq:alpha-beta-cases} remains finite. This is the reason why we
use the backward Riccati approach for all numerical computations in this
subsection.} from
Proposition~\ref{thm:pdf-orefac-degen-final}. The result for
the PDF of $F({\cal Q}^\eps(\cdot, 1))$ as $\eps \downarrow 0$ is given by
\begin{align}
\rho_F^\eps(z) \overset{\eps \downarrow 0}{\sim}
\begin{cases}
\left(2 \pi \eps \right)^{-1/2} R_z \left[ \dv{}{z}
\lambda_z^{\text{hom}}\right]^{1/2} \exp \left\{-\frac{1}{\eps}
S\left[h_z^{\text{hom}} \right] \right\} \,, \quad & z < z_{\text{c}}\,,\\
\left(2 \pi \eps \right)^{-1} \tilde{R}_{z} \left[ \dv{}{z}
\lambda_z^{\text{loc}}\right]^{1/2}
\exp \left\{-\frac{1}{\eps} S\left[h_z^{\text{loc}, u_0} \right] \right\}\,,
\quad &  z > z_{\text{c}}\,.
\end{cases}
\label{eq:pdf-reparam-kpz}
\end{align}
In~\eqref{eq:pdf-reparam-kpz}, the prefactor components
\begin{align}
R_z = \exp \left\{\frac{1}{2} \int_0^1 \dd t \int_0^{l} \dd x
\left(q_z^{\text{hom}}(x,t)\right)^2 W_z(x,x,t) \right\}
\end{align}
and
\begin{align}
\tilde{R}_z &= l V\left(z; q_z^{\text{loc}, u_0}, \psi_z^{u_0}\right)
\exp \left\{\frac{1}{2} \int_0^1 \dd t \int_0^{l} \dd x
\left(q_z^{\text{loc}, u_0}(x,t)\right)^2 W_z^{u_0}(x,x,t) \right\}
\end{align}
with volume factor
\begin{align}
V\left(z; q_z^{\text{loc}, u_0}, \psi_z^{u_0}\right):=
\left[\int_0^{l} \left(\psi_z^{u_0}(x,1) \right)^2
\left(1 - \frac{\lambda_z^{\text{loc}}}{l}
\frac{1}{\left(q_z^{\text{loc},
u_0}(x, 1) \right)^2} \right)
\dd x \right]^{1/2}
\end{align}
depend on the backward Riccati field $W_z \colon [0, l]^2
\times [0,1] \to \RR$ solving
\begin{align}
\partial_t W_z(x,y,t) &= -\left(p_z(x,t)\right)^2 \delta(x-y) -
\left(\partial_{xx} + \partial_{yy}\right) W_z(x,y,t) \nonumber \\
&\quad - 2 \left(q_z(x,t) p_z(x,t) + q_z(y,t) p_z(y,t) \right) W_z(x,y,t)
- \int_0^{l} W_z(x,x',t) \left(q_z(x',t)\right)^2 W_z(x',y,t) \dd x' \,,
\label{eq:riccati-she}
\end{align}
for both cases along the respective instantons, and with final condition
\begin{align}
W_z(x, y, 1) = -\frac{\lambda_z^{\text{hom}}}{l} \frac{\delta(x-y)}{\left( q_z^{\text{hom}}(x, 1) \right)^2}
\label{eq:w-final-kpz-wo-zero-mode}
\end{align}
for the homogeneous instanton and
\begin{align}
W_z^{u_0}(x, y, 1) &=  -\frac{\lambda_z^{\text{loc}}}{l} \frac{\delta(x-y)}{ \left(
q_z^{\text{loc}, u_0}(x, 1) \right)^2} - \left(V\left(z; q_z^{\text{loc}, u_0}, \tilde{\psi}_z^{u_0}\right)
\right)^2 \tilde{\psi}_z^{u_0}(x,1) \tilde{\psi}_z^{u_0}(y,1)
\label{eq:w-final-kpz-zero-mode}
\end{align}
for the spatially localized instanton.
In all of these expressions, the zero mode is given by
\begin{align}
\psi_z^{u_0}(x,t) = \partial_x q_z^{\text{loc}, u_0}(x, t)
\end{align}
with $u_0 \in [0, l)$ denoting the reference position
of the localized instanton, and the normalized zero mode is
defined by
\begin{align}
\tilde{\psi}_z^{u_0}(x,1) = \frac{\psi_z^{u_0}(x,1)}{\left[\int_0^{l} \left(\psi_z^{u_0}(x',1)\right)^2 \dd x' \right]^{1/2}}\,.
\end{align}\\


Numerically evaluating the prefactor by solving the Riccati equation and
differentiating~$\lambda_z^{\text{loc}}$ with respect to $z$ using finite differences,
we obtain the results shown in the right panel of Figure~\ref{fig:kpz-prefac}
for the leading order prefactor
\begin{align}
\lim_{\eps \downarrow  0} \left[ \left(2 \pi \eps \right)^{\frac{r(z) + 1}{2}}
\rho_f^\eps(z) \exp \left\{+\frac{1}{\eps} I_f(z) \right\}\right]\,,
\end{align}
where $r(z) = 0$ for $z < z_{\text{c}}$ and $r(z) = 1$ for $z > z_{\text{c}}$.
For the solution of the Riccati equation~\eqref{eq:riccati-she}, we also used
a pseudo-spectral, anti-aliased code at spatial resolution $n_x = 128$ with the
Cole-Hopf transformed KPZ instantons as an input. For the time stepping, the same
Heun integrator with an appropriate integrating factor in Fourier space was used,
but we had to choose a different time resolution for numerical stability reasons.
It turned out that the final condition~\eqref{eq:w-final-kpz-zero-mode} requires
extremely small time steps in the vicinity of $t = 1$, and accordingly, we
divided the time interval $[0,1]$ into two subintervals $I_1 = [0, t_0]$
and $I_2 = [t_0, 1]$ with time steps of a different, smaller
size $\Delta t_2$ within $I_2$ compared to $\Delta t_1$ within $I_1$.
All results shown in Figure~\ref{fig:kpz-prefac} were generated
using $t_0 = 0.99995$, $\Delta t_1 \approx 1.1 \cdot 10^{-5}$
and $\Delta t_2 = 5 \cdot 10^{-9}$ with $n_t = 10^5$ time steps in total.
Further increasing the resolution would allow to extend the dashed curve in 
Figure~\ref{fig:kpz-prefac} to higher values of $z$, the relevant influence being the
size of $\Delta t_2$ here. We made sure that the results shown are invariant under
modifications of $\Delta t_1$, $\Delta t_2$ and $t_0$ as long as these yield
finite results.\\

\begin{figure}
\centering
\includegraphics[width = \textwidth]{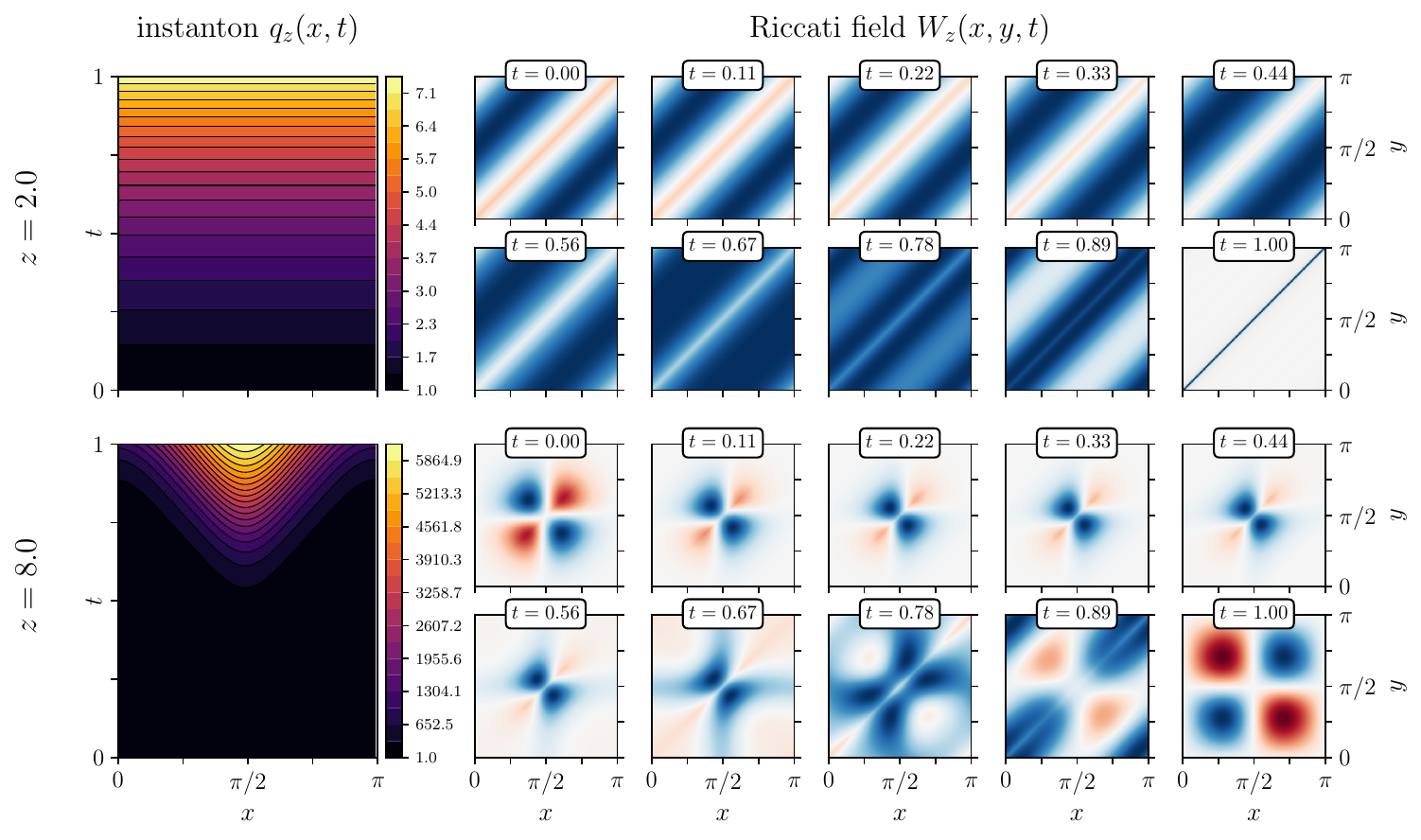}
\caption{KPZ Instantons in SHE variables~\eqref{eq:ch-trafo-inst} (left)
and corresponding backward Riccati solutions of~\eqref{eq:riccati-she}
(right) for two different observable values~$z = 2 < z_c$ (top row) and~$z = 8 > z_c$
(bottom row). All fields were computed at~$n_x = 128$ and with~$n_t = 20000$ time
steps (with uniformly spaced time steps for the instantons,
and $\Delta t_1 \approx 5.6 \cdot 10^{-5}$, $\Delta t_2 = 2.5 \cdot 10^{-8}$
and $t_0 = 0.99995$ for the Riccati fields). The color bar ranges for all
Riccati snapshots are adjusted to the current maximum absolute value of the
field and chosen to be symmetric around 0, with red symbolizing positive field
values. For the spatially homogeneous case~$z = 2$, we see that the Riccati
field remains Toeplitz for all times, starting from a Dirac~$\delta$ final
condition~\eqref{eq:w-final-kpz-wo-zero-mode}. For the localized case~$z = 8$
the
final condition for~$W_z(\cdot, \cdot, 1)$ is clearly dominated by the zero
mode dependent part of~\eqref{eq:w-final-kpz-zero-mode}, leading to a
quadrupole-like structure. For decreasing~$t \downarrow 0$, the field
transitions into a similar structure but with a flipped sign.}
\label{fig:kpz-riccati-solutions}
\end{figure}

From the left subplot of Figure~\ref{fig:kpz-prefac}, we see that for
the spatially homogeneous instanton, the numerical results from solving the
backward Riccati equation~\eqref{eq:riccati-she} closely match the analytical
calculations from Appendix~\ref{app:kpz-prefac-hom}. Further, the prefactor
beyond the critical observable value $z_{\text{c}}$ only has a weak dependence
on~$z$, and the behavior at $z > 9$ is only due to the fact that a higher
time resolution would be needed there. Furthermore, we show
the instanton and the corresponding solution of the Riccati equation
at different times for observable values $z = 2 < z_{\text{c}}$ and
$z = 8 > z_{\text{c}}$ in Figure~\ref{fig:kpz-riccati-solutions}.
All in all, we have demonstrated with this example that the formalism developed
in this paper can indeed be employed to analyze nontrivial, spatially extended
non-equilibrium systems in the presence of phase transitions.


\section{Discussion and Outlook}
\label{sec:discussion-outlook}

Going beyond large deviation estimates and obtaining sharp limits for
rare events in stochastic systems is important for many applications,
including nonequilibrium phase transitions. Importantly, one obtains
the full limiting rare event probability or probability density
instead of merely its exponential scaling, in regimes where direct
sampling methods are completely intractable. In this paper, we have
first set out to rederive such prefactor formulas at leading
order for unique instantons~\cite{schorlepp-grafke-grauer:2021,grafke-schaefer-vanden-eijnden:2021,ferre-grafke:2021,bouchet-reygner:2022},
expressed in terms of Riccati matrices, explicitly using tools from field theory,
i.e.\ by evaluating the appearing functional determinants using Forman's
theorem~\cite{forman:1987}. The resulting derivations are
short and conceptually simple. We stressed the role of
the MGF for a vast simplification of the computations, which in particular simplifies
the boundary conditions of the second variation operator in path space. Secondly,
writing the prefactor in terms of operator determinants allowed us to extend the
Riccati formalism to situations where the second variation around the instanton path that
is used for the expansion is only positive semi-definite due to the presence of zero modes, i.e.\ degenerate
submanifolds of instantons. We have demonstrated, using boundary-type
regularizations~\cite{falco-fedorenko-gruzberg:2017},
that the Riccati approach remains feasible in this case, i.e.\ that the reduced functional
determinant with removed zero eigenvalues can still be expressed through the solution of
the same matrix Riccati differential equation, only with modified initial/final conditions
or evaluations involving knowledge of the zero modes. Afterwards, we have verified our
results in four different examples involving linear and nonlinear, reversible and
irreversible SDEs as well as a nonlinear irreversible SPDE, the KPZ equation,
exhibiting spontaneous symmetry breaking of the instantons for the average surface
height.\\

With the general treatment of zero modes completed, it is now theoretically possible to
compute leading order large deviation prefactors even for multi-dimensional SPDEs
such as the two-dimensional or three-dimensional
Navier-Stokes equations where spontaneous symmetry breaking of the rotational symmetry
of instantons has indeed been
observed~\cite{falkovich-lebedev:2011,schorlepp-etal:2022}. The remaining
complication for numerical computations is the high dimensionality of the involved
Riccati matrices, and it would be interesting future work to consider low-rank
approximations of the Riccati differential equations~\cite{breiten-dolgov-stoll:2020} in
this regard, that could e.g.\ make use of the sparsity of the large-scale forcing
typically used in turbulence simulations. Alternatively, an approach based on computing
only the dominant eigenvalues of a Carleman-Fredholm determinant expression for the
prefactor~\cite{ben-arous:1988} could be used for numerical
computations, which will be the subject of a future publication. Another interesting
project would be the development of efficient importance sampling algorithms for rare
events as e.g.\ in~\cite{ebener-margazoglou-friedrich-etal:2019} for systems with
non-unique instantons due to symmetry breaking. 


\begin{acknowledgments}
The authors wish to thank Baruch Meerson, Pavel Sasorov
and Naftali Smith for pointing out important literature and sharing
their insights on the dynamical phase transition for the spatially
averaged surface height of the KPZ equation.
T.S.\ and R.G.\ benefited from support through the DFG collaborative research center SFB-1491. 
T.G.\ acknowledges the support received from the EPSRC projects
EP/T011866/1 and EP/V013319/1.
\end{acknowledgments}

%


\appendix
\counterwithin{mythm}{section}

\section{Forman's theorem for the second variation of a
general action functional}
\label{app:forman-lagrange-to-hamilton}

\begin{mythm}[Forman's theorem for second order ordinary differential
    operators~\cite{forman:1987,falco-fedorenko-gruzberg:2017}]
    \label{thm:forman-general}
   
  Let $\Omega^{(i)}$, $i \in \{1,2\}$ be two second order
  differential operators
  \begin{align}
    \Omega^{(i)} = P_0 \dv[2]{}{t} + P_1^{(i)} \dv{}{t} + P_2^{(i)}
  \end{align}
  acting on functions $\gamma: [0,T] \to \RR^n$. Here, $P_0,
  P_1^{(i)}, P_2^{(i)}:[0,T] \to \RR^{n \times n}$ are matrix-valued
  functions. The highest-order coefficient $P_0$, which is identical
  for both $\Omega^{(1)}$ and $\Omega^{(2)}$, is assumed to be
  invertible. We impose boundary conditions
  \begin{align}
    M^{(i)} \left(\begin{array}{c}
      \gamma(0)\\
     \dot{\gamma}(0)
    \end{array} \right) + N^{(i)} \left(\begin{array}{c}
      \gamma(T)\\
      \dot{\gamma}(T)
    \end{array} \right) = 0
  \end{align}
  on the functions~$\gamma$ on which~$\Omega^{(i)}$ acts,
  where~$M^{(i)}, N^{(i)} \in \RR^{2n \times 2n}$. Then the quotient of the
  functional determinants of~$\Omega^{(1)}$ and~$\Omega^{(2)}$ under
  these boundary conditions, assuming $\Omega^{(2)}$ has no zero
  eigenvalues, is well defined and can be computed as
  \begin{align}
    \frac{\Det_{M^{(1)}, N^{(1)}} \left( \Omega^{(1)} \right)}
         {\Det_{M^{(2)}, N^{(2)}} \left( \Omega^{(2)} \right)} =
         \frac{{\det}_{2n} \left(M^{(1)} \Upsilon^{(1)} (0)
         + N^{(1)} \Upsilon^{(1)}(T) \right)}
         {{\det}_{2n} \left(M^{(2)} \Upsilon^{(2)} (0) + N^{(2)}
         \Upsilon^{(2)}(T) \right)}
         \left[ \frac{{\det}_{2n} \Upsilon^{(2)}(0) {\det}_{2n}
         \Upsilon^{(2)}(T)}{{\det}_{2n} \Upsilon^{(1)}(0)
         {\det}_{2n} \Upsilon^{(1)}(T)} \right]^{1/2}\,. 
         \label{eq:forman}
  \end{align}
  Here, $\Upsilon^{(i)}:[0,T] \to \RR^{2n \times 2n}$ is any
  fundamental system of solutions of the homogeneous first order equation
  \begin{align}
    \Omega^{(i)} \gamma = 0 \; \Leftrightarrow \;
    \dv{}{t}\left(\begin{array}{c}
      \gamma\\
      \dot{\gamma}
    \end{array} \right) = \tilde{\Gamma}\left[\Omega^{(i)}\right]
    \left(\begin{array}{c}
       \gamma\\
      \dot{\gamma}
    \end{array} \right) = \left(\begin{array}{c|c}
      0_{n \times n} & 1_{n \times n}\\
      \hline
      -\left(P_0 \right)^{-1} P_2^{(i)} & -\left(P_0 \right)^{-1} P_1^{(i)}
    \end{array} \right) \left(\begin{array}{c}
      \gamma\\
      \dot{\gamma}
    \end{array} \right) \,.
    \label{eq:forman-first-order-ode-general}
  \end{align}
\end{mythm}


In the remainder of this appendix, we focus on operators
originating from the second variation of a generic action functional
\begin{align}
S[\phi] = \int_0^T L(\phi,\dot{\phi}) \dd t
\end{align}
for paths $\phi:[0,T] \to \RR^n$ with boundary conditions that we
do not specify in this section. Expanding the action to second order
around a stationary path $\phi$ yields the following quadratic form:
\begin{align}
\delta^2 S[\phi][\gamma] = \frac{1}{2} \int_0^T \left \langle
\left( \begin{array}{c}
\gamma\\
\dot{\gamma}
\end{array} \right), \left( \begin{array}{cc}
\nabla^2_\phi L & \nabla_\phi \nabla_{\dot{\phi}} L\\
\nabla_{\dot{\phi}} \nabla_\phi L & \nabla_{\dot{\phi}}^2 L
\end{array} \right)  \left( \begin{array}{c}
\gamma\\
\dot{\gamma}
\end{array} \right) \right \rangle_{2n} \dd t\,,
\end{align}
with the convention
\begin{align}
\left(\nabla_\phi \nabla_{\dot{\phi}} L \right)_{ij} :=
\frac{\partial^2 L}{\partial \phi_i \partial \dot{\phi}_j}
\end{align}
and all derivatives of $L$ evaluated along $\phi$.
We transform this expression into the form $\tfrac{1}{2}\int_0^T
\left \langle \gamma, \Omega \gamma \right \rangle_n \dd t$ via
partial integration:
\begin{align}
\delta^2S[\phi][\gamma] &= \frac{1}{2} \int_0^T \underbrace{\left 
\langle \dot{\gamma}, \left(\nabla^2_{\dot \phi} L \right)
\dot{\gamma} \right \rangle_n}_{=:I_1} + \underbrace{\left \langle
\gamma, \left(\nabla_\phi\nabla_{\dot \phi} L \right) \dot{\gamma}
\right \rangle_n + \left \langle \dot{\gamma}, \left(\nabla_{\dot \phi}
\nabla_\phi L \right) \gamma \right \rangle_n}_{=:I_2} + \left \langle
\gamma, \left(\nabla^2_\phi L \right) \gamma \right \rangle_n \dd t
\nonumber\\
I_1 &= \frac{1}{2} \left. \left \langle \gamma, \left(
\nabla^2_{\dot \phi} L \right) \dot{\gamma} \right \rangle_n \right
\vert^T_0 - \frac{1}{2} \int_0^T \left \langle \gamma, \frac{\dd}
{\dd t} \left( \left(\nabla^2_{\dot \phi} L \right) \frac{\dd}{\dd t}
\right) \gamma \right \rangle_n \dd t \nonumber\\
I_2&= \frac{1}{2}\int_0^T \frac{\dd}{\dd t} \left( \left \langle
\gamma,  \left( \nabla_{\dot{\phi}} \nabla_\phi  L \right) \gamma
\right \rangle_n \right) - \left \langle \gamma, \frac{\dd}{\dd t}
\left( \nabla_{\dot{\phi}}  \nabla_\phi L \right) \gamma \right 
\rangle_n + \left \langle \gamma, \left( \left[\nabla_{\phi},
\nabla_{\dot{\phi}} \right] L \right) \dot{\gamma} \right \rangle_n
\dd t \nonumber \\
&= \frac{1}{2} \left. \left \langle \gamma,  \left(
\nabla_{\dot{\phi}} \nabla_\phi L \right) \gamma \right \rangle_n
\right \vert^T_0 + \int_0^T \left \langle \gamma, \left( \left[
\nabla_\phi, \nabla_{\dot{\phi}} \right] L \right) \frac{\dd}{\dd t}
\gamma \right \rangle_n - \left \langle \gamma, \frac{\dd}{\dd t}
\left(\nabla_{\dot{\phi}} \nabla_\phi L \right)\gamma \right 
\rangle_n \dd t\,.
\end{align}
Here, $\left[\cdot, \cdot \right]$ denotes the commutator of
two operators. With the definition $\theta:= \nabla_{\dot{\phi}} L$
for the conjugate momentum and hence
\begin{align}
\zeta := \left( \nabla^2_{\dot{\phi}} L \right) \dot{\gamma}
+ \left( \nabla_{\dot{\phi}} \nabla_\phi L \right) \gamma
\end{align}
for the momentum fluctuations, the additional boundary term that
we obtain and that needs to vanish through the imposition of suitable
boundary conditions (cf.\ main text) for the fluctuations
is $ \tfrac{1}{2} \left. \left \langle \gamma, \zeta \right \rangle_n
\right \vert^T_0$, leaving us with
\begin{align}
\delta^2S[\phi][\gamma] = \frac{1}{2} \int_0^T \left \langle \gamma,
\left[- \frac{\dd}{\dd t} \left( \left(\nabla^2_{\dot \phi} L \right)
\frac{\dd}{\dd t} \right) +  \left( \left[\nabla_\phi,
\nabla_{\dot{\phi}} \right] L \right) \frac{\dd}{\dd t}  +
\left(\nabla_\phi^2 L \right) - \frac{\dd}{\dd t}
\left(\nabla_{\dot{\phi}} \nabla_\phi L \right) \right] \gamma
\right \rangle_n \dd t\,.
\end{align} 
Written in this way, the Jacobi operator~\cite{evans:2021}, i.e.\ the
second order linear differential operator
\begin{align}
\Omega = \Omega[\phi] = \left[- \frac{\dd}{\dd t} \left(
\left(\nabla^2_{\dot \phi} L \right) \frac{\dd}{\dd t} \right) +
\left( \left[\nabla_\phi, \nabla_{\dot{\phi}} \right] L \right)
\frac{\dd}{\dd t}  + \left(\nabla_\phi^2 L \right) - \frac{\dd}{\dd t}
\left(\nabla_{\dot{\phi}} \nabla_\phi L \right) \right]
\end{align}
realizing the second variation is $L^2([0,T],\RR^n)$-self-adjoint,
i.e.\ $\left \langle \gamma_1, \Omega \gamma_2 \right \rangle = \left 
\langle  \Omega \gamma_1, \gamma_2 \right \rangle$ for all fluctuation
paths with boundary conditions such that $ \left. \left \langle
\zeta_1, \gamma_2  \right \rangle_n \right \vert^T_0 - \left. \left
\langle \gamma_1, \zeta_2 \right \rangle_n \right \vert^T_0 = 0$.\\


For the first order equation in Forman's theorem, we read off
\begin{align}
P_0 &= - \nabla^2_{\dot{\phi}} L\\
P_1 &= \left[\nabla_\phi, \nabla_{\dot{\phi}} \right] L -
\frac{\dd}{\dd t} \left( \nabla^2_{\dot{\phi}} L \right) \\
P_2 &= \nabla^2_\phi L - \frac{\dd}{\dd t} \left(\nabla_{\dot{\phi}}
\nabla_\phi L \right)\,.
\end{align} 
The first order version of the Jacobi equation
\begin{align}
\Omega[\phi] \gamma = 0 \label{eq:jacobi}
\end{align}
appearing in Forman's theorem hence becomes
\begin{align}
\frac{\dd}{\dd t} \left(\begin{array}{c}
\gamma \\
\dot{\gamma}
\end{array} \right) &= \tilde{\Gamma}[\Omega] \left(\begin{array}{c}
\gamma \\
\dot{\gamma}
\end{array} \right) = \left(\begin{array}{c|c}
      0_{n \times n} & 1_{n \times n}\\
      \hline
      -\left(P_0 \right)^{-1} P_2 & -\left(P_0 \right)^{-1} P_1
    \end{array} \right) \left(\begin{array}{c}
      \gamma\\
      \dot{\gamma}
    \end{array} \right) \nonumber\\
    &= \left(\begin{array}{c|c}
      0_{n \times n} & 1_{n \times n}\\
      \hline
      \left(\nabla_{\dot{\phi}}^2 L \right)^{-1} \left(
      \nabla^2_\phi L -\frac{\dd}{\dd t} \left(\nabla_{\dot{\phi}}
      \nabla_\phi L \right) \right) & \left(\nabla_{\dot{\phi}}^2 L
      \right)^{-1} \left( \left[\nabla_\phi, \nabla_{\dot{\phi}} \right]
      L - \frac{\dd}{\dd t} \left( \nabla^2_{\dot{\phi}} L \right) \right)
    \end{array} \right) \left(\begin{array}{c}
      \gamma\\
      \dot{\gamma}
    \end{array} \right)\,. \label{eq:forman-general-lagrange}
\end{align}
In many application, such as in this paper, it is more natural to
switch to a Hamiltonian instead of a Lagrangian formulation of the
Jacobi equation. In fact, we have already seen above that the natural
boundary conditions for the fluctuations include the conjugate
momentum fluctuations. Due to this reason, we associate to the
fundamental system of solutions $\Upsilon$ of $\Omega$,
understood as a first
order differential equation~\eqref{eq:forman-general-lagrange}
in~$(\gamma,\dot{\gamma})$, the following fundamental system of solutions
\begin{align}
\tilde{\Upsilon} := \Lambda \Upsilon\,, \quad \Lambda =
\left(\begin{array}{c|c}
      1_{n \times n} & 0_{n \times n}\\
      \hline
      \nabla_{\dot{\phi}} \nabla_\phi L & \nabla^2_{\dot{\phi}} L
    \end{array} \right)\,. 
\end{align}
The transformation is invertible iff $P_0 = - 
\nabla^2_{\dot{\phi}} L$ is invertible (which is exactly an
assumption of Forman's theorem) with
\begin{align}
\Lambda^{-1} = \left(\begin{array}{c|c}
      1_{n \times n} & 0_{n \times n}\\
      \hline
      -\left( \nabla^2_{\dot{\phi}} L \right)^{-1}
      \nabla_{\dot{\phi}} \nabla_\phi L & \left(
      \nabla^2_{\dot{\phi}} L \right)^{-1}
    \end{array} \right)\,.
\end{align}
A straightforward calculation then shows that
\begin{align}
\frac{\dd}{\dd t} \tilde{\Upsilon} = \left(\dot{\Lambda}
\Lambda^{-1}  + \Lambda \dot{\Upsilon} \Lambda^{-1} \right)
\tilde{\Upsilon} =: \Gamma[\Omega] \tilde{\Upsilon}
\end{align}
with
\begin{align}
 \Gamma[\Omega] &= \left(\begin{array}{c|c}
      -\left( \nabla^2_{\dot{\phi}} L \right)^{-1} \left(
      \nabla_{\dot{\phi}} \nabla_\phi L \right) & \left(
      \nabla^2_{\dot{\phi}} L \right)^{-1}\\
      \hline
       \nabla^2_\phi L - \left(\nabla_\phi \nabla_{\dot{\phi}}
       L\right) \left(\nabla_{\dot{\phi}}^2 L\right)^{-1}
       \left(\nabla_{\dot{\phi}} \nabla_\phi  L\right)& \left(
       \nabla_\phi \nabla_{\dot{\phi}} L \right)
       \left(\nabla^2_{\dot{\phi}} L \right)^{-1}
    \end{array} \right) = \left(\begin{array}{c|c}
      \nabla_\theta \nabla_\phi H & \nabla_\theta^2 H\\
      \hline
      - \nabla_\phi^2 H & - \nabla_\phi \nabla_\theta H
    \end{array} \right)\nonumber\\
    &= 
    \left( \begin{array}{c|c}
    0_{n \times n} & 1_{n \times n}\\
    \hline
    -1_{n \times n} & 0_{n \times n} \end{array} \right)
    \left( \begin{array}{c|c}
      \nabla_\phi^2 H & \nabla_\phi \nabla_\theta H\\
      \hline
      \nabla_\theta \nabla_\phi H & \nabla_\theta^2 H
    \end{array} \right) =: J \; \nabla^2 H\,,
    \label{eq:first-order-operator-general}
\end{align}
where~$J$ is the standard $2n \times 2n$ symplectic matrix.
The second equality in~\eqref{eq:first-order-operator-general}
holds if $\phi$ is a critical point of the action functional,
or, equivalently, $(\phi, \theta)$ is a solution of the canonical
equations of motion
\begin{align}
\dv{}{t} \left(\begin{array}{c}
\phi\\\theta
\end{array} \right) = J\; \nabla H(\phi,\theta) =
\left(\begin{array}{c}
\nabla_\theta H(\phi,\theta)\\-\nabla_\phi H(\phi,\theta)
\end{array} \right)
\end{align}
The Hamiltonian $H$ is defined via
\begin{align}
H(\phi, \theta) = \sup_y \left(\left \langle \theta, y
\right \rangle_n - L(\phi, y) \right) = \left \langle \theta,
\dot{\phi}(\phi, \theta) \right \rangle -
L(\phi,\dot{\phi}(\phi, \theta))\,,
\end{align}
where the second equality follows by assuming strict convexity
of $L$ in $\dot{\phi}$ and solving the implicit equation $\theta
= \partial L(\phi, \dot{\phi}) / \partial \dot{\phi}$ for
$\dot{\phi}$. Let us summarize the results of the transformation
in the following proposition.


\begin{mythm}[Forman's theorem for second variations in
Hamiltonian formulation (cf.\ \citep{corazza-fadel:2020})]
\label{thm:forman-hamiltonian}
  Let
  \begin{align}
  S[\phi] = \int_0^T L(\phi,\dot{\phi}) \dd t
  \end{align}
  be an action functional with $\nabla^2_{\dot{\phi}} L$ independent
  of $\phi$ and $\dot{\phi}$,
  and consider two paths $\phi_1, \phi_2: [0,T] \to \RR^n$ that are
  critical points of the action. Then the quotient of functional
  determinants of~$\Omega[\phi_i]$, realizing the second variation
  of~$S$ along~$\phi_i$ as
  \begin{align}
  \delta^2S[\phi_i][\gamma] = \frac{1}{2} \int_0^T \left \langle
  \gamma(t), (\Omega[\phi_i] \gamma)(t) \right \rangle_n \dd t
  \end{align}
  with boundary conditions ${\cal A}_i$ imposed on the fluctuations,
  can be computed as
    \begin{align}
    \frac{\Det_{M_1, N_1} \left( \Omega[\phi_1] \right)}
         {\Det_{M_2, N_2} \left( \Omega[\phi_2] \right)} =
         \frac{{\det}_{2n} \left(M_1 \Upsilon_1 (0) + N_1
         \Upsilon_1(T) \right)}
              {{\det}_{2n} \left(M_2 \Upsilon_2 (0) + N_2
              \Upsilon_2(T) \right)}
              \left[ \frac{{\det}_{2n} \Upsilon_2(0)
              {\det}_{2n} \Upsilon_2(T)}{{\det}_{2n} \Upsilon_1(0)
              {\det}_{2n} \Upsilon_1(T)} \right]^{1/2}\,.
  \end{align}
Here, $M_i, N_i \in \RR^{2n \times 2n}$ impose the boundary
conditions~${\cal A}_i$ for $(\gamma, \zeta) :=
(\gamma, \left( \nabla^2_{\dot{\phi}} L \right) \dot{\gamma}
+ \left( \nabla_{\dot{\phi}} \nabla_\phi L \right) \gamma)$
as
\begin{align}
M_i \left(\begin{array}{c}
\gamma(0)\\
\zeta(0)
\end{array} \right) + N_i \left(\begin{array}{c}
\gamma(T)\\
\zeta(T)
\end{array} \right) = 0\,,
\end{align}
which we assume to guarantee the condition
\begin{align}
\left. \left \langle \zeta_1, \gamma_2  \right \rangle_n
\right \vert^T_0 - \left. \left \langle \gamma_1, \zeta_2
\right \rangle_n \right \vert^T_0 = 0
\end{align}
for all variations $(\gamma_1, \zeta_1)$ and $(\gamma_2,
\zeta_2)$. Further, $\Upsilon_i:[0,T] \to \RR^{2n \times 2n}$
is any fundamental system of solutions of the Jacobi equation
\begin{align}
&\dv{}{t} \left(\begin{array}{c}
\gamma\\
\zeta
\end{array} \right) = J \cdot \nabla^2 H(\phi_i,\theta_i)
\cdot \left(\begin{array}{c}
\gamma\\
\zeta
\end{array} \right)\,.
\label{eq:forman-hamiltonian-ode}
\end{align}
\end{mythm}


\begin{myrmk}
If the second order coefficient matrix $\nabla^2_{\dot{\phi}} L$ does depend
on the path around which the expansion is performed, as is the case for
multiplicative noise in the main text, then considering variations
$\left( \nabla^2_{\dot{\phi}} L\right)^{-1/2} \gamma$ instead of $\gamma$
naturally leads
to the computation of the ratio
\begin{align}
\frac{\Det_{M_1, N_1} \left(  \left(\nabla^2_{\dot{\phi}} L (\phi_1, \dot{
\phi}_1)\right)^{-1} \Omega[\phi_1] \right)}
         {\Det_{M_2, N_2} \left(  \left(\nabla^2_{\dot{\phi}} L (\phi_2, \dot{
\phi}_2)\right)^{-1} \Omega[\phi_2] \right)}
\end{align}
instead, to which the proposition can then be applied without any further
changes (note that for these new operators, the second order coefficient
matrix will be negative unity, and the other coefficients are multiplied
by $\left(\nabla^2_{\dot{\phi}} L (\phi_i, \dot{
\phi}_i)\right)^{-1}$, which yields the same equation
in~\eqref{eq:forman-first-order-ode-general} as before, thereby
leaving~\eqref{eq:forman-hamiltonian-ode} invariant).
\end{myrmk}


\begin{myexpl}
For the Freidlin-Wentzell Lagrangian
\begin{align}
L(\phi,\dot{\phi}) = \frac{1}{2} \left \langle \dot{\phi} -
b(\phi), a^{-1}(\phi) \left(\dot{\phi} - b(\phi) \right) \right \rangle_n\,,
\end{align}
the corresponding Hamiltonian is given by
\begin{align}
H(\phi,\theta) = \left \langle b(\phi), \theta \right
\rangle_n + \frac{1}{2} \left \langle \theta, a(\phi) \theta
\right \rangle_n\,.
\end{align}
The derivatives of $L$ and $H$ are
\begin{align}
\nabla_{\dot{\phi}} L = a^{-1}(\phi) \left(\dot{\phi} - b(\phi)
\right) = \theta\,, \quad \nabla_\phi L = - \nabla b(\phi)^\top
\theta - \tfrac{1}{2} \left \langle \theta, \nabla a(\phi) \theta
\right \rangle_n\,, \quad \nabla^2_{\dot{\phi}} L = a^{-1}(\phi) \nonumber\\
\nabla^2_\phi L = \nabla b(\phi)^\top a^{-1}(\phi)
\nabla b(\phi) - \left \langle \nabla^2 b(\phi), \theta \right
\rangle_n + \nabla b(\phi)^\top a^{-1}(\phi) \left( \nabla a(\phi) \theta \right)^\top
+ \dots \nonumber \\
\dots + \left( \nabla a(\phi) \theta \right) a^{-1}(\phi) \nabla b(\phi)
+ \left \langle \theta, \nabla a(\phi) a^{-1}(\theta) \nabla a(\phi)
\theta \right \rangle_n - \tfrac{1}{2} \left \langle \theta, \nabla^2 a(\phi)
\theta \right \rangle_n  \nonumber\\
\nabla_\phi \nabla_{\dot{\phi}} L = - \nabla
b(\phi)^\top a^{-1}(\phi) - \left(\nabla a(\phi) \theta \right)^\top a^{-1}
(\phi)\,, \quad \nabla_{\dot{\phi}} \nabla_\phi L
= -a^{-1}(\phi) \nabla b(\phi) - a^{-1}(\phi) \left( \nabla a(\phi) \theta \right)\,,
\end{align}
and 
\begin{align}
\nabla_\theta H = b(\phi) + a \theta\,, \quad \nabla_\phi H
= \nabla b(\phi)^\top \theta + \tfrac{1}{2} \left \langle \theta, \nabla
a(\phi) \theta \right \rangle_n\,, \nonumber\\
\quad \nabla^2_\theta H = a(\phi)\,, \quad 
\nabla^2_\phi H = \left \langle \nabla^2 b(\phi), \theta \right
\rangle_n + \tfrac{1}{2} \left \langle \theta, \nabla^2 a(\phi) \theta \right
\rangle_{n}\,, \nonumber\\
\nabla_\phi \nabla_\theta H = \nabla
b(\phi)^\top + \left( \nabla a(\phi) \theta \right)^\top\,,
\quad \nabla_\theta \nabla_\phi H
= \nabla b(\phi) + \left(\nabla a(\phi) \theta \right)\,,
\end{align}
where we use the notation
\begin{align}
\left[ \nabla a(\phi) \theta \right]_{ij} = \partial_j a_{ik}(\phi) \theta_k\,.
\end{align}
Hence
\begin{align}
\Omega[\phi] &= \left(- \dv{}{t} - \nabla b(\phi)^\top - \left(\nabla a(\phi) \theta\right)^\top \right)
a^{-1}(\phi) \left( \dv{}{t} - \nabla b(\phi) - \left(\nabla a(\phi)
\theta\right) \right) \nonumber \\
& \quad - \left \langle
\nabla^2 b(\phi), \theta \right \rangle_n - \frac{1}{2} \left \langle
\theta, \nabla ^2 a(\phi) \theta \right \rangle_n
\end{align}
and
\begin{align}
\Gamma[\phi] = \left(\begin{array}{c|c}
      \nabla b(\phi) + \left(\nabla a(\phi) \theta \right) & a(\phi)\\
      \hline
      - \left \langle \nabla^2 b(\phi), \theta \right
      \rangle_n - \frac{1}{2} \left \langle \theta, \nabla^2 a(\phi) \theta \right
\rangle_{n} & - \nabla b(\phi)^\top - \left(\nabla a(\phi) \theta\right)^\top
    \end{array} \right)\,.
\end{align}
\end{myexpl}

\begin{myexpl}
For the Lagrangian
\begin{align}
L(\phi, \dot{\phi}) = \frac{1}{2} \norm{\dot{\phi}}_n^2 + V(\phi)
\end{align}
appearing in quantum mechanics in imaginary time, with Hamiltonian
\begin{align}
H(\phi, \theta) = \frac{1}{2} \norm{\dot{\phi}}_n^2 - V(\phi)\,,
\end{align}
Jacobi's equation becomes
\begin{align}
\dv{}{t} \left(\begin{array}{c}
\gamma\\\zeta
\end{array} \right) = \left( \begin{array}{c|c}
0_{n \times n} & 1_{n \times n}\\
\hline
\nabla^2 V(\phi) & 0_{n \times n}
\end{array} \right) \left(\begin{array}{c}
\gamma\\\zeta
\end{array} \right)
\end{align}
or
\begin{align}
\left[-\dv[2]{}{t} + \nabla^2 V(\phi) \right] \gamma = 0\,,
\end{align}
which is the classical Gel'fand-Yaglom
formula~\cite{gelfand-yaglom:1960,coleman:1979}.
\end{myexpl}

It is well known from the calculus of variations that, if the Jacobi
equation~\eqref{eq:jacobi} has no
\textit{conjugate points}~\cite{levi:2014} in
$[0,T]$ (\enquote{Jacobi condition}), then it is possible to
construct a solution of a certain
symmetric matrix Riccati differential equation~\cite{reid:1972},
either forward or backward in time,
out of solutions $(\gamma, \zeta):[0,T] \to \RR^{2n \times n}$ (if,
depending on the solution, either $\gamma(t_0)$ or $\zeta(t_0)$ is
invertible for any $t_0 \in [0,T]$ and hence for all $t \in
[0,T]$~\cite{clarke-zeidan:1986}):
\begin{itemize}
\item $W := \zeta \gamma^{-1} :[0,T] \to \RR^{n \times n}$
satisfies the backward Riccati equation
\begin{align}
\dot{W} = - \nabla^2_\phi H - W \nabla_\theta \nabla_\phi H -
\left( \nabla_\phi \nabla_\theta H \right) W - W \left(
\nabla^2_\theta H \right) W\,.
\label{eq:riccati-w-general}
\end{align}
\item $Q = W^{-1} = \gamma \zeta^{-1}:[0,T] \to \RR^{n
\times n}$ solves the forward Riccati equation
\begin{align}
\dot{Q} = \nabla^2_\theta H + Q \nabla_\phi \nabla_\theta H +
\left( \nabla_\theta \nabla_\phi H \right) Q + Q \left(
\nabla^2_\phi H\right) Q\,.
\label{eq:riccati-q-general}
\end{align}
\end{itemize}

\begin{myrmk}
If it exists, the solution of the backward matrix Riccati
equation $W$ can naturally be connected to the positive definiteness
of $\delta^2 S$~\cite{evans:2021}, which is why the fact that Riccati matrix
differential equations appear in the functional determinant computations
is not very surprising from a calculus of variations perspective:
Observing that
\begin{align}
\nabla^2_\phi L = \dot{W} +
\left(W - \nabla_\phi \nabla_{\dot{\phi}} L \right)
\left(\nabla^2_{\dot{\phi}} L\right)^{-1}
\left(W - \nabla_{\dot{\phi}} \nabla_\phi L \right)
\end{align}
and inserting this expression for $\nabla^2_\phi L$ into the second
variation, we obtain, assuming that $\nabla^2_{\dot{\phi}} L$ is
positive definite (\enquote{Legendre condition}),
\begin{align}
\delta^2S[\phi][\gamma] = \frac{1}{2} \left. \left \langle \gamma,
W \gamma \right \rangle_n  \right \rvert_0^T + \frac{1}{2} \int_0^T
\norm{ \left(\nabla^2_{\dot{\phi}} L \right)^{1/2} \left[\dot{\gamma}
- \left( \nabla^2_{\dot{\phi}} L \right)^{-1} \left(W -
\nabla_{\dot{\phi}} \nabla_\phi L \right) \gamma \right]}^2_n \dd t\,,
\end{align}
which can be used to show that $\delta^2 S[\phi][\gamma] > 0$ for
all $\gamma \neq 0$ under appropriate boundary conditions.
\end{myrmk}


\section{Sharp moment-generating function estimate for nondegenerate instantons from WKB analysis for a general Hamiltonian}
\label{app:mgf-ldt-estimate}

As a reference, we state a general sharp estimate for the MGF of a
final-time observable $f:\RR^n \to \RR$
\begin{align}
A_f^\eps(\lambda) = \EE_x \left[ \exp \left\{ \frac{\lambda}{\eps}
f \left( X_T^\eps \right) \right\} \right]
\end{align}
for a $(\eps > 0)$-indexed family of continuous-time Markov processes
$\left(X_t^\eps \right)_{t \in [0,T]}$ with state space~$\RR^n$, deterministic
initial value $X^0 = x \in \RR^n$ and generator~$L_\eps$ which
we assume to satisfy a large deviation principle as~$\eps \downarrow 0$.
Defining~(see e.g.\ \cite{feng-kurtz:2006})
\begin{align}
H_\eps \varphi := \eps \exp \left\{-\frac{\varphi}{\eps}\right\} L_\eps \exp
\left\{\frac{\varphi}{\eps}\right\}
\end{align}
for test functions $\varphi \colon \RR^n \to \RR$ as well as the LDT
Hamiltonian $H \colon \RR^n \times \RR^n \to \RR$, $(\phi, \theta)
\mapsto H(\phi, \theta)$ via
\begin{align}
H(\cdot, \nabla \varphi) = \lim_{\eps \downarrow 0} H_\eps \varphi,
\end{align}
we have the following result, obtained via WKB analysis of the Kolmogorov
backward equation for $A_f^\eps(\lambda)$:


\begin{mythm}[Sharp MGF estimate for nondegenerate instantons from WKB analysis for a general Hamiltonian]
\label{thm-general-mgf-prefac-any-hamiltonian}
The MGF $A_f^\eps$ for $\eps \downarrow 0$ satisfies
\begin{align}
A_f^\eps(\lambda) \overset{\eps \downarrow 0}{\sim} R_\lambda \exp
\left\{\frac{1}{\eps} \left(\lambda f \left(\phi_\lambda(T) \right) -
\int \left \langle \theta_\lambda, \dd \phi_\lambda \right \rangle_n
+ H(\phi_\lambda, \theta_\lambda) T \right) \right\}
\end{align}
with leading-order prefactor
\begin{align}
R_\lambda = \exp \left\{\int_0^T  \left(\left. \dv{}{\eps}
\right|_{\eps = 0} H_\eps S \right)_{(t, \phi_\lambda(t))} \dd t \right\}.
\end{align}
Here, $(\phi_\lambda, \theta_\lambda)$ solve the instanton equations
\begin{align}
\dv{}{t} \left(\begin{array}{c}
\phi_\lambda \\ \theta_\lambda
\end{array} \right) = J \nabla H(\phi_\lambda, \theta_\lambda)
= \left(\begin{array}{c} \nabla_\theta H(\phi_\lambda, \theta_\lambda) \\
- \nabla_\phi H(\phi_\lambda, \theta_\lambda)
\end{array} \right) \,, \quad \phi_\lambda(0) = x\,, \quad
\theta_\lambda(T) = \lambda \nabla f(\phi_\lambda(T))\,,
\label{eq:inst-general}
\end{align}
which are the relevant characteristic for $S \colon [0,T] \times
\RR^n \to \RR$, solving the Hamilton-Jacobi equation
\begin{align}
\partial_t S(t,x) + H(x, \nabla S(t,x)) = 0\,, \quad S(T, x) = \lambda f(x)\,,
\end{align}
such that
\begin{align}
\nabla S\left(t, \phi_\lambda(t)\right) = \theta_\lambda(t)\,.
\end{align}
For the evaluation of the prefactor~$R_\lambda$, the second derivative
of $S$ along the characteristic
\begin{align}
\nabla^2 S\left(t, \phi_\lambda(t)\right) =: W_\lambda(t)
\end{align}
with $W_\lambda \colon [0, T] \to \RR^{n \times n}$ can be found by solving
the backward Riccati equation
\begin{align}
\begin{cases}
\dot{W}_\lambda = - \nabla^2_\phi H - W_\lambda \nabla_\theta \nabla_\phi H -
\left( \nabla_\phi \nabla_\theta H \right) W_\lambda - W_\lambda \left(
\nabla^2_\theta H \right) W_\lambda\,,\\
W_\lambda(T) = \lambda \nabla^2 f(\phi_\lambda(T))\,.
\end{cases}
\label{eq:riccati-general-hamiltonian}
\end{align}
\end{mythm}


\begin{myrmk}
As remarked in~\cite{grafke-schaefer-vanden-eijnden:2021}, it is possible to
transfer the backward to the forward Riccati equation in general
solely on the level of Riccati equations (if both are well-posed for
the problem at hand), the general link being
\begin{align}
\frac{{\det}_n \left(1_{n \times n} - W(T) Q(T) \right)}{{\det}_n
\left(1_{n \times n} - W(0) Q(0) \right)} = \exp \left\{\int_0^T \trace
\left[\left( \nabla^2_\phi H \right) Q - \left( \nabla^2_\theta H \right)
W \right] \dd t \right\}
\end{align}
with $Q$ solving
\begin{align}
\dot{Q} = \nabla^2_\theta H + Q \nabla_\phi \nabla_\theta H +
\left( \nabla_\theta \nabla_\phi H \right) Q + Q \left(
\nabla^2_\phi H\right) Q\,.
\end{align}
\end{myrmk}


\begin{myproof}{Proposition~\ref{thm-general-mgf-prefac-any-hamiltonian}}

Analogously to~\cite{grafke-schaefer-vanden-eijnden:2021}, we define
\begin{align}
u_\eps(T - t, x) = \EE_x \left[\exp \left\{\frac{\lambda}{\eps}
f(X_t^\eps) \right\} \right]\,, \quad u_\eps(T, x) = \exp
\left\{\frac{\lambda}{\eps} f(x) \right\}
\end{align}
such that $A_f^\eps(\lambda) = u_\eps(0, x)$ and $u_\eps$ solves
\begin{align}
\partial_t u_\eps + L_\eps u_\eps = 0\,.
\end{align}
The WKB ansatz
\begin{align}
u_\eps(t, x) = Z_\eps(t,x) \exp \left\{\frac{S(t,x)}{\eps} \right\}\,,
\quad S(T, x) = \lambda f(x)\,, \quad Z_\eps(T, x) = 1\,,
\end{align}
where we later assume that $Z_\eps = Z + {\cal O}(\eps)$, leads to
\begin{align}
\exp \left\{\frac{S}{\eps} \right\} \left[ \frac{1}{\eps} Z_\eps
\partial_t S + \partial_t Z_\eps + \frac{Z_\eps}{\eps} H_\eps \left(S +
\eps \log Z_\eps \right) \right] = 0\,.
\end{align}
Expanding $H_\eps$ yields
\begin{align}
H_\eps \left(S + \eps \log Z_\eps \right) = H(\cdot, \nabla S) + \eps
\left[\left. \dv{}{\eps}
\right|_{\eps = 0} H_\eps S + \left \langle \nabla_\theta H(\cdot, \nabla S),
\frac{\nabla Z}{Z} \right \rangle_n\right] + {\cal O} \left(\eps^2 \right)\,,
\end{align}
so, at order $\eps^{-1}$, we obtain the Hamilton-Jacobi equation
\begin{align}
\partial_t S + H(\cdot, \nabla S) = 0
\label{eq:hjb-general}
\end{align}
for $S$ as expected, which can be solved by the method of characteristics,
yielding the instanton equations~\eqref{eq:inst-general}.
Differentiating~\eqref{eq:hjb-general}
twice and plugging in the characteristics results in the Riccati
equation~\eqref{eq:riccati-general-hamiltonian}.
For the determination of the leading order prefactor~$Z$, we note
that at order $\eps^0$,
\begin{align}
\partial_t Z + \left \langle \nabla_\theta H(\cdot, \nabla S), \nabla Z
\right \rangle_n +  \left(\left. \dv{}{\eps}
\right|_{\eps = 0} H_\eps S \right) Z = 0\,. 
\end{align}
so evaluating $Z(t, x)$ along the characteristic $\phi_\lambda$ where
$\nabla_\theta H(\cdot, \nabla S) = \dot{\phi}_\lambda$ results in
\begin{align}
\dv{}{t} Z(t, \phi_\lambda(t)) = - \left(\left. \dv{}{\eps}
\right|_{\eps = 0} H_\eps S \right)_{(t, \phi_\lambda(t))} Z(t, \phi_\lambda(t))\,,
\end{align}
which can then directly be integrated to get $Z(0, x)$.
\end{myproof}


\begin{myexpl}
For an It{\^o} diffusion
\begin{align}
\begin{cases}
\dd X^\eps_t = b(X^\eps_t) \dd t + \sqrt{\eps} \sigma(X_t^\eps) \dd B_t\,,\\
X^\eps_0 = x
\end{cases}
\end{align}
with the generator $L_\eps$ acting via
\begin{align}
\left(L_\eps f\right)(x) = \left \langle b(x), \nabla f(x) \right \rangle_n
+ \frac{\eps}{2} \trace \left[ a(x) \nabla^2 f(x) \right]\,,
\end{align}
we have
\begin{align}
\left(H_\eps f\right)(x) = \left \langle b(x), \nabla f(x) \right \rangle_n
+ \frac{1}{2} \left \langle \nabla f(x), a(x) \nabla f(x) \right \rangle_n +
\frac{\eps}{2} \trace \left[ a(x) \nabla^2 f(x) \right]\,,
\end{align}
so the Hamiltonian is of course given by
\begin{align}
H(\phi, \theta) = \left \langle b(\phi), \theta \right \rangle_n + \frac{1}{2}
\left \langle \theta, a(\phi) \theta \right \rangle_n\,.
\end{align}
Furthermore, since
\begin{align}
\left(\left. \dv{}{\eps}
\right|_{\eps = 0} H_\eps S \right)_{(t, \phi_\lambda(t))} = \tfrac{1}{2}
\trace \left[ a(\phi_\lambda(t)) W_\lambda(t) \right] = \tfrac{1}{2}
\trace \left[ \nabla_\theta^2 H(\phi_\lambda(t), \theta_\lambda(t)) W_\lambda(t) \right]
\end{align}
and
\begin{align}
\left \langle \theta, \dot{\phi} \right \rangle_n - H(\phi, \theta) =
\frac{1}{2} \left \langle \theta, a(\phi) \theta \right \rangle_n\,,
\end{align}
we indeed arrive at the MGF estimate
\begin{align}
A_f^\eps(\lambda) \overset{\eps \downarrow 0}{\sim} R_\lambda \exp
\left\{\frac{1}{\eps} \left(\lambda f \left(\phi_\lambda(T) \right) - \frac{1}{2}
\int_0^T \left \langle \theta_\lambda, a(\phi_\lambda) \theta_\lambda \right
\rangle_n \dd t  \right) \right\}
\end{align}
with prefactor
\begin{align}
R_\lambda = \exp \left\{ \frac{1}{2} \int_0^T \trace \left[ a(\phi_\lambda)
W_\lambda \right] \dd t \right\} = \frac{\exp \left\{ \frac{1}{2} \int_0^T
\trace \left[ \left( \left \langle \nabla^2 b(\phi_\lambda), \theta_\lambda \right
\rangle_n + \tfrac{1}{2} \left \langle \theta_\lambda, \nabla^2 a(\phi_\lambda)
\theta_\lambda \right
\rangle_{n} \right) Q_\lambda \right] \dd t \right\}}{\left[ {\det}_n \left(1_{n \times n}
- \lambda \nabla^2 f(\phi_\lambda(T)) Q_\lambda(T)\right) \right]^{1/2}}
\end{align}
where the Riccati matrices $W_\lambda, Q_\lambda \colon
[0,T] \to \RR^{n \times n}$ solve
\begin{align}
\begin{cases}
\dot{W}_\lambda &= -W_\lambda a(\phi_\lambda) W_\lambda -
\left[ \nabla b\left(\phi_\lambda
\right)^\top + \left( \nabla a(\phi_\lambda) \theta_\lambda \right)^\top
\right]W_\lambda \\
& \quad- W_\lambda
\left[ \nabla b\left(\phi_\lambda \right) + \left(\nabla a(\phi_\lambda)
\theta_\lambda \right)\right]- \left< \nabla^2
b(\phi_\lambda), \theta_\lambda\right>_n - \tfrac{1}{2} \left \langle
\theta_\lambda, \nabla^2 a(\phi_\lambda) \theta_\lambda \right
\rangle_{n} \,,\\
W_\lambda(T) &= \lambda \nabla^2 f(\phi_\lambda(T))
\in \RR^{n \times n}\,,
\end{cases}
\end{align}
and
\begin{align}
\begin{cases}
\dot{Q}_\lambda &= a(\phi_\lambda) + Q_\lambda \left[ \nabla b\left(\phi_\lambda
\right)^\top + \left( \nabla a(\phi_\lambda) \theta_\lambda \right)^\top \right] \\
& \quad+
\left[ \nabla b\left(\phi_\lambda \right) + \left(\nabla a(\phi_\lambda)
\theta_\lambda \right)\right] Q_\lambda + Q_\lambda \left[ \left< \nabla^2
b(\phi_\lambda), \theta_\lambda\right>_n + \tfrac{1}{2} \left \langle
\theta_\lambda, \nabla^2 a(\phi_\lambda) \theta_\lambda \right
\rangle_{n} \right] Q_\lambda \,,\\
Q_\lambda(0) &= 0_{n \times n}\,.
\end{cases}
\end{align}
\end{myexpl}


\begin{myexpl}
Since previous papers~\cite{schorlepp-grafke-grauer:2021,grafke-schaefer-vanden-eijnden:2021,ferre-grafke:2021,bouchet-reygner:2022} have mostly dealt with
additive noise, we test the more general case of multiplicative noise that is included
here in a simple toy example. Consider the one-dimensional It{\^o} SDE
\begin{align}
\begin{cases}
\dd X_t^\eps = -\beta X_t^\eps \dd t + \sqrt{2 \eps} X_t^\eps \dd B_t\,,\\
X_0^\eps = 1
\end{cases}
\end{align}
describing geometric Brownian motion. Using It{\^o}'s lemma, this SDE can be solved
explicitly to get
\begin{align}
X_t^\eps = \exp \left\{-(\beta + \eps) t + \sqrt{2 \eps} B_t \right\}\,,
\end{align}
and hence the distribution of $X_T^\eps$ is log-normal with PDF
\begin{align}
\rho^\eps(x) = \frac{1}{\sqrt{4 \pi \eps T}} \frac{1}{x}
\exp \left\{- \frac{\left(\log x - (\beta + \eps)T \right)^2}{4 \eps T} \right\}\,.
\end{align}
Choosing
\begin{align}
f(x) = \tfrac{1}{2} \left(\log x\right)^2
\end{align}
as our observable, we can explicitly evaluate the MGF $A_f^\eps(\lambda)$
for $\lambda < 1/(2T)$ by integration of the PDF, obtaining
\begin{align}
A_f^\eps(\lambda) = \underbrace{\left[1 - 2 \lambda T\right]^{-1/2} \exp
\left\{\frac{\beta \lambda T^2}{1 - 2 \lambda T} \right\}}_{= R_\lambda}
\underbrace{\exp \left\{\frac{\eps \lambda T^2}{2 \left( 1 - 2 \lambda T \right)}
\right\}}_{= 1 + {\cal O}(\eps)}  \exp \left\{ \frac{\lambda}{\eps} \frac{\beta^2
T^2}{2 \left( 1 - 2 \lambda T \right)} \right\}\,.
\label{eq:mgf-geom-bm-ex}
\end{align}
We will now reproduce this result at leading order using the general theory
stated above. For the Hamiltonian
\begin{align}
H(\phi, \theta) = (\phi \theta)^2 - \beta \phi \theta\,,
\end{align}
the instanton equations become
\begin{align}
\begin{cases}
\dot{\phi}_\lambda = \pdv{H}{\theta} (\phi_\lambda, \theta_\lambda) =
- \beta \phi_\lambda + 2 \phi_\lambda^2 \theta_\lambda\,, \quad &\phi_\lambda(0) = 1\\
\dot{\theta}_\lambda = -\pdv{H}{\phi} (\phi_\lambda, \theta_\lambda)
= + \beta \theta_\lambda - 2 \phi_\lambda \theta_\lambda^2\,,
\quad &\theta_\lambda(T) = \lambda \frac{\log \phi_\lambda(T)}{\phi_\lambda(T)}\,.
\end{cases}
\end{align}
In addition to the Hamiltonian $H$ being conserved along the instanton,
we can read off that the quantity
\begin{align}
c_\lambda:= \phi_\lambda \theta_\lambda = \theta_\lambda(0)
= \lambda \log \phi_\lambda(T) 
\end{align}
is also conserved. We obtain
\begin{align}
\dot{\phi}_\lambda = \frac{2 c_\lambda^2 - \beta c_\lambda}{\theta_\lambda}
= \left(2 c_\lambda - \beta \right) \phi_\lambda \quad \Rightarrow \quad
\phi_\lambda(t) = \exp\left\{ (2 c_\lambda - \beta) t \right\}
\end{align}
and hence
\begin{align}
c_\lambda = - \frac{\beta \lambda T}{1 - 2 \lambda T}
\end{align}
from the final time condition, the instanton trajectories then being
\begin{align}
\phi_\lambda(t) = \exp \left\{- \frac{\beta t}{1 - 2 \lambda T} \right\}\,,
\quad \theta_\lambda = c_\lambda \exp \left\{\frac{\beta t}{1 - 2 \lambda T} \right\}\,.
\end{align}
The ${\cal O}\left(\eps^{-1} \right)$-contribution of the instanton
in the exponent becomes
\begin{align}
\frac{1}{\eps} \left(\lambda f(\phi_\lambda(T)) - \int_0^T \phi_\lambda^2
\theta_\lambda^2 \dd t \right) = \frac{1}{\eps} \left(\frac{c_\lambda^2}{2\lambda}
- c_\lambda^2 T \right) = \frac{\lambda}{\eps} \frac{\beta^2 T^2}{2 \left( 1
- 2 \lambda T \right)}
\end{align}
as expected. The prefactor at leading order in $\eps$ is
\begin{align}
R_\lambda = \exp \left\{\int_0^T \phi_\lambda^2 W_\lambda \dd t \right\} =:
\exp \left\{\int_0^T \tilde{W}_\lambda \dd t \right\}
\label{eq:prefac-geom-bm}
\end{align}
for
\begin{align}
\begin{cases}
\dot{W}_\lambda = -2 \theta_\lambda^2 -2 (4 c_\lambda - \beta) W_\lambda
- 2 \phi_\lambda^2 W_\lambda^2\,, \\
W_\lambda(T) = \lambda \frac{1 - \log \phi_\lambda(T)}{\left( \phi_\lambda(T) \right)^2}
\end{cases}
\end{align}
and hence, for the transformed Riccati solution $\tilde{W}_\lambda =
\phi_\lambda^2 W_\lambda$, 
\begin{align}
\begin{cases}
\dot{\tilde{W}}_\lambda = -2 c_\lambda^2 - 4 c_\lambda \tilde{W}_\lambda
- 2 \tilde{W}_\lambda^2\,, \\
\tilde{W}_\lambda(T) = \lambda \left(1 - \frac{c_\lambda}{\lambda}\right)\,.
\end{cases}
\end{align}
The solution of this Riccati equation with constant coefficients can easily
be integrated to get
\begin{align}
\tilde{W}_\lambda(t) = -\frac{1 + c_\lambda (C_\lambda -2t)}{C_\lambda - 2t}
= -c_\lambda + \frac{1}{2} \dv{}{t} \log \left(2t - C_\lambda \right)
\end{align}
where the integration constant $C_\lambda$, determined through the final condition, is
\begin{align}
C_\lambda = -\frac{1 - 2 \lambda T}{\lambda}\,.
\end{align}
Evaluating~\eqref{eq:prefac-geom-bm} then reproduces $R_\lambda$ as found
in~\eqref{eq:mgf-geom-bm-ex}.
\end{myexpl}


\section{Prefactor for spatially homogeneous KPZ instantons}
\label{app:kpz-prefac-hom}

In this section, we want to evaluate the term
\begin{align}
R_z &= 
\EE\left[e^{ \tfrac{1}{2} \lambda_z^{\text{hom}} \int_0^{l} \dd x \; Y(x,1)
\nabla^2 f \left( q_z^{\text{hom}}(T) \right) Y(x,1)} e^{\tfrac{1}{4}
\int_0^1 \dd t \int_0^{l} \dd x \int_0^{l} \dd y \; Y(x,t) 2
\delta(x-y) \left( p_z^{\text{hom}}(y, t) \right)^2 Y(x,t)}\right]\nonumber \\
&=\EE \left[ \exp \left\{-\frac{z}{2} \int_0^{l} \dd x \left(\exp
\left\{-z \right\} Y(x, 1)
\right)^2 + \frac{z^2}{2} \int_0^1 \dd t \int_0^{l} \dd x \;
\left(\exp \left\{-zt \right\} Y(x, t) \right)^2\right\} \right]
\end{align}
for the Gaussian fluctuations $Y = \left(Y(x,t) \right)_{x \in [0, l], \; t
\in [0,1]}$ around the spatially homogeneous KPZ instanton~\eqref{eq:gaussian-inst}
for the PDF prefactor in~\eqref{eq:kpz-gauss-pdf}, where we consider the fluctuations
in the Cole-Hopf transformed fields. These fluctuations satisfy the
linear SPDE
\begin{align}
\partial_t Y(x, t) &= \partial_{xx} Y(x,t) + 2 q_z(x,t) p_z(x,t) Y(x,t) +
q_z(x,t) \eta(x, t) \nonumber \\
&= \partial_{xx} Y(x,t) + 2 z Y(x,t) + \exp \left\{zt \right\} \eta(x, t)
\end{align}
with initial condition $Y(\cdot, 0) \equiv 0$. We define the Fourier transform of $Y$ as
\begin{align}
\hat{Y}_k(t) := \frac{1}{l} \int_0^{l} \dd x \; Y(x, t) \exp \left\{-2 \pi i \frac{kx}{l} \right\}
\end{align}
for $k \in \ZZ$, such that
\begin{align}
Y(x,t) = \sum_{k \in \ZZ} \hat{Y}_k(t) \exp \left\{2 \pi i \frac{kx}{l} \right\}\,.
\end{align}
Then $R_z$ becomes
\begin{align}
R_z = \EE \left[ \exp \left\{-\frac{l z}{2} \sum_{k \in \ZZ} \abs{\exp
\left\{-z \right\} \hat{Y}_k(1)}^2 + \frac{l z^2}{2} \int_0^1 \dd t  \sum_{k \in \ZZ} \abs{\exp
\left\{-zt \right\} \hat{Y}_k(t)}^2 \right\} \right]
\end{align}
in terms of the Fourier modes $\left(\hat{Y}_k(t) \right)_{k \in \ZZ, \;
t \in [0,1]}$ solving
\begin{align}
\dv{}{t}\hat{Y}_k(t) = - \left[\left(\frac{2 \pi k}{l} \right)^2 - 2 z\right] \hat{Y}_k(t) + \exp\left\{zt\right\} \hat{\eta}_k(t)\,,
\quad \hat{Y}_k(0) = 0
\end{align}
with white in time and uncorrelated complex Gaussian noise
\begin{align}
\EE \left[\hat{\eta}_k(t) \left(\hat{\eta}_{k'}(t')\right)^*
\right] = l^{-1} \delta_{k,k'}
\delta(t - t')\,,
\end{align}
i.e.\ for $k \neq 0$ the real and imaginary parts of $\hat{\eta}_k$ are independent real
Gaussian variables with variance $(2l)^{-1}$,
and $\hat{\eta}_{-k} = \hat{\eta}_{k}^*$ due to $\eta$ being real.
For $k = 0$, $\text{Im}\, \hat{\eta}_0 \equiv 0$ and $\text{Re}\, \hat{\eta}_0$ has variance $l^{-1}$. Hence (simultaneously rescaling all $\text{Re}\, \hat{\eta}_k$ to unit variance)
\begin{align}
R_z &= \EE \left[ \exp \left\{-\frac{z}{2} \abs{\exp
\left\{-z \right\} \text{Re} \, \hat{Y}_0(1)}^2 + \frac{z^2}{2} \int_0^1 \dd t
\abs{\exp
\left\{-zt \right\} \text{Re} \, \hat{Y}_0(t)}^2 \right\} \right]
\times \nonumber \\
& \quad \times \left[ \EE \left[ \exp \left\{-\frac{z}{2} \sum_{k = 1}^\infty \abs{\exp
\left\{-z \right\} \text{Re} \, \hat{Y}_k(1)}^2 + \frac{z^2}{2} \int_0^1 \dd t
\sum_{k = 1}^\infty \abs{\exp
\left\{-zt \right\} \text{Re} \, \hat{Y}_k(t)}^2 \right\} \right] \right]^2\\
&= \EE \left[ \exp \left\{-\frac{z}{2} \left( \abs{\exp
\left\{-z \right\} \text{Re} \, \hat{Y}_0(1)}^2-z \int_0^1 \dd t
\abs{\exp
\left\{-zt \right\} \text{Re} \, \hat{Y}_0(t)}^2 \right) \right\} \right]
\times \nonumber \\
& \quad \times \left[\prod_{k = 1}^\infty \EE \left[ \exp \left\{-\frac{z}{2} \left( \abs{\exp
\left\{-z \right\} \text{Re} \, \hat{Y}_k(1)}^2 -z \int_0^1 \dd t \, \abs{\exp
\left\{-zt \right\} \text{Re} \, \hat{Y}_k(t)}^2 \right) \right\} \right] \right]^2
\end{align}
For $z = 0$, we have $R_z = 1$ of course, and we start by considering the
case $z < 0$ now where the spatially homogeneous instanton remains the global
minimizer of the action functional for all $z$. Then, rescaling to a standard
real Ornstein-Uhlenbeck process via
\begin{align}
Z = \left[\left(\frac{2 \pi k}{l} \right)^2 + \abs{z}\right]^{1/2} \exp
\left\{-zt \right\} \text{Re} \, \hat{Y}_k\,, \quad s =
\left[\left(\frac{2 \pi k}{l} \right)^2 + \abs{z}\right] t\,, \quad \tilde{\eta} = \left[\left(\frac{2 \pi k}{l} \right)^2 + \abs{z}\right]^{-1/2}\text{Re} \,
\hat{\eta}_k
\end{align}
yields
\begin{align}
R_z &=  \EE \left[ \exp \left\{\frac{1}{2} \left( Z_{\abs{z}}^2 + \int_0^{\abs{z}} \dd s
\, Z_s^2 \right) \right\} \right] \times \nonumber \\
& \quad \times \left[ \prod_{k = 1}^\infty \EE
\left[ \exp \left\{\frac{\abs{z}}{2 \left[\left(\frac{2 \pi k}{l} \right)^2 +
\abs{z} \right]} \left( Z_{\left(\frac{2 \pi k}{l} \right)^2 +
\abs{z}}^2 + \frac{\abs{z}}{\left(\frac{2 \pi k}{l} \right)^2 +
\abs{z}} \int_0^{\left(\frac{2 \pi k}{l} \right)^2 +
\abs{z}} \dd s
\, Z_s^2 \right) \right\} \right] \right]^2
\end{align}
with
\begin{align}
\dd Z_s = -Z_s \dd s + \dd W_s\,, \quad Z_0 = 0\,.
\end{align}
Hence, the problem reduces to the computation of the expectation
\begin{align}
\EE \left[ \exp \left\{\alpha \left( Z_T^2 + 2 \alpha
\int_0^T \dd s \, Z_s^2 \right) \right\} \right]
\end{align}
of a standard one-dimensional Ornstein-Uhlenbeck process with $\alpha, T > 0$.
This problem can be solved using the same functional integration methods as in
the main text, or e.g. by using the Feynman-Kac formula. We follow the latter
strategy here. In order to cover all cases that will appear for positive $z$ as well,
where coefficients $0$ and $+1$ for the Ornstein-Uhlenbeck drift are possible,
we consider 
\begin{align}
E(\alpha,\beta, T) := \EE \left[ \exp \left\{\alpha \left(
\left( Z_T^\beta \right)^2 + 2 \alpha
\int_0^T \dd s \, \left( Z_s^\beta \right)^2 \right) \right\} \right]
\end{align}
with
\begin{align}
\dd Z_s^\beta = - \beta Z_s^\beta \dd s + \dd W_s\,, \quad Z_0^\beta = 0\,.
\end{align}
in the following where $\beta \in \{-1, -0, +1 \}$. Then we know that
\begin{align}
E(\alpha, \beta, T) = \int_{-\infty}^\infty \dd y \; \exp \left\{\alpha y^2 \right\}
K_{\alpha, \beta}(y, T; 0, 0)
\end{align}
where the propagator $K_{\alpha, \beta}(y,s;x,t)$ from point $x$ at time $t$ to point $y$ at time $s$ solves
\begin{align}
\begin{cases}
\partial_s K_{\alpha, \beta}(y, s; 0, 0) = \beta \partial_y
\left(y K_{\alpha, \beta}(y, t; 0, 0)
\right) + \frac{1}{2} \partial_{yy} K_{\alpha, \beta}(y, t; 0, 0) + 2 \alpha^2
y^2 K_{\alpha, \beta}(y, t; 0, 0)\,,\\
K_{\alpha, \beta}(y, 0; 0, 0) = \delta(y)\,.
\end{cases}
\end{align}
A Gaussian ansatz for $K_{\alpha, \beta}$ leads to
\begin{align}
K_{\alpha, \beta}(y, s; 0, 0) = \left[2 \pi Q_{\alpha, \beta}(s) \right]^{-1/2} \exp \left\{
2 \alpha^2 \int_0^s \dd s' \; Q_{\alpha, \beta}(s')\right\} \exp \left\{-\frac{y^2}{2
Q_{\alpha, \beta}(s)} \right\}
\end{align}
with
\begin{align}
\begin{cases}
\dv{}{s} Q_{\alpha, \beta}(s) = 1 - 2 \beta Q_{\alpha, \beta}(s)
+ 4 \alpha^2 Q_{\alpha, \beta}(s)^2\,,\\
Q_{\alpha, \beta}(0) = 0\,.
\end{cases}
\end{align}
The solution of the Riccati equation in the relevant cases that we need are:
\begin{align}
Q_{\alpha, \beta}(s) = \begin{cases}
\frac{\sinh \left(\sqrt{\beta^2 - 4 \alpha^2} \, s \right)}{
\beta \sinh \left(\sqrt{\beta^2 -
4 \alpha^2} \, s \right) + \sqrt{\beta^2 - 4 \alpha^2} \cosh \left(\sqrt{\beta^2 -
4 \alpha^2} \, s \right)}\,, \quad & 4 \alpha^2 < \beta^2\,,\\
\frac{s}{1 + \beta s}\,, \quad & 4 \alpha^2 = \beta^2\,,\\
\frac{\sin \left(\sqrt{4 \alpha^2 - \beta^2} \, s \right)}{
\beta \sin \left(\sqrt{4 \alpha^2 - \beta^2} \,
s \right) + \sqrt{4 \alpha^2 - \beta^2} \cos \left(\sqrt{4 \alpha^2 -
\beta^2} \, s \right)}\,, \quad & 4 \alpha^2 > \beta^2\,,\\
\frac{\tan \left(2 \alpha s \right)}{2 \alpha}\,, \quad & 4 \alpha^2 > \beta^2 = 0\,.
\end{cases}
\label{eq:ric-alpha-beta}
\end{align}
Hence, the expectation is
\begin{align}
E(\alpha, \beta, T) &= \int_{-\infty}^\infty \dd y \; \exp \left\{\alpha y^2 \right\}
K_{\alpha, \beta}(y, T; 0, 0) = \frac{\exp \left\{2 \alpha^2 \int_0^T \dd s' \;
Q_{\alpha, \beta}(s') \right\}}{\sqrt{1 - 2 \alpha Q_{\alpha, \beta}(T)}} \nonumber\\
&=
\begin{cases} 
\left[\frac{\exp \left\{\beta T\right\} \sqrt{\beta^2 - 4 \alpha^2}}{\sqrt{\beta^2
- 4 \alpha^2} \cosh \left(\sqrt{\beta^2 - 4 \alpha^2} \, T\right) + (\beta - 2 \alpha)
\sinh \left(\sqrt{\beta^2 - 4 \alpha^2} \, T \right)} \right]^{1/2}\,,
\quad & 4 \alpha^2 < \beta^2\,,\\
\left[ \frac{\exp\left\{\beta T\right\}}{1 + (\beta - 2 \alpha) T}
\right]^{-1/2}\,, \quad & 4 \alpha^2 = \beta^2\,,\\
\left[\frac{\exp \left\{\beta T\right\} \sqrt{4 \alpha^2 - \beta^2}}{
\sqrt{4 \alpha^2 - \beta^2} \cos \left(\sqrt{4 \alpha^2 - \beta^2} \,
T\right) + (\beta - 2 \alpha)
\sin \left(\sqrt{4 \alpha^2 - \beta^2} \, T \right)} \right]^{1/2}\,,
\quad & 4 \alpha^2 > \beta^2\,,\\
\left[\sqrt{2} \sin \left(\frac{\pi}{4} - 2 \alpha T \right)\right]^{-1/2}\,,
\quad & 4 \alpha^2 > \beta^2 = 0\,.\\
\end{cases}
\label{eq:alpha-beta-cases}
\end{align}
For negative $z$, the cases that appear are
\begin{align}
\begin{cases}
k^2 = 0: \quad \alpha = \frac{1}{2}\,, \; \beta = 1\,, \; T = \abs{z}\,, \;
\text{case } 4 \alpha^2 = \beta^2 \text{ in~\eqref{eq:alpha-beta-cases}}\\
k^2 > 0:  \quad \alpha = \abs{z} / \left( 2\left[\left(\frac{2 \pi k}{l} \right)^2 +
\abs{z}\right] \right)\,,
\; \beta = 1\,, \;  T = \left(\frac{2 \pi k}{l} \right)^2 +
\abs{z}\,, \;
\text{case } 4 \alpha^2 < \beta^2 \text{ in~\eqref{eq:alpha-beta-cases}}\\
\end{cases}
\end{align}
and we thus find
\begin{align}
R_z &= \exp{+\frac{\abs{z}}{2}} \prod_{k= 1}^\infty \left[\exp \left\{\left(\frac{2
\pi k}{l} \right)^2 +
\abs{z}\right\}
\sqrt{1 - \frac{\abs{z}^2}{\left[\left(\frac{2 \pi k}{l} \right)^2 +
\abs{z} \right]^2}}
\left(\sqrt{1 -
\frac{\abs{z}^2}{\left[\left(\frac{2 \pi k}{l} \right)^2 +
\abs{z} \right]^2}} \right. \right. \times \nonumber \\
&\quad\times
\cosh \left(\sqrt{1 -
\frac{\abs{z}^2}{\left[\left(\frac{2 \pi k}{l} \right)^2 +
\abs{z} \right]^2}} \, \left[\left(\frac{2 \pi k}{l} \right)^2 +
\abs{z} \right]\right) + \left( 1 - \frac{\abs{z}}{\left(\frac{2 \pi k}{l} \right)^2 +
\abs{z}} \right) \times \nonumber \\
& \quad \times\left. \left.
\sinh \left(\sqrt{1 - \frac{\abs{z}^2}{\left[\left(\frac{2 \pi k}{l} \right)^2 +
\abs{z} \right]^2}}
\, \left[\left(\frac{2 \pi k}{l} \right)^2 +
\abs{z} \right] \right) \right)^{-1} \right]
\label{eq:she-res-rz}
\end{align}
for the prefactor at negative $z$, which increases monotonically with
increasing absolute value of $z$ and can be seen to be finite for all
$z < 0$. For $z>0$, a similar analysis leads to the following cases:
\begin{align}
\begin{cases}
0 = \left( \frac{2 \pi k}{l} \right)^2 < z: \quad \alpha = -\frac{1}{2}\,, \; \beta = -1\,, \; T = z\,, \;
\text{case } 4 \alpha^2 = \beta^2 \text{ in~\eqref{eq:alpha-beta-cases}}\\
0 < \left( \frac{2 \pi k}{l} \right)^2 = z:  \quad \alpha = -z/2\,, \; \beta = 0\,, \;  T = 1\,, \;
\text{case } 4 \alpha^2 > \beta^2 = 0 \text{ in~\eqref{eq:alpha-beta-cases}}
\\
0 < \left( \frac{2 \pi k}{l} \right)^2 < z:  \quad \alpha = -z / \left(2 \left[z -\left(\frac{2 \pi k}{l} \right)^2 \right] \right)\,,
\; \beta = -1\,, \;  T = z - \left(\frac{2 \pi k}{l} \right)^2\,, \;
\text{case } 4 \alpha^2 > \beta^2 \text{ in~\eqref{eq:alpha-beta-cases}}
\\
0 < z < \left( \frac{2 \pi k}{l} \right)^2:  \quad \alpha = -z / \left(2 \left[\left(\frac{2 \pi k}{l} \right)^2 - z \right] \right)\,,
\; \beta = 1\,, \;  T = \left(\frac{2 \pi k}{l} \right)^2 - z \,, \;
\text{cases } 4 \alpha^2 \gtreqqless \beta^2
\text{ in~\eqref{eq:alpha-beta-cases} possible}\,,
\end{cases}
\label{eq:prefac-cases-pos-z}
\end{align}
and the corresponding $E(\alpha, \beta, T)$'s need to be multiplied together
for each $z$ to get the prefactor.\\

In particular, we can use this result to explicitly find the
critical point $z_{\text{c}} = z_{\text{c}}(l)$ if the dynamical
phase transition is second order. At this point the first
factor, namely for $k = 1$,
diverges and becomes negative; i.e. at the critical observable value the
spatially homogeneous instanton ceases to be a minimizer and transitions
into a saddle. Setting the denominator in the third case
of~\eqref{eq:alpha-beta-cases} to zero for $k = 1$ and $z > 0$,
we find that the critical point is determined via the equation
\begin{align}
\tan \left(\frac{2 \pi}{l} \sqrt{2 z_{\text{c}}(l) -
\left(\frac{2 \pi}{l} \right)^2} \right) + \left(\frac{2 \pi}{l}
\right)^{-1} \sqrt{2 z_{\text{c}}(l) - \left(\frac{2 \pi}{l}
\right)^2} = 0\,. \label{eq:zc-condition}
\end{align}
Focusing on $l = \pi$ as in the main text and numerically determining the smallest nontrivial real solution to~\eqref{eq:zc-condition} yields
\begin{align}
z_{\text{c}}(l = \pi) \approx 2.82588980079639\,.
\end{align}
We remark that for other domain sizes, it is possible that
the transition is first order and hence the point where the prefactor for
the homogeneous instanton diverges is a priori unrelated to the critical
point, or that other modes than $k = 1$ become unstable first.

\bibliography{bib}
\end{document}